\documentclass[12pt,preprint]{aastex}
\usepackage{natbib}
\usepackage{epsf}

\newcommand{\gsim}{\gtrsim}
\newcommand{\lsim}{\lesssim}

\newenvironment{inlinefigure}{
\def\@captype{figure}
\noindent\begin{minipage}{0.999\linewidth}\begin{center}}
{\end{center}\end{minipage}\smallskip}

\slugcomment{ApJ, in prep.}
\shorttitle{Hercules SFH and Structure}
\shortauthors{Sand et al.}

\begin{document}
 \title{The Star Formation History and Extended Structure of the Hercules Milky Way Satellite$\!$\altaffilmark{1}}

\author{David J. Sand,$\!$\altaffilmark{2,3}  Edward W. Olszewski,$\!$\altaffilmark{4} Beth Willman,$\!$\altaffilmark{5} Dennis Zaritsky,$\!$\altaffilmark{4} Anil Seth,$\!$\altaffilmark{3} Jason Harris,$\!$\altaffilmark{6} Slawomir Piatek$\!$\altaffilmark{7} and Abhijit Saha$\!$\altaffilmark{8}} \email{dave.j.sand@gmail.com}

\begin{abstract}
We present imaging of the recently discovered Hercules Milky Way
satellite and its surrounding regions to study its structure, star
formation history and to thoroughly search for signs of disruption.
We robustly determine the distance, luminosity, size and morphology of
Hercules utilizing a bootstrap approach to characterize our
uncertainties.  We derive a distance to Hercules of $133 \pm 6$ kpc
via a comparison to empirical and theoretical isochrones.  As previous
studies have found, Hercules is very elongated, with $\epsilon=0.67\pm0.03$
and a half light radius of $r_{h} \simeq 230$ pc.  Using the color
magnitude fitting package StarFISH, we determine that Hercules is old
($>12$ Gyr) and metal poor ($[Fe/H]\sim-2.0$), with a spread in
metallicity, in agreement with previous spectroscopic work.  We infer
a total absolute magnitude of $M_V=-5.3\pm0.4$.  Our innovative search
for external Hercules structure both in the plane of the sky and along
the line of sight yields some evidence that Hercules is embedded in a
larger stream of stars.  A clear stellar extension is seen to the
Northwest with several additional candidate stellar overdensities
along the position angle of Hercules out to $\sim$35' ($\sim$1.3 kpc).
While the association of any of the individual stellar overdensities
with Hercules is difficult to determine, we do show that the summed
color magnitude diagram of all three is consistent with Hercules'
stellar population.  Finally, we estimate that any change in the
distance to Hercules across its face is at most $\sim$6 kpc; and the
data are consistent with Hercules being at the same distance
throughout.

\end{abstract}

\altaffiltext{1}{Based on data acquired using the
Large Binocular Telescope (LBT).  The LBT is an international
collaboration among institutions in the US, Italy, and Germany.  LBT
Corporation partners are the University of Arizona, on behalf of the
Arizona university system; Instituto Nazionale do Astrofisica, Italy;
LBT Beteiligungsgesellschaft, Germany, representing the Max Planck
Society, the Astrophysical Institute of Postdam, and Heidelberg
University; Ohio State University; and the Research Corporation, on
behalf of the University of Notre Dame, the University of Minnesota,
and the University of Virginia.}
\altaffiltext{2}{Harvard Center for Astrophysics and Las Cumbres Observatory Global Telescope Network Fellow}
\altaffiltext{3}{Harvard-Smithsonian Center for Astrophysics, 60 Garden Street, Cambridge MA 02138}
\altaffiltext{4}{Steward Observatory, University of Arizona, Tucson, AZ 85721}
\altaffiltext{5}{Haverford College, Department of Astronomy, 370 Lancaster Avenue, Haverford PA 19041}
\altaffiltext{6}{Previous address: National Optical Astronomical Observatories, 950 North Cherry Avenue, Tucson, AZ 85726}
\altaffiltext{7}{Department of Physics, New Jersey Institute of Technology, Newark, NJ 07102}
\altaffiltext{8}{National Optical Astronomical Observatories, 950 North Cherry Avenue, Tucson, AZ 85726}

\section{Introduction}

The Sloan Digital Sky Survey has proved a fruitful database for
uncovering extremely low surface brightness satellites of the Milky
Way (MW).  Spectroscopic studies have confirmed ten recently
discovered satellites to be the least luminous ($-6.5 \lsim M_V \lsim
-2$, not including Canes Venatici I; \citealt{sdssstruct}), most dark
matter dominated ($(M/L)_0 \gsim 100$) galaxies known, based on mass
models that assume dynamical equilibrium
\citep{munoz06,simongeha,geha09}.  Recent evidence that all of the MW
dwarf spheroidals occupy a similar mass scale
\citep[e.g.][]{strigari08,Walker09}, despite their very different
luminosities and sizes, suggests that the MW satellites will provide
unique clues to basic astrophysics in simple dark matter potential
wells, and to the formation of the Galactic halo.

The uncertain extent to which the MW's tidal field has played a role
in shaping both the kinematics and luminosity of individual MW
ultra-faint (UF) satellites and their spatial distribution as a
population complicates interpretation of these data.  Direct
morphological arguments, as well as indirect arguments based on the
mass-metallicity relationship, hint that most or all of the five UF
satellites with $d < 50$ kpc may have been affected by tides
\citep[for Willman 1, Segue 1, Bo\"otes II, Ursa Major II, Coma
Berenices, respectively; but see Walker et
al. 2009]{Willman06,Zucker06,Belokurov07,simongeha,sdssstruct}.
Because these five objects are both the nearest and the very least
luminous ($M_V > -4$) UF satellites, even if tides did shape all five
of these they did not necessarily shape the UFs as a population.  It
is important to carefully investigate whether the more distant and
relatively more luminous MW UFs have lost stars to the MW's tidal
field to determine the degree to which tides may have affected the UFs
as a population.

Initial attempts at studying the parameterized structure of the
population as a whole \citep{sdssstruct}, along with their star
formation history (SFH) via color magnitude diagram (CMD) fitting
techniques \citep{sdsssfh}, have provided a basic overview of these
new systems. The SFHs are broadly consistent with old ($>$10 Gyr) and
metal-poor ([Fe/H] $<-2$) stellar populations, with only UMa II, CVn I
and Leo T showing evidence for extended star formation.  Stellar
population studies of UFs can also be used to investigate the extent
to which the MW's tidal field has influenced their structural
properties.  For example, \citet{sdssstruct} showed that the apparent
deviation in some of the UFs from a symmetric distribution can be
explained by shot noise, rather than requiring truly distorted
morphologies.  The UFs do have more elliptical morphologies on average
than the MW dwarf spheroidals known prior to 2003. However, these
studies are arguably limited due to their SDSS-level magnitude limit
-- a level at which many of the new satellites are barely detectable
in the first place.

Deep, follow-up imaging studies of individual satellites, with their
ability to detect many more stars than the discovery data, have
provided tighter constraints on their stellar populations and
structure \citep[e.g.][]{Coleman07,Walsh08,Martin08,leot,Okamoto08}.
However, to date these deeper studies have covered only the central
regions of the UFs, leaving their outer properties unexplored.  For
instance, there has been no deep and wide-field mosaics around the new
systems to characterize their outer structure or to search for
extremely low surface brightness extensions and hyper-faint
companions.  Both observational and theoretical studies of the Local
Group suggest that this is a potentially rich vein of research, with
some MW satellites exhibiting clear substructure
\citep[e.g.][]{Coleman05}.  Others may have their own faint satellites
\citep{leov}, and stellar streams are found throughout the Local Group
\citep[e.g.][]{Ibata94,Belokurov06,Grillmair09}. These low surface
brightness phenomena are expected based on simulations of structure
formation in a cold dark matter-dominated universe, and can be used as
tests of galaxy formation plus dark matter models
\citep[e.g.][]{Kravtsov04,Bullock05}.

The Hercules dwarf galaxy is an excellent candidate for further deep
and wide-field study.  An initial Large Binocular Telescope (LBT)
study by \citet{Coleman07} found Hercules has an ellipticity of
$\epsilon=0.67$, with some indication of tidal debris directly to the
West of the satellite's center.  Its ellipticity is remarkable, given
that spectroscopic work by Simon \& Geha (2007) has shown that
Hercules shows no sign of internal rotation, and a velocity dispersion
of $\sigma\sim$5 km/s.  It is difficult to understand how a stellar
system can have an ellipticity this large and no rotational support.
One solution to this apparent paradox may be that Hercules is not in
dynamical equilibrium; it may instead be severely tidally distorted.  As
with many of the new MW satellites, Hercules appears to be metal poor
($[Fe/H]_{Herc} \sim -2.6$), with an intrinsic $\sigma_{ [Fe/H]}$ of
0.5 dex \citep{Kirby08}.  Hercules is distant, with $d=132$ kpc
\citep{Coleman07}, and is racing away from the Milky Way at 145 km/s
\citep{simongeha}, the highest radial velocity of the new satellites.

Here we present deep photometry of Hercules and surrounding regions
with the Large Binocular Telescope (LBT).  The goal is to perform a
detailed analysis of both the structure and SFH of Hercules.
Additionally, we systematically search for signs of extended structure
both along the line of sight and in the plane of the sky via our
multiple pointings.  The outline of the paper is as follows.  In
\S~\ref{sec:observations} we describe the observations, data reduction
and photometry.  In \S~\ref{sec:hercprop}, we derive basic properties
of Hercules, including its distance, structure and SFH.  We describe
our techniques for searching for extended structure associated with
Hercules in \S~\ref{sec:extendsearch}.  Finally, we discuss our
results and conclude in \S~\ref{sec:discuss}.

\section{Observations and Data Reduction}\label{sec:observations}

\begin{figure*}
\begin{center}

\mbox{\epsfysize=5.0cm \epsfbox{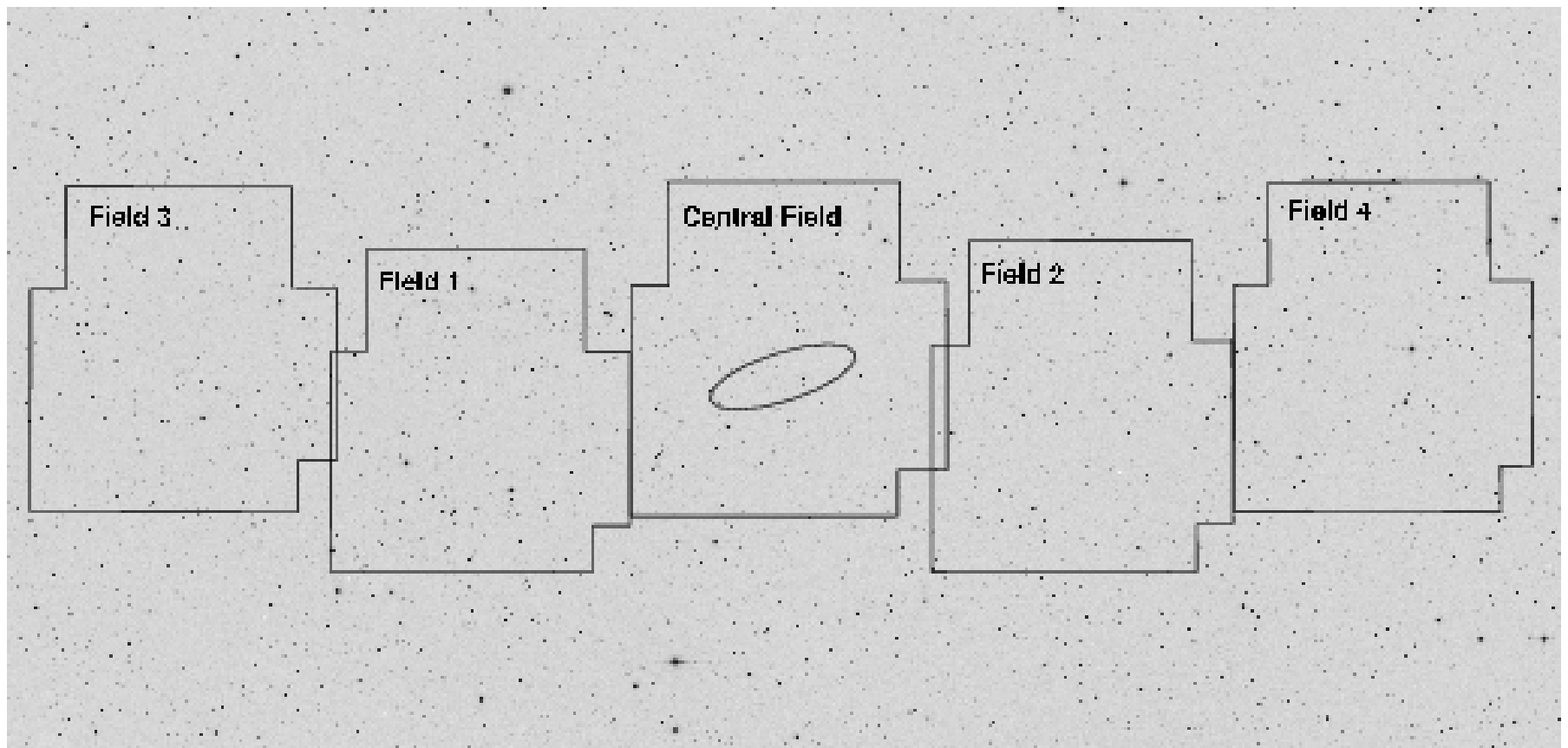}}

\caption{An outline of our five LBT pointings on a Digital Sky Survey
image backdrop.  The ellipse in the central pointing shows the
orientation and half light radius of Hercules, as determined for an
exponential profile in \S~\ref{sec:structparams}.  For a sense of
scale, each LBT pointing is roughly 23' along its base.  North is up
and East is to the left.
\label{fig:LBTlayout}}
\end{center}
\end{figure*}

Our observing strategy was to get deep, wide field $B$ and $r$ band
imaging of the Hercules dwarf spheroidal in order to study its
extended structure and SFH.  For our central
pointing of Hercules, we have also used $V$ band imaging, as presented
in \citet{Coleman07}.  We split our imaging between fields on and
adjacent to Hercules, situated roughly along the major axis.  In all
we obtained five fields, whose orientation is shown in
Figure~\ref{fig:LBTlayout}.  

Observations of Hercules were, with one exception, taken during May
and June 2008 during normal operations of the Large Binocular
Telescope, fitted with the red and blue channel of the Large Binocular
Camera \citep[LBC;][]{Ragazzoni06}.  The $B$ and $V$ band images for the
central Hercules field were taken during Science Demonstration Time
(SDT) and were presented in \citet{Coleman07}.  During this period,
only the blue channel of the LBC was employed, but otherwise the
camera set up was identical.  LBC consists of two nearly identical
prime focus imagers, one for each of the LBT's 8.4 meter mirrors, with
one optimized for blue and one for red wavelengths.  Each camera has
four 2048$\times$4608 pixel CCDs, sampled at 0.23 arcsec/pixel and a
$\sim$23'$\times$23' field of view.

For each of our fields, we sought six 300s dithered exposures in both
bands.  After experimentation, we found that we could improve our
point spread function (PSF) photometry via DAOPHOT \citep{Stetson94}
by including only the best four or five out of six frames in the
analysis, and did so when necessary.  This was due either to a
strongly variable PSF during an imaging sequence, or a slightly out of
focus frame.  Neither of these issues effected the ultimate quality of
our reduced data, once it was properly culled.  A summary of
observations can be found in Table~\ref{table:obsinfo}.

\subsection{Data Reduction}

Basic image reductions were performed in two parts and the process was
identical for the $B$, $V$ and $r$ bands.  First, initial reductions
were executed using the {\it mscred} mosaic data reduction system in
{\sc IRAF}.  This initial script trims and subtracts the overscan
region, applies an additional bias subtraction to remove structure
seen in the bias exposures, flat-fields the data and rejects cosmic
rays in each individual exposure using the {\sc LACOSMIC} task
\citep{vandokkum01}.  Saturated objects were masked along with a
growth radius of three pixels.  A $\sim$1k$\times$1k region in the
extreme southwest corner of the red channel image array was excised
due to poorer image quality, greatly improving overall point source
photometry fits while only negligibly impacting the total area
studied.  Flat fields were generated by median combining
flux-scaled twilight flats.  An exposure weight map is calculated by
combining the normalized flat field and the bad pixel map generated
via cosmic ray rejection.  This weight map is fed into the script
which performs the next steps of the reduction process, and is
specifically used by both the astrometric correction software {\sc
scamp} and the program {\sc SWarp}, which resamples and coadds the
images.  The weight maps are not used in the determination of point
source photometry, which is discussed in \S~\ref{sec:instphotom}.

These flat-fielded images along with their accompanying weight maps
are fed into the publicly available code {\sc scamp} \citep{scampbib}
to determine the astrometric solution.  Given the $f/1.14$ focal
ratio, there is significant image distortion across the field of view,
and a 3rd-degree polynomial fit is utilized to correct this.  The
astrometric catalog used was from the sixth data release of the Sloan
Digital Sky Survey (SDSS-DR6) \citep{sdsscite}, and the $g$ band was
used as the reference band for the $B$ band exposures.  Final
astrometric solutions were good to $\sim$0\farcs1 rms.

Once a good astrometric solution was found and placed into the image
headers, the image resampling and coaddition software {\sc
SWarp}\footnote{version 2.15.7; http://terapix.iap.fr/soft/swarp} is
employed.  The {\sc lanczos3} interpolation function was used for
image resampling, which preserves source signal while minimizing
artifacts near image discontinuities, such as saturation trails.  For
the image coaddition we used a weighted average of the input images,
which is most appropriate for detection of faint sources.

\subsection{Instrumental Photometry}\label{sec:instphotom}

Stellar photometry was performed on the final image outputs from {\sc
SWarp} similarly to \citet{Harris07}, using the command line version
of the {\sc DAOPHOTII/Allstar} package \citep{Stetson94}.  We allowed
for a quadratically varying PSF across the field when determining our
model PSF.  Similar to \citet{Harris07}, we ran {\sc Allstar} in two
passes; once on the image and then again on the image with the first
round's stars subtracted, so that fainter sources can be recovered.
The {\sc Allstar} catalogs for each imaging band were culled of
outliers in $\chi^{2}$ vs. magnitude, magnitude error versus magnitude
and sharpness versus magnitude space to remove objects that were
erroneously selected as point sources in {\sc DAOPHOT}.  The point
source $B$ and $r$ catalogs (along with the $V$ band for the central
pointing) were positionally matched with a maximum match radius of
0\farcs5.  Only those sources detected in both bands (or all three
bands in the central pointing) are placed into our final catalog.

\subsection{Photometric Calibration and inclusion of SDSS data}

Calibrating the instrumental magnitudes output by our stellar
photometry analysis onto a standard photometric system was done using
stars in common with SDSS-DR6\footnote{http://cas.sdss.org/DR6/en/}.
For the $r$-band, this calibration was done by matching to point
sources with $19.5 < r < 21.0$.  We fit a zero-point and a color term,
with a total photometric uncertainty of $\delta r \sim 0.03 - 0.04$
mag depending on the pointing.  For the $B$ band, we used the
relations found by \citet{Jordi06} to convert from SDSS magnitudes,
again over point sources with $19.5 < r < 21.0$ and $19.5 < g < 21.0$.
The total photometric uncertainty is $\delta B \sim 0.05$ mag.  Color
terms were found and eliminated in $(B-B_{LBT})$ versus
$(B_{LBT}-r_{LBT})$, $(V-V_{LBT})$ versus $(B_{LBT}-V_{LBT})$ and
$(r-r_{LBT})$ versus $(B_{LBT}-r_{LBT})$ space; the linear slope of
these terms was 0.09, $-0.03$ and $-0.02$ mag, respectively.  There were
slight residual zeropoint gradients across the field of view, and it
was necessary to fit a quadratic function to the zeropoint as a
function of chip position to achieve the above zero-point
uncertainties. This was done for each individual pointing.

When necessary, we adopted SDSS photometry directly for stars brighter
than and near the saturation limit of a given LBT field (which depends
both on the observing conditions and point spread function).  In this
case, $g$ and $r$ magnitudes are once again converted to $B$
magnitudes via the relations found by \citet{Jordi06}.

All reported magnitudes are corrected for Galactic extinction with the
values from the \citet{Schlegel98} dust maps, using the IDL routine
{\sc dust\_getval}.  Specifically, we used $A_{B}=4.315E(B-V)$ and
$A_{r}=2.751E(B-V)$.  Unless stated otherwise, all magnitudes reported
in the remainder of this paper will be extinction corrected.

\subsection{Artificial Star Tests}\label{sec:ASTs}

Artificial star tests were used to measure both our photometric errors
and completeness as a function of magnitude and color, with a
methodology analogous to that presented by \citet{Walsh08}.  First,
artificial stars are injected into the original images based on the
point spread function (PSF) measured by {\sc DAOPHOT} with the routine
{\sc ADDSTAR}.  Artificial stars were placed on a regular grid with
spacing between ten and twenty times the full width at half maximum,
so that the overlap between artificial stars is negligible.  Given the
geometry of the LBC field of view, this allows for $\sim$5000-20000
artificial stars per iteration.  In order to build up sufficient
statistics, 10 iterations were performed per field for a total of
$\sim$100000 artificial stars each.  The $r$ magnitude of the
artificial stars is drawn randomly from 18 to 29 mag, with an
exponentially increasing probability toward fainter magnitudes.  The
$B-r$ color is then randomly assigned over the range $-$0.5 to 1.5 mag,
with uniform probability.  The artificial star frames are run through
the same photometry pipeline as the science frames, with identical
$\chi^{2}$, sharpness and error on the magnitude cuts.  Also, a given
star must be detected in both the $B$ and $r$ band ($B$,$V$ and $r$ in
the central field) to be considered a true detection.  The 50\%, 90\%
and 95\% completeness limits for each field is detailed in
Table~\ref{table:obsinfo}.

\subsection{Final Hercules Catalog}

The final step in preparing our Hercules photometric catalog was the
combination of the individual catalogs from the five separate image
pointings.  Since there was some overlap between the pointings, we
chose the photometry with the lower formal photometric error to put in
the final catalog.  We directly compared the photometry of objects
detected in more than one pointing and found them to be consistent
with the uncertainty in the recovered magnitudes found in our
artificial star tests.

We present our full Hercules catalog in Tables~\ref{table:centphot}
and \ref{table:fieldphot}.  Table~\ref{table:centphot} focuses on our
central pointing and includes our $B,V,$ and $r$ band magnitudes
(uncorrected for extinction) with their uncertainty.  We also include
the extinction values derived for each star and whether or not the
star was taken from the SDSS catalog and converted to $B$ and $V$ via
the relations of \citet{Jordi06}, rather than from our LBT data.
Table~\ref{table:fieldphot} is similar, and includes our data from our
adjacent fields, which is noted in its own separate column.

\section{Hercules Properties}\label{sec:hercprop}

Figure~\ref{fig:cent_cmds} shows the CMD of stars within 5.9
arcminutes of the center of Hercules, the half light radius for an
exponential profile parameterization (\S~\ref{sec:structparams}).
Using this data, we will measure the distance, structural properties,
and SFH of Hercules.

\subsection{Distance}\label{sec:dist}

Most recently, the distance to Hercules has been calculated in the LBT
study of \citet{Coleman07}, and was found to be $132\pm12$ kpc
($m-M=20.6\pm0.2$).  For completeness, along with our study of
Hercules' extended structure and SFH, we
reinvestigate the distance to Hercules.

We proceed by comparing Hercules' CMD with empirical globular cluster
fiducials and theoretical isochrones utilizing a bootstrap technique
analogous to \citet{Walsh08}. We use four empirical fiducials from
\citet{Clem08} recently imaged in Sloan $g'$ and $r'$: M92, M3, M13
and M71 which span a range of metallicity $-2.4 < [Fe/H] < -0.7$.  We
take $m-M$=14.60, 15.14, 14.42 and 13.71
\citep{Paust07,Kraft03,Cho05,Grundahl02} and
$E(B-V)$=0.022,0.013,0.017,0.308 for the four clusters respectively.
The \citet{Clem08} fiducials were converted from $g'$,$r'$ to $g,r$
using the transformation of \citet{Rider04} and then to $B,r$ using
the transformation of \citet{Jordi06}.  Besides these four empirical
fiducials, theoretical isochrones were taken from \citet{Dotter08} and
\citet{Girardi04}.  The two Dotter isochrones used were for a
$[Fe/H]=-2.49$ and $[Fe/H]=-1.5$, 15 Gyr stellar population, while the
Girardi isochrones were for a $[Fe/H]=-2.3$ and $[Fe/H]=-1.7$ 15 Gyr
stellar population.  The following technique is robust only if the
underlying stellar population of Hercules is old, which we discuss
further in \S~\ref{sec:starform}.  Note that if Hercules has a spread
in metallicities, as indicated by \citet{Kirby08} and for which we
present evidence in \S~\ref{sec:starform}, then the distance modulus
will have an additional uncertainty, which we will try to quantify
later in this section.

We include all stars within $r_h$=5.9' of the centroid of Hercules,
taken via the best-fitting exponential profile (see
\S~\ref{sec:structparams} and Table~\ref{table:paramfits}), down to
$r=25.5$ in our analysis.  Restricting ourselves to stars with $r <
24.5$ does not change the result.  To determine the best fit distance
modulus, each fiducial is stepped through 0.025 magnitude intervals in
($m-M$) from 19.5 to 21.5, noting the number of stars consistent with
(taking into account photometric uncertainties) that of the fiducial.
To account for background/foreground contamination, we then calculate
the same numbers for stars in an equal area box 12 arcminutes north of
the centroid of Hercules (since Hercules is oriented nearly East-West
and is highly elongated, there should be little Hercules contamination
at this position) and subtract it from the Hercules-centered result.
The best fit distance modulus is that which maximizes the number of
Hercules stars.

The best-fit distance moduli for the M92, M3, M13, and M71 fiducials
are 20.625, 20.375, 20.25 and 20.925, with 894, 843, 850 and 877
stars, respectively.  Note that if we perform a similar analysis on
the $B-V$ versus $V$ CMD, we achieve a nearly identical result, with
the M92 CMD being the best fit with a distance modulus of 20.60.  For
the theoretical isochrones, the Dotter $[Fe/H]=-2.49$ 15 Gyr isochrone
yields a 20.60 distance modulus with 895 stars while the Dotter
$[Fe/H]=-1.5$ 15 Gyr isochrone has a distance modulus of 20.2 and 819
stars.  Likewise, the Girardi $[Fe/H]=-2.3$, 15 Gyr isochrone is at a
distance modulus of 20.65 with 892 stars and the Girardi
$[Fe/H]=-1.7$, 15 Gyr isochrone is at 20.2 with 842 stars.  

\begin{figure*}
\begin{center}
\mbox{
\mbox{\epsfysize=6.0cm \epsfbox{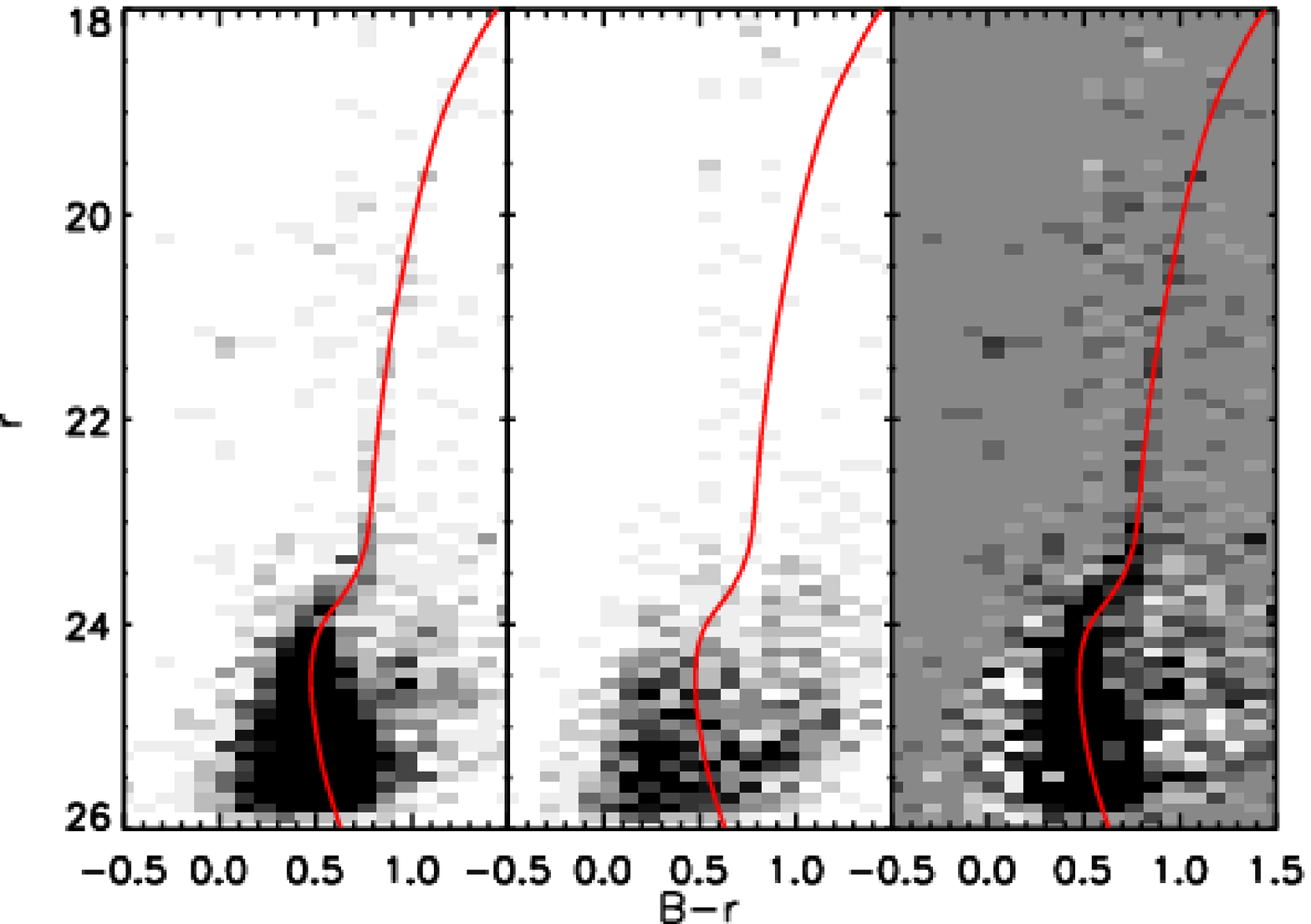}}
\mbox{\epsfysize=6.0cm \epsfbox{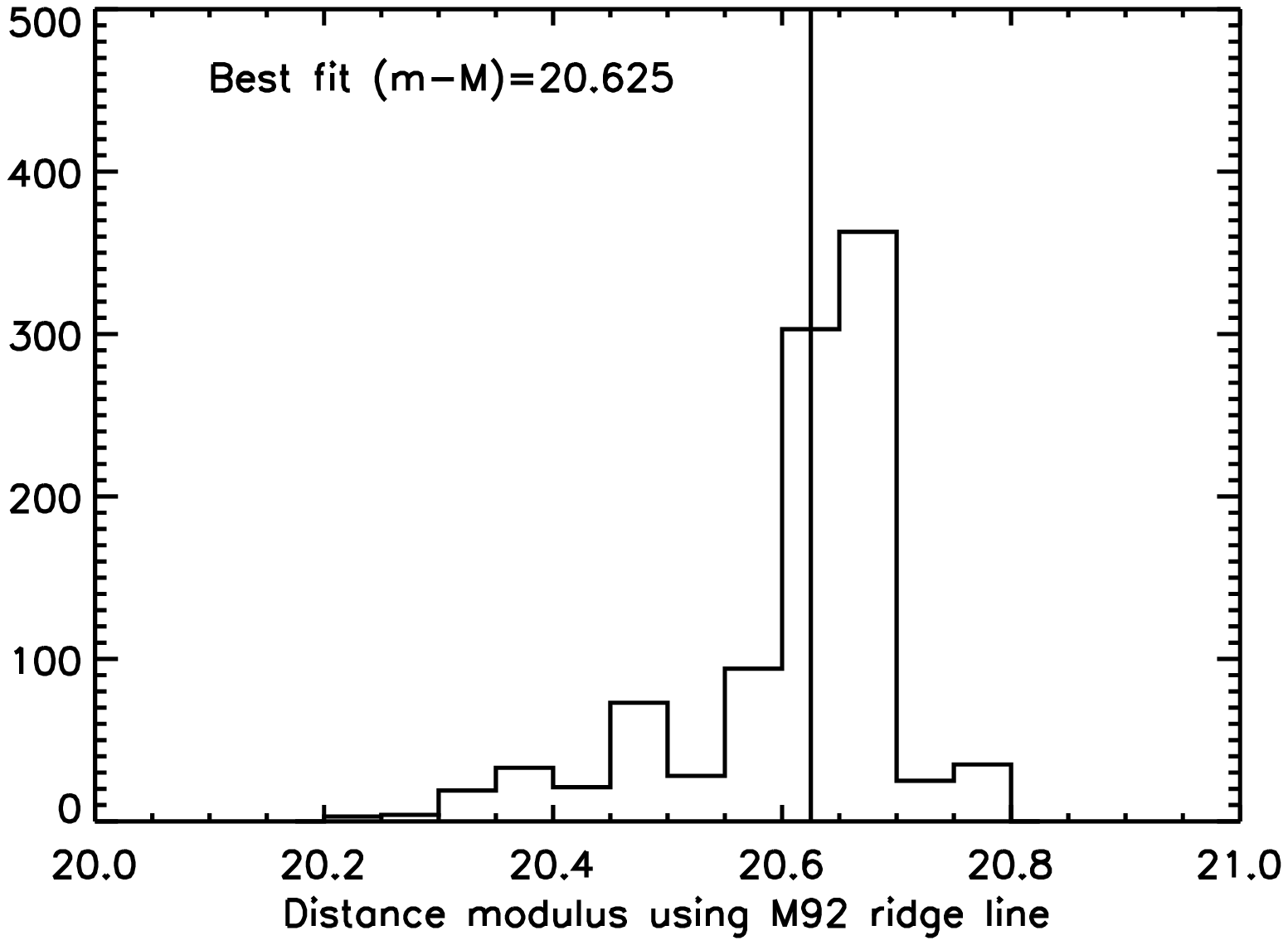}}
}

\caption{Results of our empirical determination of Hercules' distance
modulus using an M92 fiducial. {\bf Left --} A three-panel Hess
diagram of the region within one half light radius of Hercules.  The
left panel shows all stars within this radius, the center panel shows
an equivalent area background Hess diagram while the right panel shows
the background subtracted Hess diagram of Hercules.  Overplotted in
red in all three panels is an M92 fiducial transformed to a distance
modulus of 20.625.  {\bf Right --} Histogram results of our bootstrap
resampling error analysis for the distance modulus of Hercules.
Eighty percent of the bootstrap resamples are within
($m-M$)=$20.625\pm0.1$ and so we adopt this as our conservative
uncertainty on the measurement.  \label{fig:bestdm}}
\end{center}
\end{figure*}

Clearly, old and metal poor isochrones provide the best fit to the
Hercules CMD, with the M92, Dotter $[Fe/H]=-2.49$ and Girardi
$[Fe/H]=-2.3$ isochrones all giving similar results.  In
Figure~\ref{fig:bestdm}, we present a Hess diagram of the central
$r_h$=5.9' of Hercules with background subtracted, along with the M92
fiducial adjusted to ($m-M$)=20.625.  The fit is excellent, and nearly
identical for the two good theoretical isochrones as well.  As an
exercise, if we force isochrones with $[Fe/H]\sim-1.5$ to the best fit
distance modulus of the M92 fit, as we do for illustrative purposes in
Figure~\ref{fig:DMwm13} using the M13 isochrone, we see that they
provide a poorer match to the Hercules CMD, but we can not rule out
that a fraction of the Hercules CMD belongs to a slightly more metal
rich population -- a fact we will return to in \S~\ref{sec:starform}.

For most of this work we choose to adopt the best-fit distance modulus
found for the empirical globular fiducial, M92, of ($m-M$)=20.625.
However, in \S~\ref{sec:starform}, where we attempt to fit the star
formation history of Hercules using a set of theoretical isochrones
from \citet{Girardi04} to the observed CMD, we use the best fit
Girardi $[Fe/H]=-2.3$ distance modulus of 20.65.

\subsection{Structural Parameters}\label{sec:structparams}

It is traditional to fit the surface density profile of both globular
clusters and dSphs to King \citep{King66}, Plummer \citep{Plummer11},
and exponential profiles.  While real MW satellites have a complexity
that is difficult to characterize with parameterized models, it is
nonetheless important to facilitate comparisons with other
observational studies and for studies of the faint MW satellites as a
population \citep[e.g.,][]{Martin08}.  We fit all three profiles to
the stellar distribution of Hercules:

\begin{equation}
\Sigma_{King}(r) = \Sigma_{0,K}\left( \left(1+\frac{r^2}{r_c^2}\right)^{-\frac{1}{2}}-\left(1+\frac{r_t^2}{r_c^2}\right)^{-\frac{1}{2}}\right)^2
\end{equation}

\begin{equation}
\Sigma_{Plummer}(r) =  \Sigma_{0,P}\left(1+\frac{r^2}{r_P^2}\right)^{-{2}}
\end{equation}

\begin{equation}
\Sigma_{exp}(r) =  \Sigma_{0,E}\exp\left(-\frac{r}{\alpha}\right)
\end{equation}

\noindent where $r_P$ and $\alpha$ are the scale lengths for the
Plummer and exponential profiles and $r_c$ and $r_t$ are the King core
and tidal radii, respectively. For the Plummer profile, $r_P$ equals
the half-light radius $r_h$, while for the exponential profile $r_h
\approx 1.668\alpha$.  For this investigation, we use all stars in the
central field of Hercules which are consistent with the ($m-M$)=20.625
M92 fiducial, taking into account our photometric uncertainties.  The
four outlying fields were not used due to their different depths and
completeness.  For the King profile, there is a degeneracy between the
truncation radius and the background surface density.  We thus fix the
background value to the average of that found for the Plummer and
exponential profiles for our King profile fits
\citep[e.g.][]{Walsh08}.

We use a maximum likelihood (ML) technique for constraining structural
parameters similar to that of \citet{Martin08}, and point the reader
to that work for further details concerning the expression of the
likelihood function.  Whereas \citet{Martin08} use an iteratively
refined grid to find the ML, we use the amoeba simplex
algorithm \citep{Press88}, restarted five times in order to ensure
that the ML is reached, although it generally
converges after three restarts.  Including the central position of
Hercules, $\alpha_{0}$ and $\delta_{0}$, position angle ($\theta$),
and ellipticity ($\epsilon$) both the exponential and Plummer profiles
have the same free parameters --
($\alpha_{0}$,$\delta_{0}$,$\theta$,$\epsilon$,$r_{half}$,$\Sigma_{b}$),
while the King profile free parameters are
($\alpha_{0}$,$\delta_{0}$,$\theta$,$\epsilon$,$r_{c}$,$r_{t}$).
Uncertainties on structural parameters are determined through 1000
bootstrap resamples, from which a standard deviation is calculated.

Our results are presented in Table~\ref{table:paramfits}.  We show our
best fit stellar profiles in Figure~\ref{fig:stellarprofile}.
Although the plotted stellar profiles are not fit to the plotted
binned data points, they do show excellent agreement.  For
illustration and comparison with the SDSS data set (see below), we
show our bootstrap histogram for our exponential profile fit for
$r_{h}$ in Figure~\ref{fig:sdsscomp}.  These Hercules parameters
are in good agreement with the LBT data of \citet{Coleman07}, with
nearly identical ellipticity, position and position angle.  There is
some confusion as to the literature value of the half light radius of
Hercules.  Originally, \citet{Coleman07} found a half light radius of
$r_{h} = 4.37' \pm 0.29'$ via their King profile fit.  Using only SDSS
data, \citet{Martin08} found $r_{h} = 8.6_{-1.1}^{+1.8}$' and reported
via a private communication that the half light radius of Hercules
derived by \citet{Coleman07} was $9.4 \pm 1.4'$ versus the originally
reported value. We believe that our half light radius is in agreement
with the SDSS-only data, which are plagued by high background and
small numbers, which we discuss now.

To illustrate the gain in parameter constraints made via the deep
Hercules photometry, we repeat our analysis with only the SDSS data,
fitting an exponential profile.  In this, we seek to mimic the
analysis of \citet{Martin08}, taking all SDSS stars within 1 degree of
Hercules, and make magnitude cuts at $r < 22.0$ and $g < 22.5$, while
selecting Hercules stars that are consistent with the M92 isochrone
shifted to $(m-M)$=20.625.  We find excellent agreement with the
Hercules results of \citet{Martin08}, albeit with larger uncertainties
than their published numbers.  In Figure~\ref{fig:sdsscomp}, we
compare the bootstrap $r_h$ histograms from the LBT and SDSS data,
although this comparison is not necessarily fair given the different
fields of view that each data set was taken from.  This clearly
illustrates the need for deep photometry of all of the new faint MW
dwarfs in order to strongly constrain their structural parameters.

As first reported by \citet{Coleman07}, the ellipticity of
$\epsilon=0.67$ is remarkable.  Additionally, kinematics results of
Hercules indicate a $\sigma \sim 5 km/s$, with no sign of rotation,
although there is some tentative evidence that there may be some
kinematic substructure \citep{simongeha}. Taken together, this may
suggest that Hercules is disrupting or a stellar enhancement in an
unidentified stream.  We search for signs of extended structure in our
Hercules fields in \S~\ref{sec:extendsearch}.

\subsection{Star Formation History}\label{sec:starform}

It is important to understand the SFH and metallicity evolution of the
new dwarfs, since they may provide important clues to the formation
and assembly of the Local Group and may serve as a comparison to
cosmological simulations.  One technique for doing this is via
CMD-fitting, which has already been employed to some degree to study
the new SDSS dwarf galaxies \citep[e.g.,][]{leot,sdsssfh}.

Here we apply the CMD-fitting package StarFISH \citep{starfish} to our
photometry of stars within the half light radius ($r_h$=5.9') of
Hercules to determine its SFH and metallicity evolution.
Conceptually, StarFISH uses theoretical isochrones, taken from
\citet{Girardi04,Girardi02} (although in practice any set of
isochrones can be used), to construct a set of artificial CMDs with
different combinations of distance, age, and metallicity.  Utilizing
the observed photometric errors and completeness (obtained from
artificial star tests in \S~\ref{sec:ASTs}), these theoretical CMDs
can be converted into realistic model CMDs which can be compared
directly to the data.  Conversion of both the data and the model into
Hess diagrams enable a pixel-to-pixel comparison.  The best fitting
linear combination of model CMDs is determined through an efficient
downhill simplex algorithm, and uncertainties are evaluated by
examining the parameter space about the best-fit.  See
\citet{starfish} for details of the algorithm.  We should note that
StarFISH has been shown to give very similar results as MATCH
\citep{leot}, another CMD-fitting software package with a slightly
different implementation \citep{match,sdsssfh}.

We include isochrones with [Fe/H]=$-$2.3,$-$1.7, $-$1.3, $-$0.7, and
$-$0.4 and ages between $\sim$10 Myr and $\sim$16 Gyr.  Age bins of
width $\Delta log (t)$=0.4 dex were adopted, except for the two oldest
age bins at $\sim$10 Gyr and $\sim$14 Gyr, where the binning was
$\Delta log (t)$=0.3 dex.  We have included a 'foreground' CMD, taken
from a region 10 arcminutes north of the center of Hercules measuring
16 by 7 arcminutes, in order to correct for contamination by
foreground stars.

Two CMDs were fit simultaneously; $B-V$ versus $V$ and $B-r$ versus
$r$.  Stars with colors in the range $-0.5 < B-V < 1.5$ and $-0.5 <
B-r < 1.5$ were fit.  The magnitude range included all stars brighter
than $r=26$ and $V=26$.  The Hess diagram bin size was 0.1 in
magnitude and 0.1 in color.  We assume a Salpeter initial mass
function and a binary fraction of 0.5.  These stars were taken from
our final three band Hercules catalog, and so have already been
corrected for foreground extinction with the dust extinction maps of
\citet{Schlegel98}.  We have chosen to fix the distance modulus in the
code to $m-M=20.65$, the best distance modulus found for the Girardi
ischrones in \S~\ref{sec:dist}, although our results our robust with
respect to this assumption (see below).

\begin{figure*}
\begin{center}
\mbox{
\epsfysize=7.0cm \epsfbox{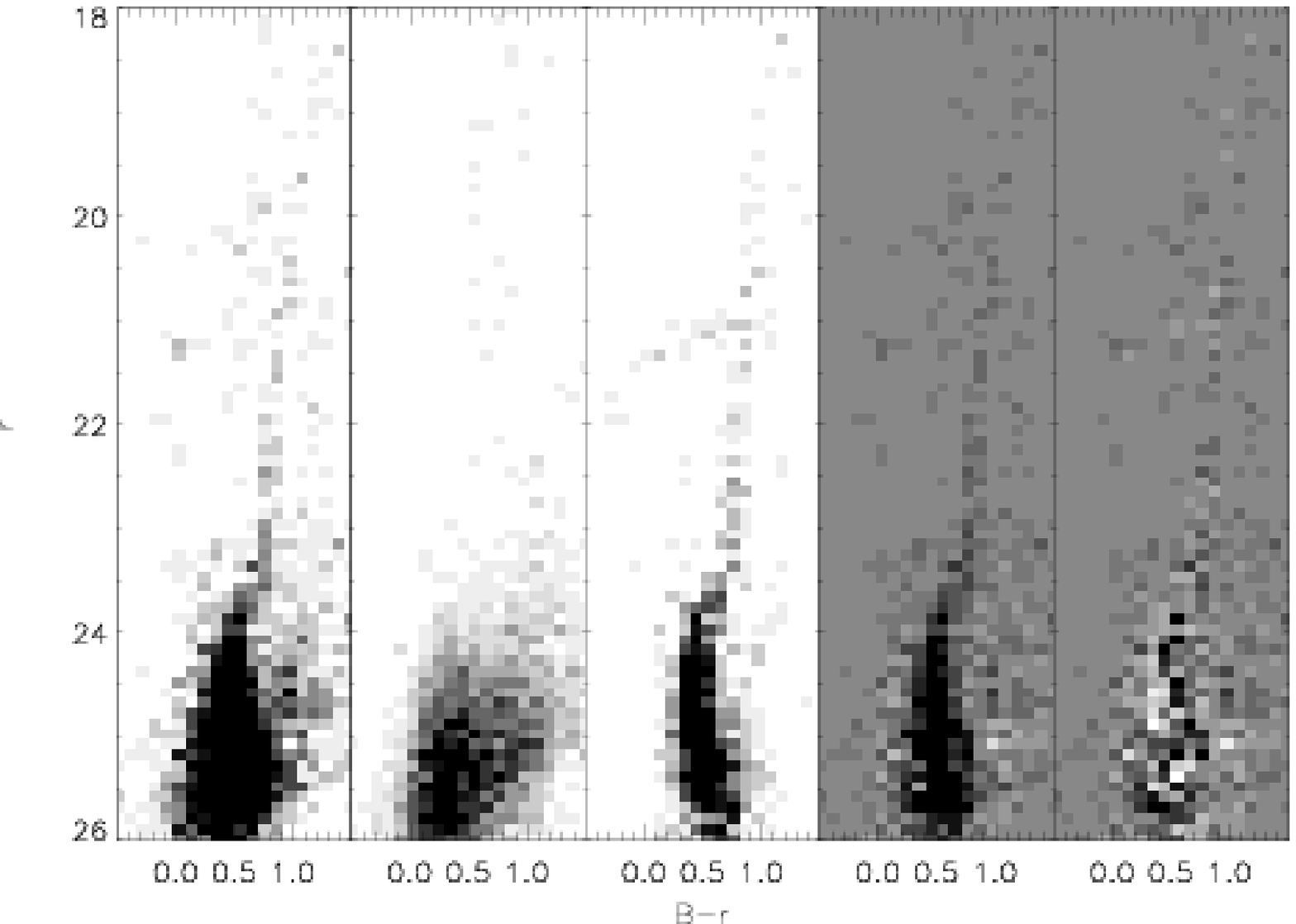}
}
\mbox{
\epsfysize=7.0cm \epsfbox{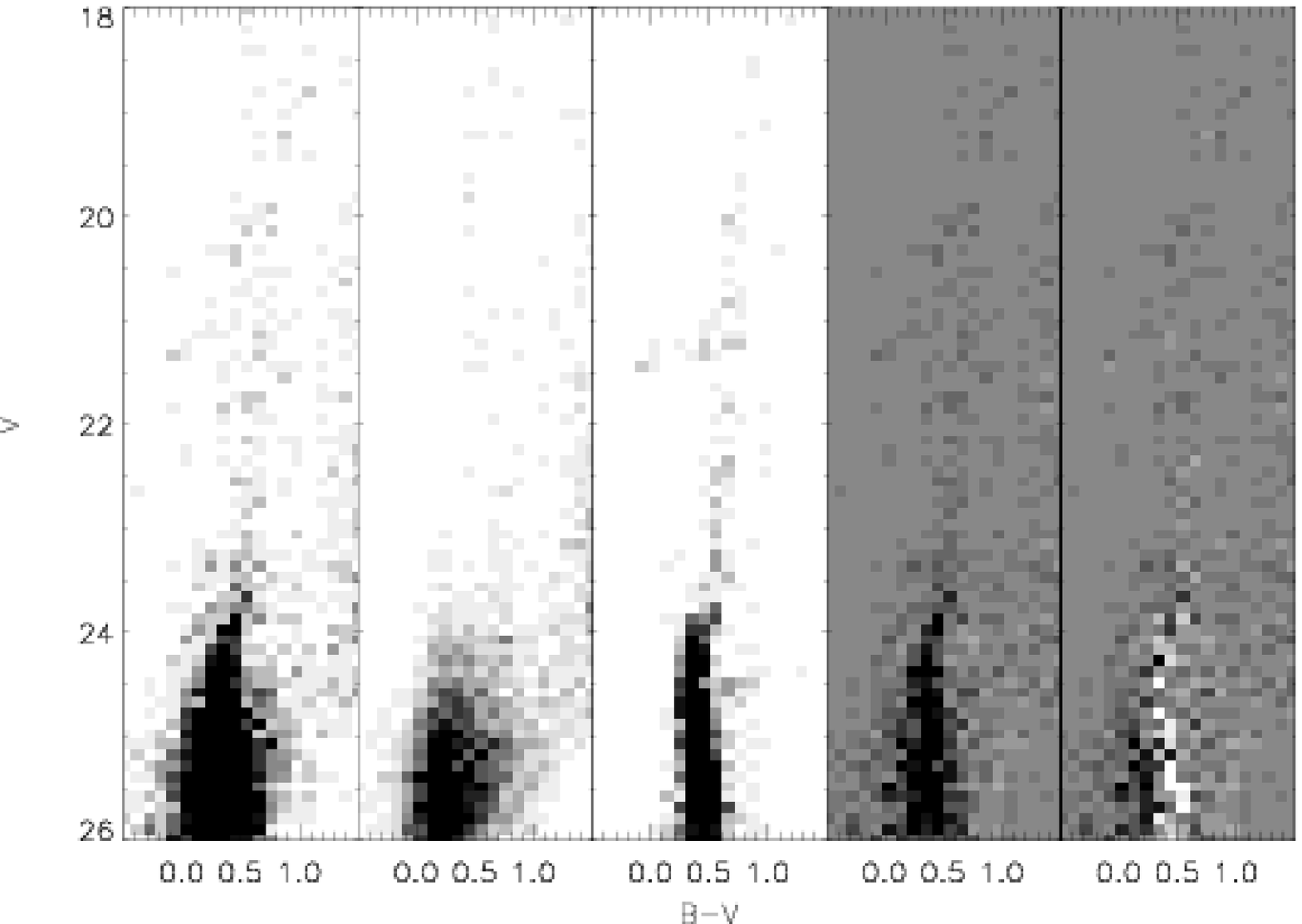}
}

\caption{ Comparison of data to best StarFISH model fit. The top panel
is for our $B-r$ versus r data, and the bottom for $B-V$ versus V.  {\bf
Left--} Raw Hess diagram of the central $r_h=$5.9' of Hercules.  {\bf
Second Left --} Background Hess diagram.  {\bf Center --} The Hess
diagram of the synthetic populations corresponding to the best-fit
StarFISH solution.  {\bf Second Right --} Background subtracted Hess
diagram of the central $r_h=$5.9' of Hercules.  {\bf Far Right --}
Residual Hess diagram after subtraction of the best-fit StarFISH model
from the background subtracted data.  \label{fig:starfishhess}}
\end{center}
\end{figure*}

The best-fit StarFISH solution is shown in Figure~\ref{fig:sfh}, with
a comparison of the best-fit model to the data shown in
Figure~\ref{fig:starfishhess}.  Note that bins with only error bars
should be considered upper limits.  We find that Hercules is old ($>12
$ Gyr) and metal poor, although there is an intrinsic spread in
metallicity, with both $[Fe/H]=-2.3$ and $-1.7$ populations
contributing to the SFH.  The best fit includes 1045 Hercules stars in
the central 5.9'.  There has been negligible star formation for the
last 12 Gyr.  This general result is robust with respect to our chosen
distance modulus.  If we fix the distance modulus to 20.55 and rerun
StarFISH, we get a similar mix of metal poor populations but with a
larger fraction of the ancient star formation coming from the
$[Fe/H]=-1.7$ bin.  Likewise, if we fix the distance modulus to 20.75,
then Hercules has a SFH which consists of only the $[Fe/H]=-2.3$
stellar population.

Recent spectroscopic results in Hercules have also measured a spread
in metallicity.  Medium resolution spectroscopy of 22 red giant branch
stars in Hercules has indicated that it is very metal poor, $\langle
[Fe/H] \rangle$=$-2.58$ with a spread of $\sigma_{[Fe/H]}=0.51$
\citep{Kirby08}.  Note that, if this is true, the Girardi isochrones
are not available for the most metal-poor half of the Hercules
distribution.  High resolution spectra of two Hercules stars was
presented in \citet{Koch08}, both of which were at $[Fe/H]$=$-2.0$,
which reinforces the spread in Hercules metallicities.

While it is safe to say that Hercules is old and metal poor, there are
several reasons that the model and observed CMDs will never match
perfectly (Figure~\ref{fig:starfishhess}).  The first has to do with
the theoretical isochrones, which are excellent for determining
general properties of stellar systems, but do have systematic
variations with respect to empirical isochrones
\citep[e.g.,][]{Girardi04}.  For instance, \citet{Girardi04} note that
there is a systematic offset in color of $\sim$0.1 mag below the main
sequence turn off between SDSS CMDs of Pal 5 and their theoretical
isochrones.  They conclude that this may be due to a real color shift
in the model.  A similar offset in our $V$ versus $B-V$ CMD could be
responsible for our residuals in the Hess diagram in
Figure~\ref{fig:starfishhess}.  In addition to mild inaccuracies of
the theoretical isochrones, the available models do not span the
metallicity range that is apparent in the new MW dwarfs.  As noted
above, \citet{Kirby08} found $\langle [Fe/H] \rangle$=$-2.58$ in
Hercules, which is more metal poor than the available
\citet{Girardi04} isochrones allow.  While the difference between a
$[Fe/H]$=$-2.6$ and $-2.3$ CMD will be very small, it nonetheless may
contribute to any systematic residuals.  Another concern for studies
of this type is proper correction for dust extinction.  We have
corrected for reddening using the extinction maps of
\citet{Schlegel98}, but this may not be perfect, given the beam size
of $\sim$6.1' in these maps.  Since this is roughly the size of
$r_{h}$, even though we account for reddening for each star, it is in
practice impossible to do so if the extinction varies significantly on
scales smaller than what we are considering here.  Additionally, any
dust associated with Hercules itself would not be accounted for in our
reddening correction.

\subsection{Absolute Magnitude}\label{sec:absmag}

As has been pointed out by \citet{Martin08} and others, measuring the
total magnitude of the new MW satellites is difficult due to their
relatively small number of stars.  To account for this 'CMD shot
noise', we mimic the luminosity measurement technique of Martin et
al. in the following way.

Given the SFH solution presented in \S~\ref{sec:starform}, we created a
well-populated CMD with $\sim$140,000 stars, including our
completeness and photometric uncertainties.  From this master CMD, we
drew one thousand random realizations of the Hercules CMD with an
identical number of stars as determined via our exponential profile
fit (which, from Figure~\ref{fig:stellarprofile} seems to best
represent the profile of Hercules) and determined the 'observed'
magnitude of each realization from the luminosity of all stars above a
limiting magnitude corresponding to our 90\% completeness limit
(switching to our 95\% completeness limit effects the total magnitude
by only $\sim$0.1 mag, which we add in quadrature to our overall
uncertainty).  Those stars fainter than this magnitude were accounted
for by using luminosity function corrections derived from
\citet{Dotter08}.  From the one thousand realizations, we take the
median as our absolute magnitude and its standard deviation as our
uncertainty (Table~\ref{table:paramfits}).  The absolute magnitude of
Hercules changes by $\sim$0.03 magnitudes depending on whether we use
$[Fe/H]=-2.5$ or $[Fe/H]=-1.7$ stellar populations with an age of 15
Gyr for this luminosity correction, and so use the $[Fe/H]=-2.5$
result.  We also calculate Hercules' central surface brightness,
$\mu_{0,V}$, assuming our exponential profile fit.

Our final $M_{V}=-5.3\pm0.4$ mag is $\sim$1.3 magnitudes fainter than that
found by \citet{Martin08}, although it is consistent with the
discovery data of \citet{Belokurov07} at the $1-\sigma$ level (who
used deeper follow-up observations to measure integrated properties).
We believe that this is due to the relatively small numbers of stars
used by Martin et al.

\section{Extended Structure Search}\label{sec:extendsearch}

In this Section, we look for evidence of tidal tails, hyper-faint
companions or other disturbances in the morphology of Hercules through
unexpected enhancements in stellar density of likely Hercules members
and by searching for systematic changes in its distance modulus along
its face.

\subsection{Morphology}\label{sec:morphology}

\begin{figure*}
\begin{center}
\mbox{
\epsfysize=3.75cm \epsfbox{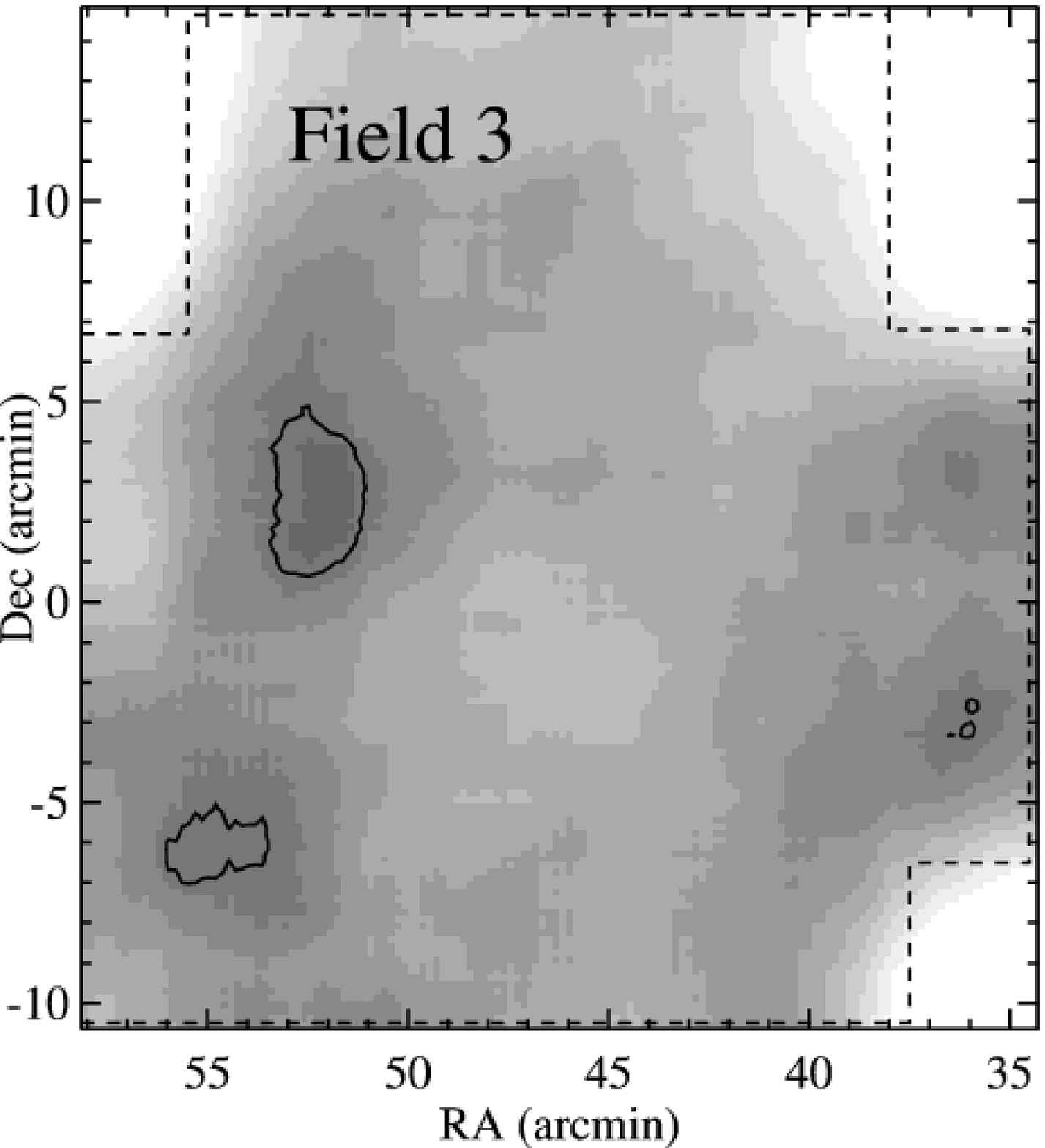}
\epsfysize=3.75cm \epsfbox{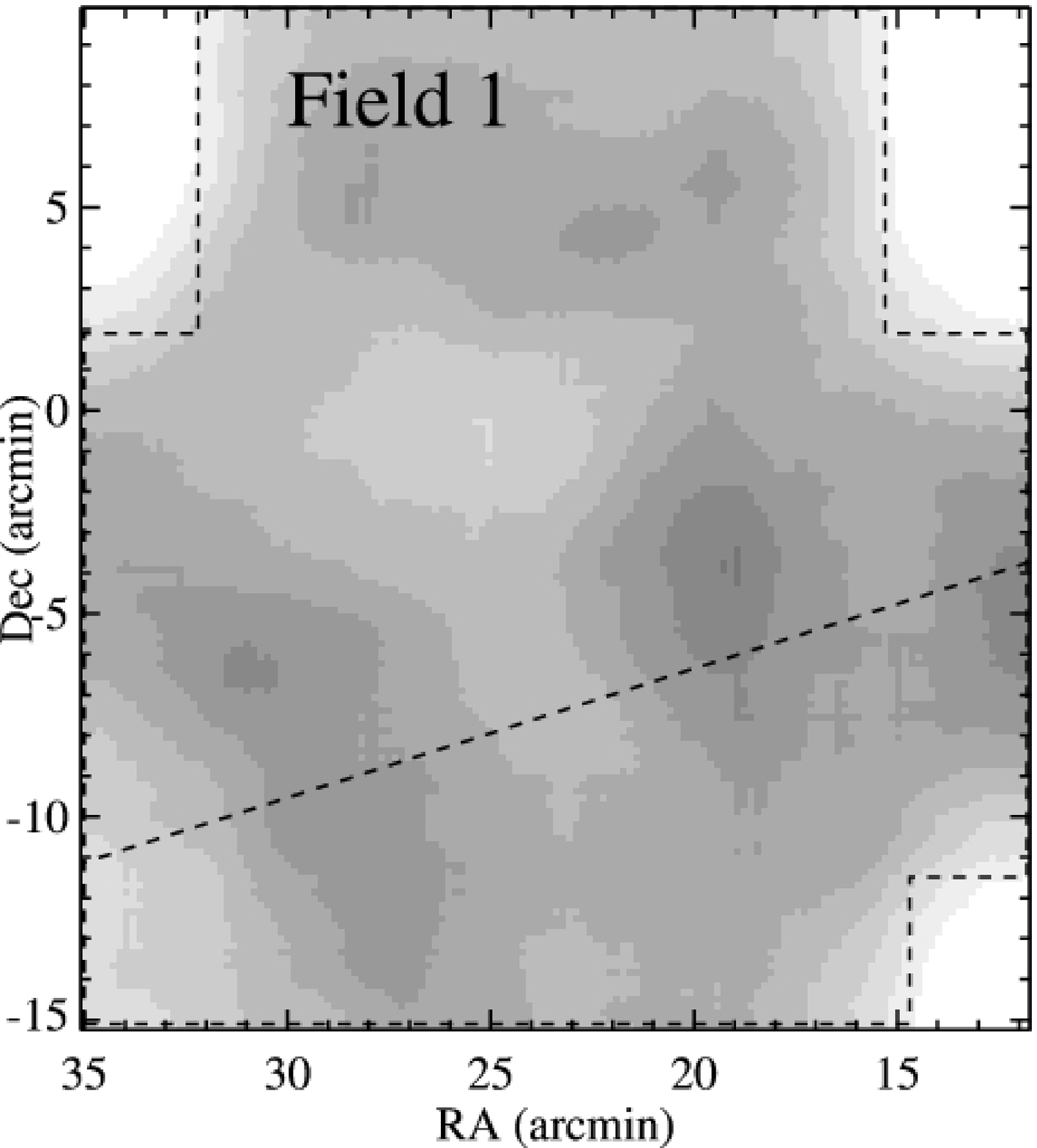}
\epsfysize=3.75cm \epsfbox{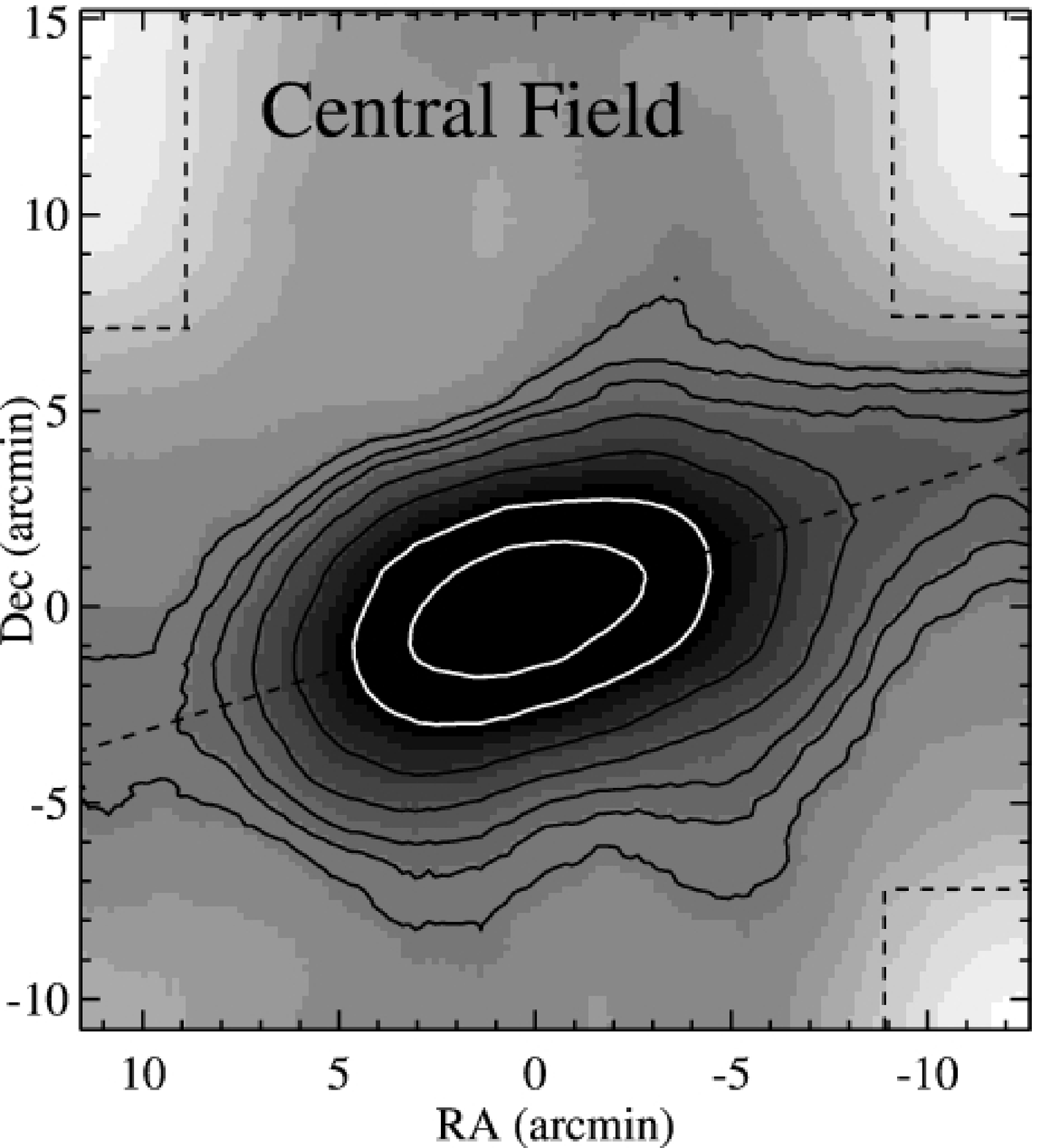}
\epsfysize=3.75cm \epsfbox{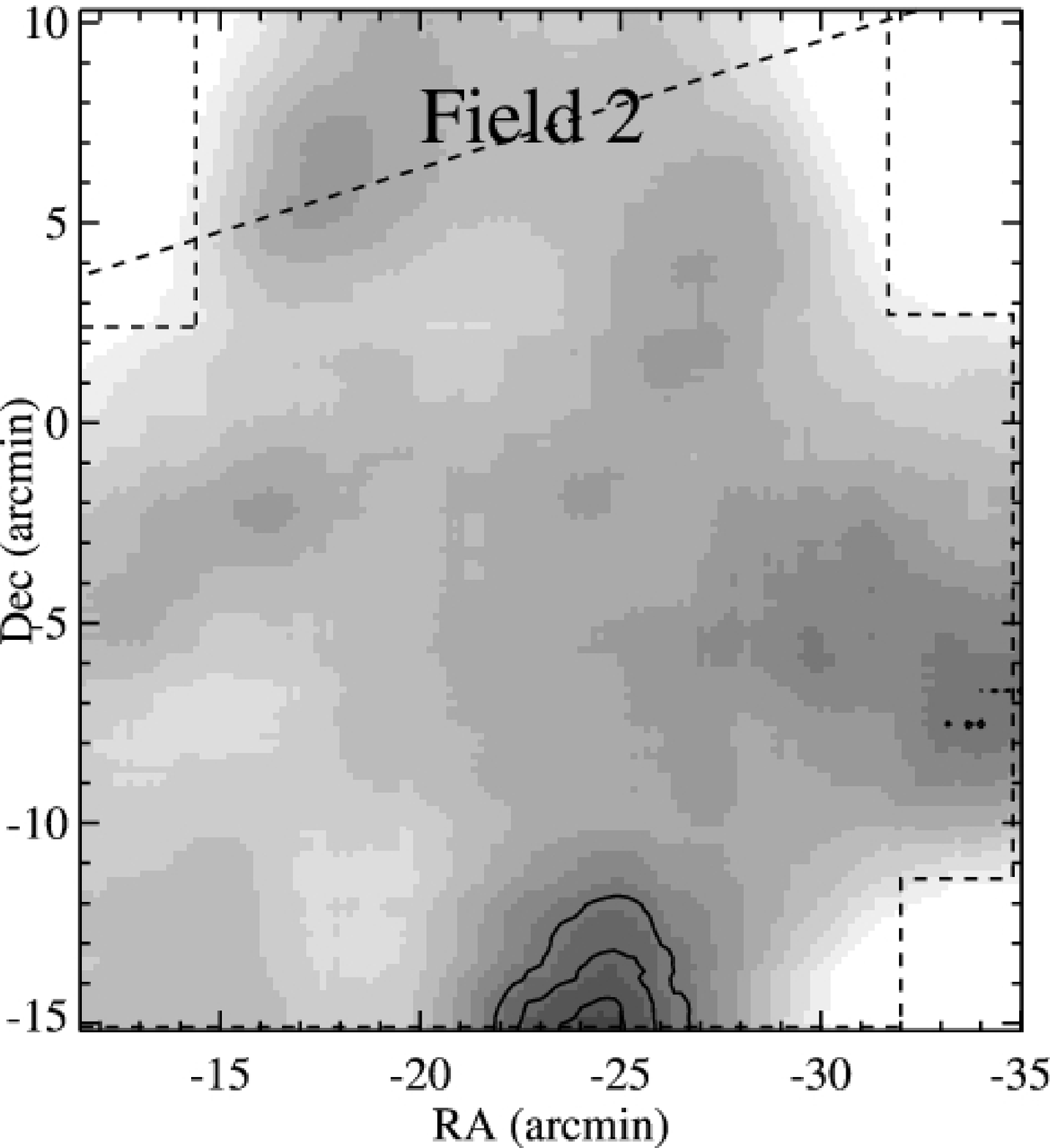}
\epsfysize=3.75cm \epsfbox{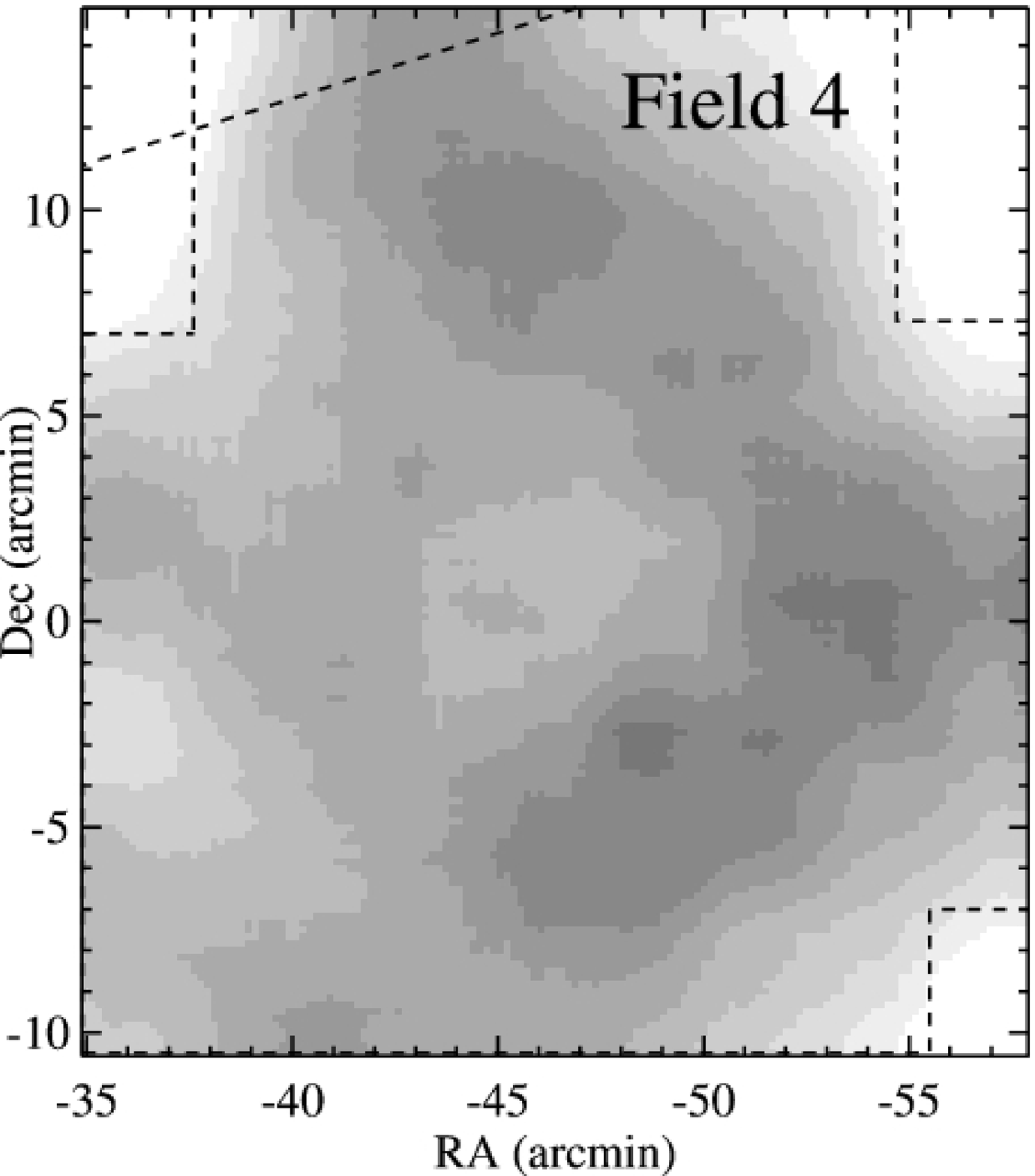}
}
\mbox{
\epsfysize=3.75cm \epsfbox{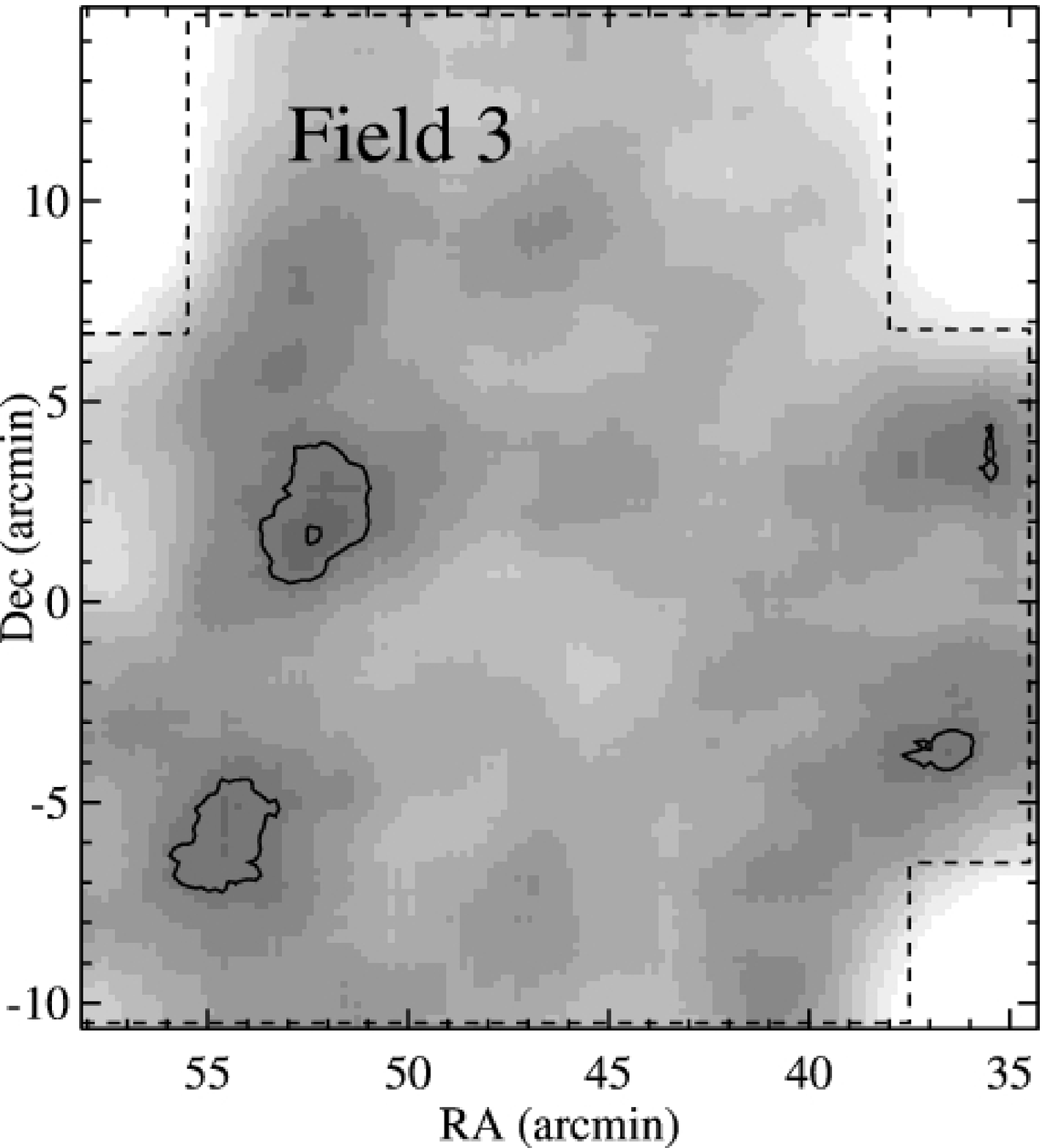}
\epsfysize=3.75cm \epsfbox{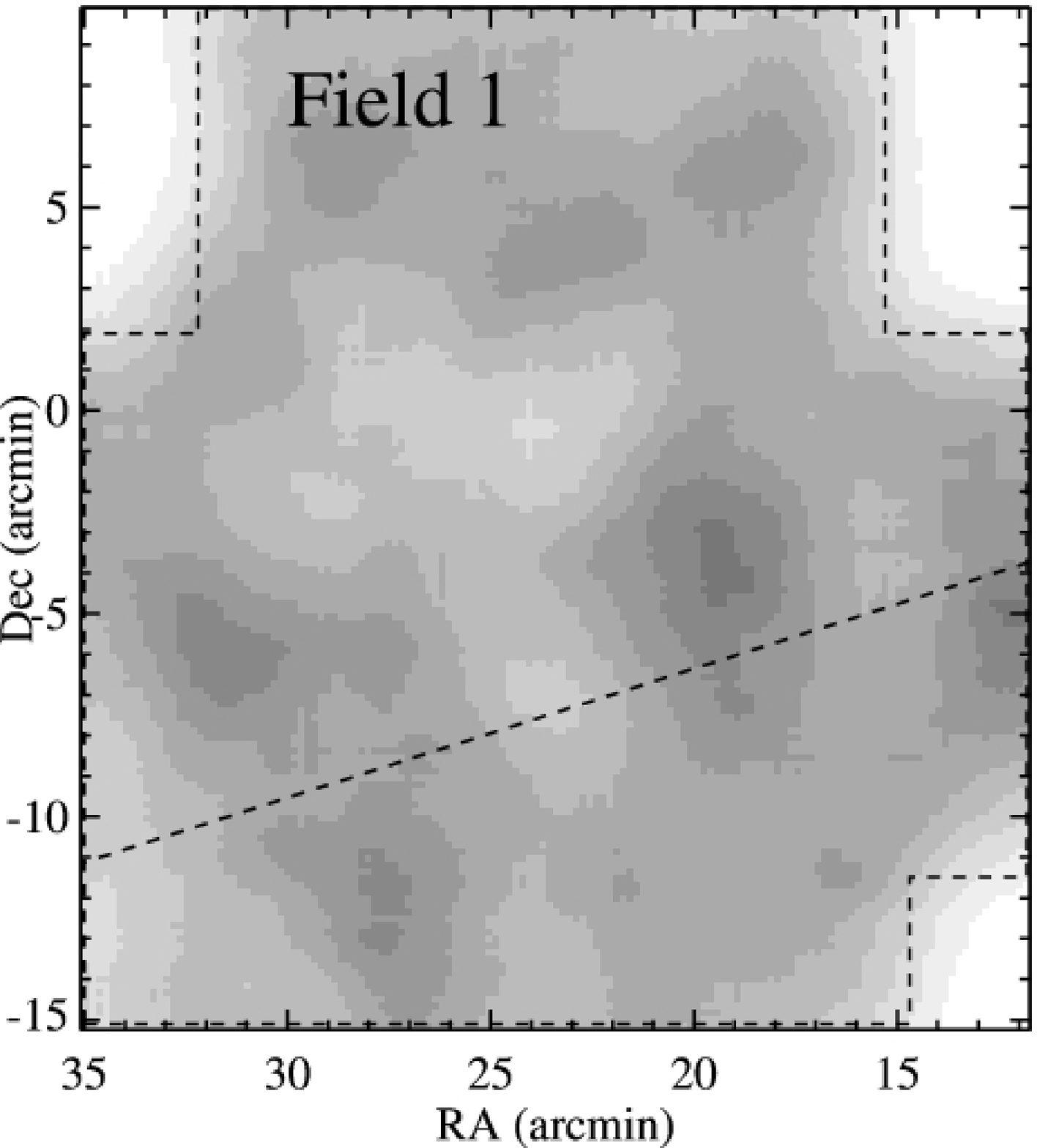}
\epsfysize=3.75cm \epsfbox{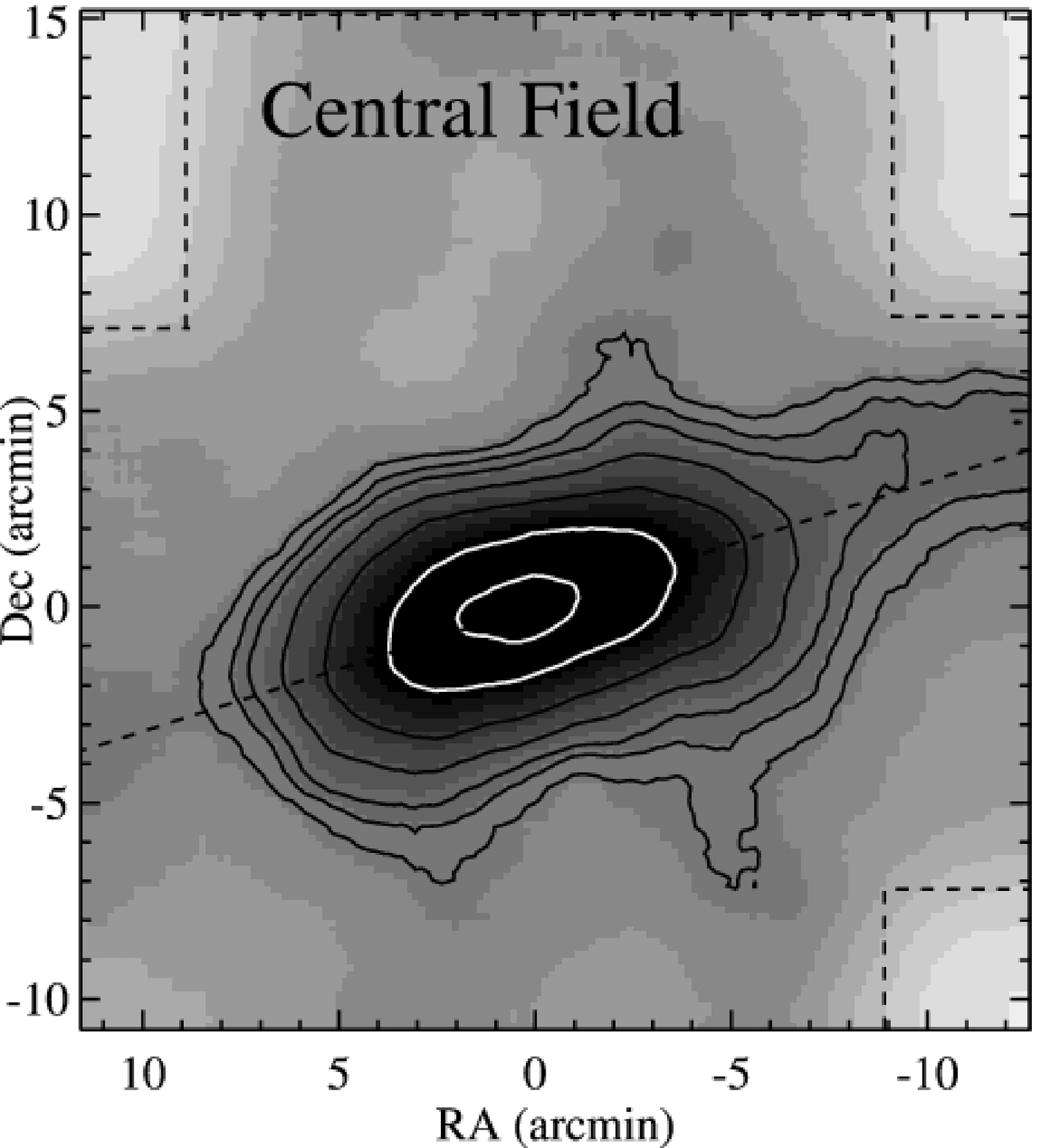}
\epsfysize=3.75cm \epsfbox{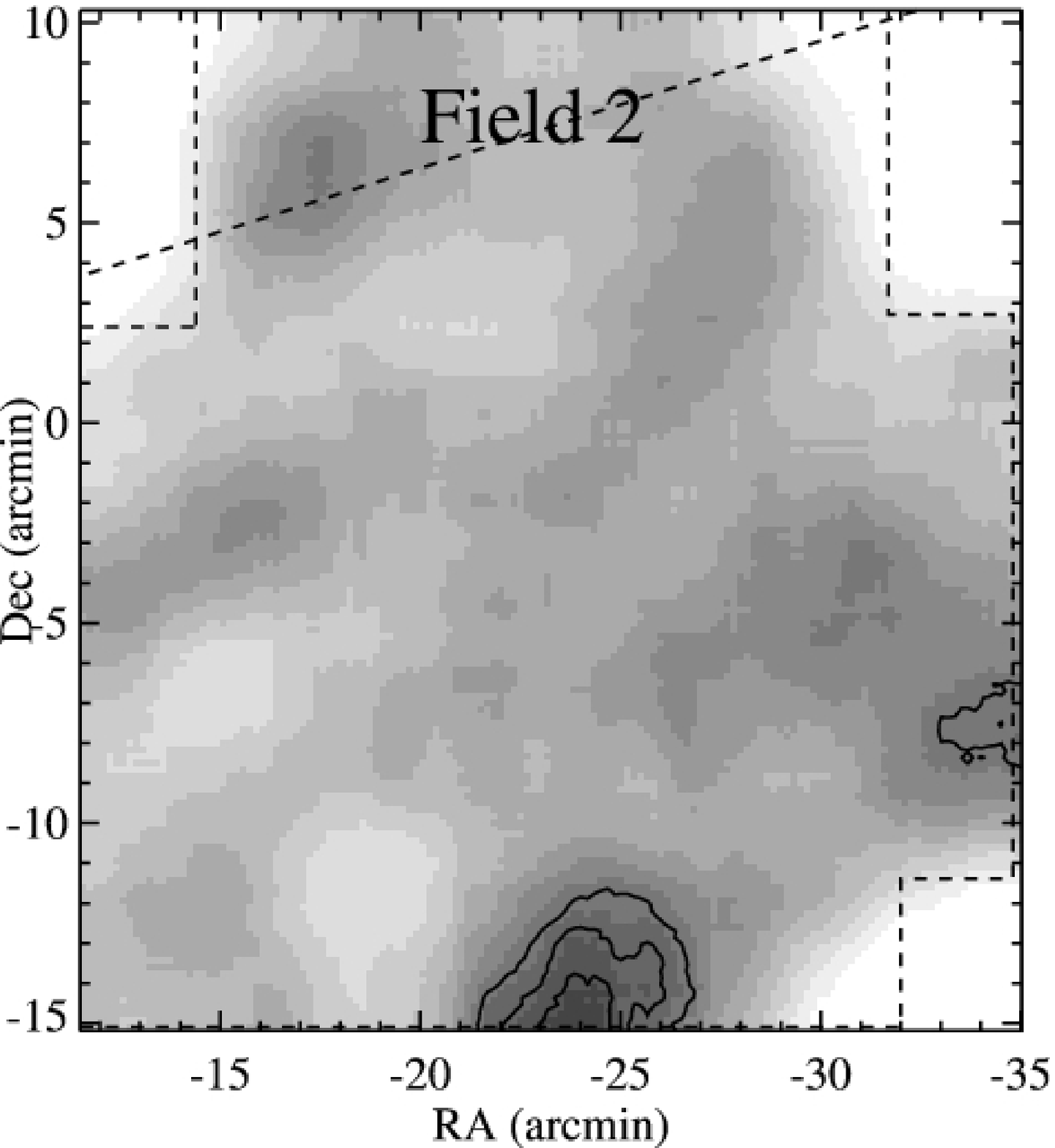}
\epsfysize=3.75cm \epsfbox{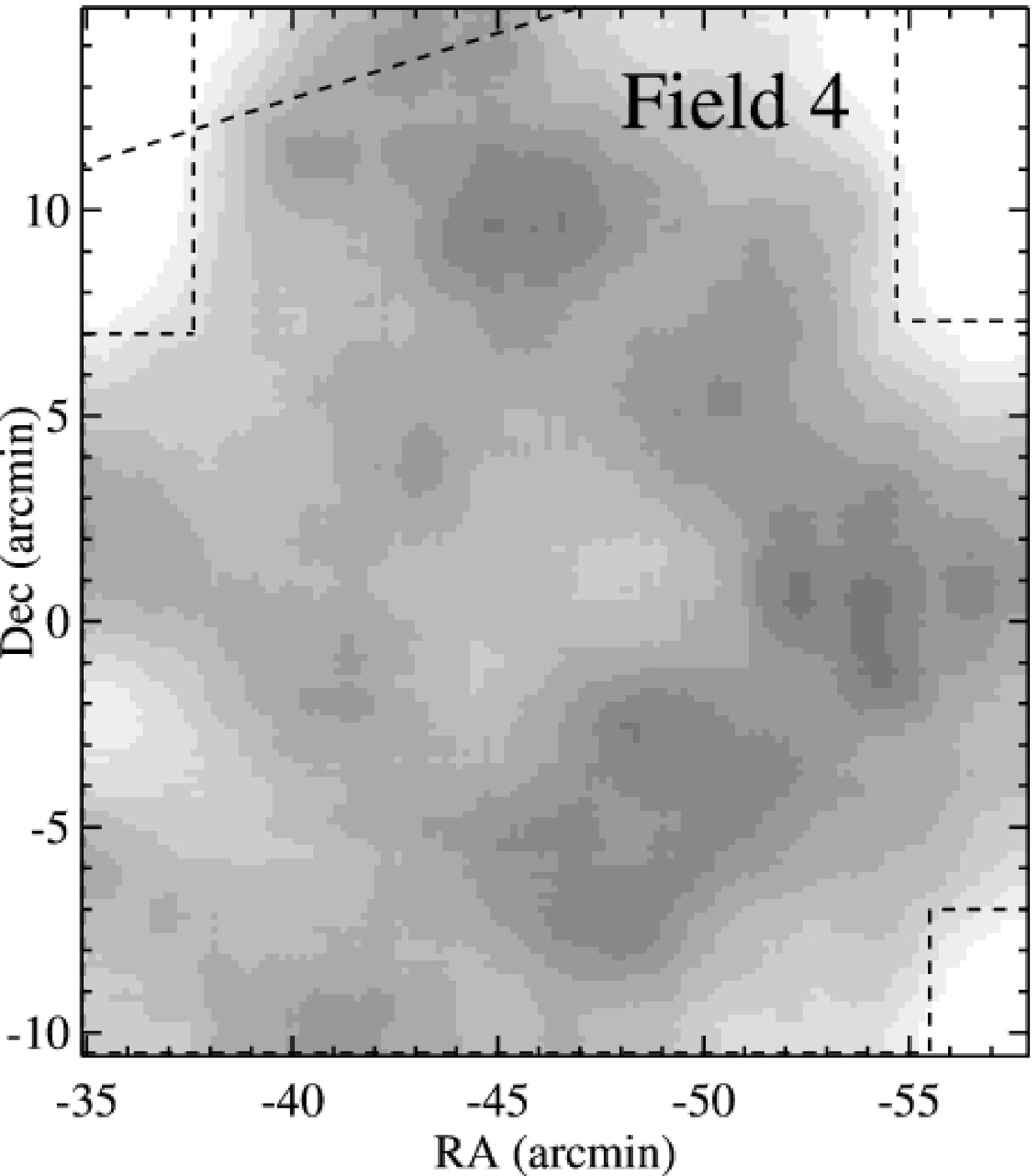}
}
\mbox{
\epsfysize=3.75cm \epsfbox{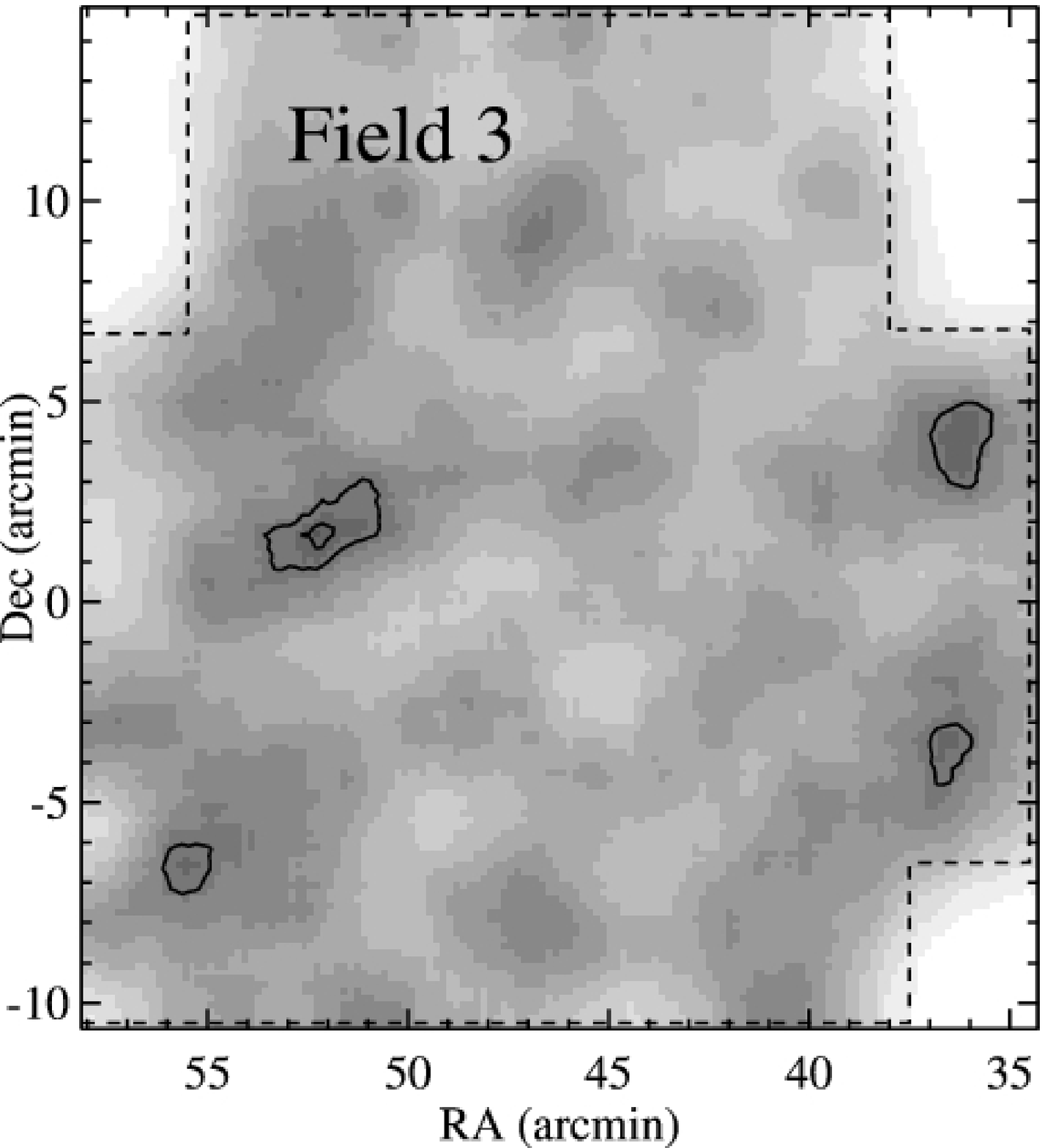}
\epsfysize=3.75cm \epsfbox{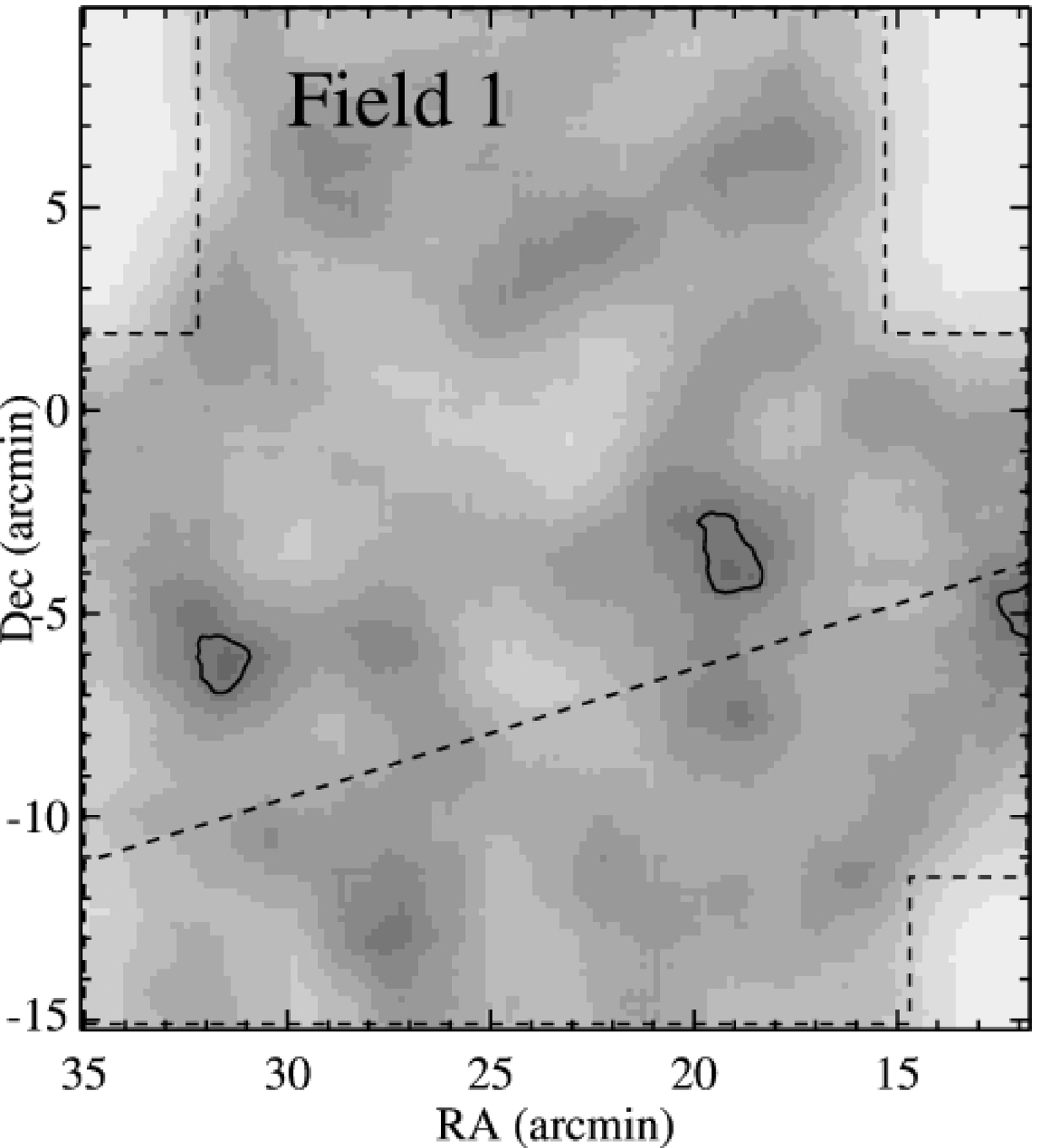}
\epsfysize=3.75cm \epsfbox{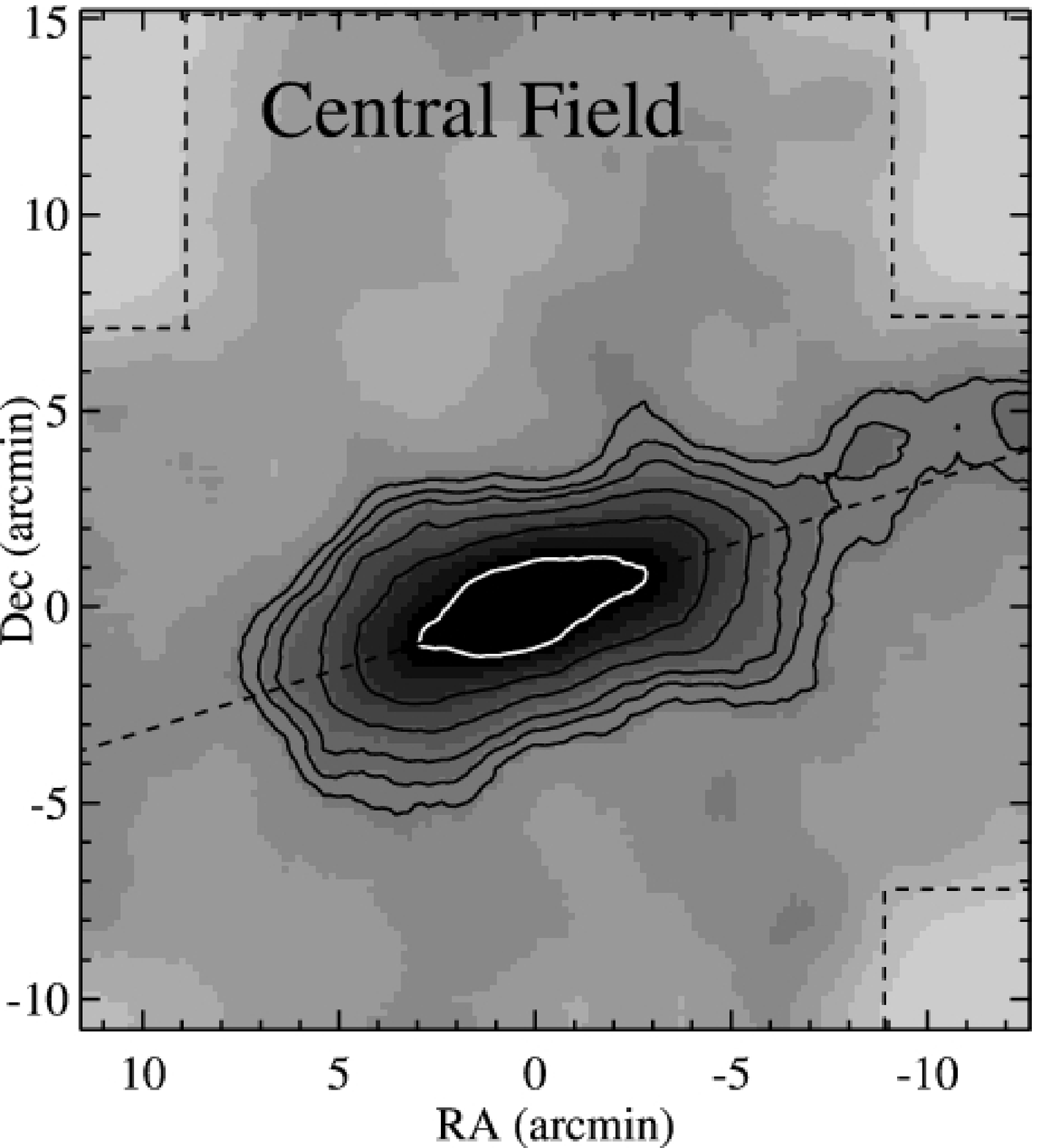}
\epsfysize=3.75cm \epsfbox{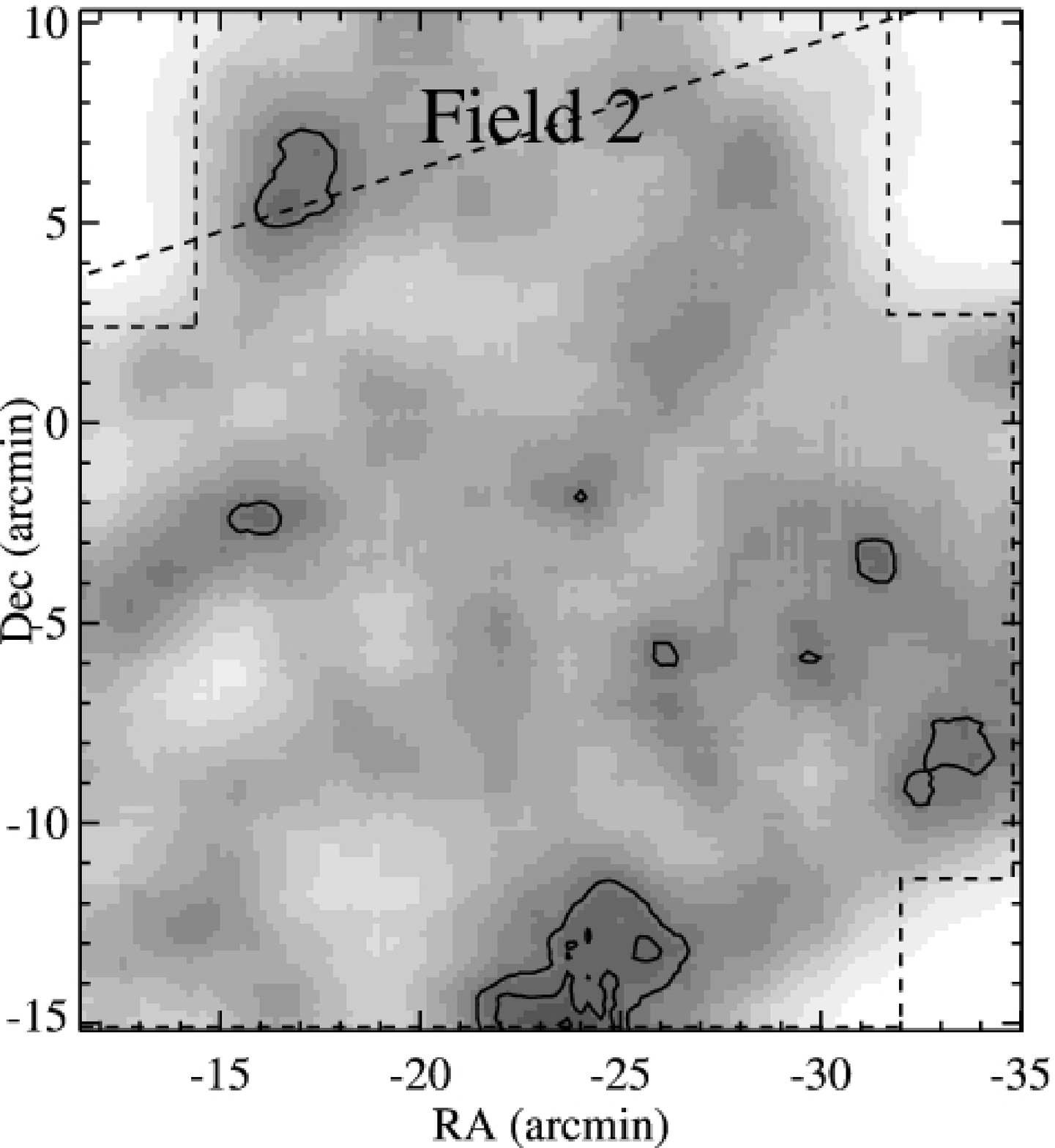}
\epsfysize=3.75cm \epsfbox{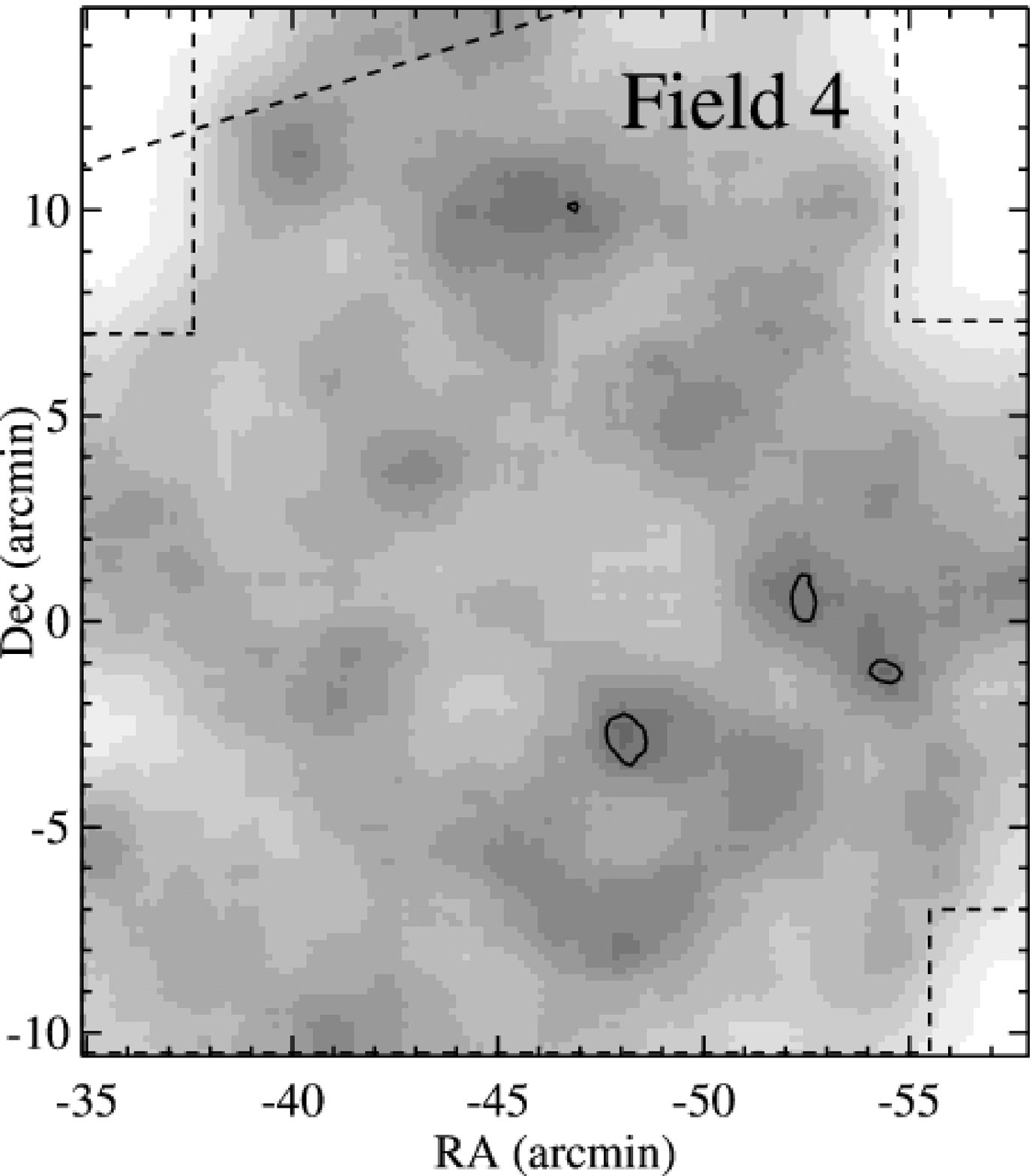}
}

\caption{ Smoothed contour plots of Hercules and adjacent fields.  The
contours show the 3, 4, 5, 7, 10, 15 and 20 $\sigma$ levels.  Each row
shows our Hercules pointings arranged from East to West, with the
middle panel roughly centered on the position of Hercules.  The top
row shows contour plots after smoothing the data by 2 arcmin.  The
middle row is smoothed by 1.5 arcmin, and the bottom row is smoothed
by 1 arcmin.  Dashed lines represent the actual LBT field of view.
The dashed line going through the center of Hercules traces the
position angle found for our exponential profile fit presented in
\S~\ref{sec:structparams}, $\theta$=$-72.4$ degrees.
\label{fig:smoothherc}}
\end{center}
\end{figure*}

First we search for signs of tidal disturbance and other Hercules
features in all five LBT fields based on the morphology of Hercules'
isodensity contours.  All stars that are consistent, within the
$1-\sigma$ photometric uncertainties, with the M92 isochrone
transformed to a distance modulus of 20.625 mag are placed in
$10''\times10''$ bins and spatially smoothed with three different
Gaussians -- with $\sigma$=1.0,1.5 and 2.0 arcminutes -- in order to
pick out structures of different scales.  Since each field has a
different depth and completeness at a given magnitude, it is difficult
to make a combined, smoothed mosaic of all fields simultaneously.  We
choose instead to present each field individually, with stars in the
central field taken down to $r=25.5$ mag and $r=24.5$ mag in the
other, adjacent fields -- corresponding roughly to the 50\%
completness limit.  The background level and variance was determined
from the entire image via the {\sc MMM} routine in {\sc IDL}, which
assumes that most contaminated pixel values are higher than the true
background.  However, since Hercules occupies the bulk of the central
pointing, the background for that field was determined in a box ten
arcminutes north of Hercules centroid that measures $4'\times10'$.  We
present these smoothed maps in Figure~\ref{fig:smoothherc}, with the
marked contours representing regions that are 3, 4, 5, 7, 10, 15 and
20 standard deviations above the background.  As can be seen, there
are no structures apparent in the $\sigma$=1.5 and 2.0 arcminute maps
that are not also in the $\sigma$=1.0 arcminute maps.  For this
reason, we will focus on the $\sigma$=1.0 arcminute case for the rest
of this work.  Although our maps are dependent on the number of
members stars above a magnitude threshold, and not directly on surface
brightness, it is informative to note that the 3-$\sigma$ contour of
the central pointing corresponds to $\mu_{r} \sim 29.3$ mag
arcsec$^{-2}$.

\begin{figure*}
\begin{center}
\mbox{
\epsfysize=7.5cm \epsfbox{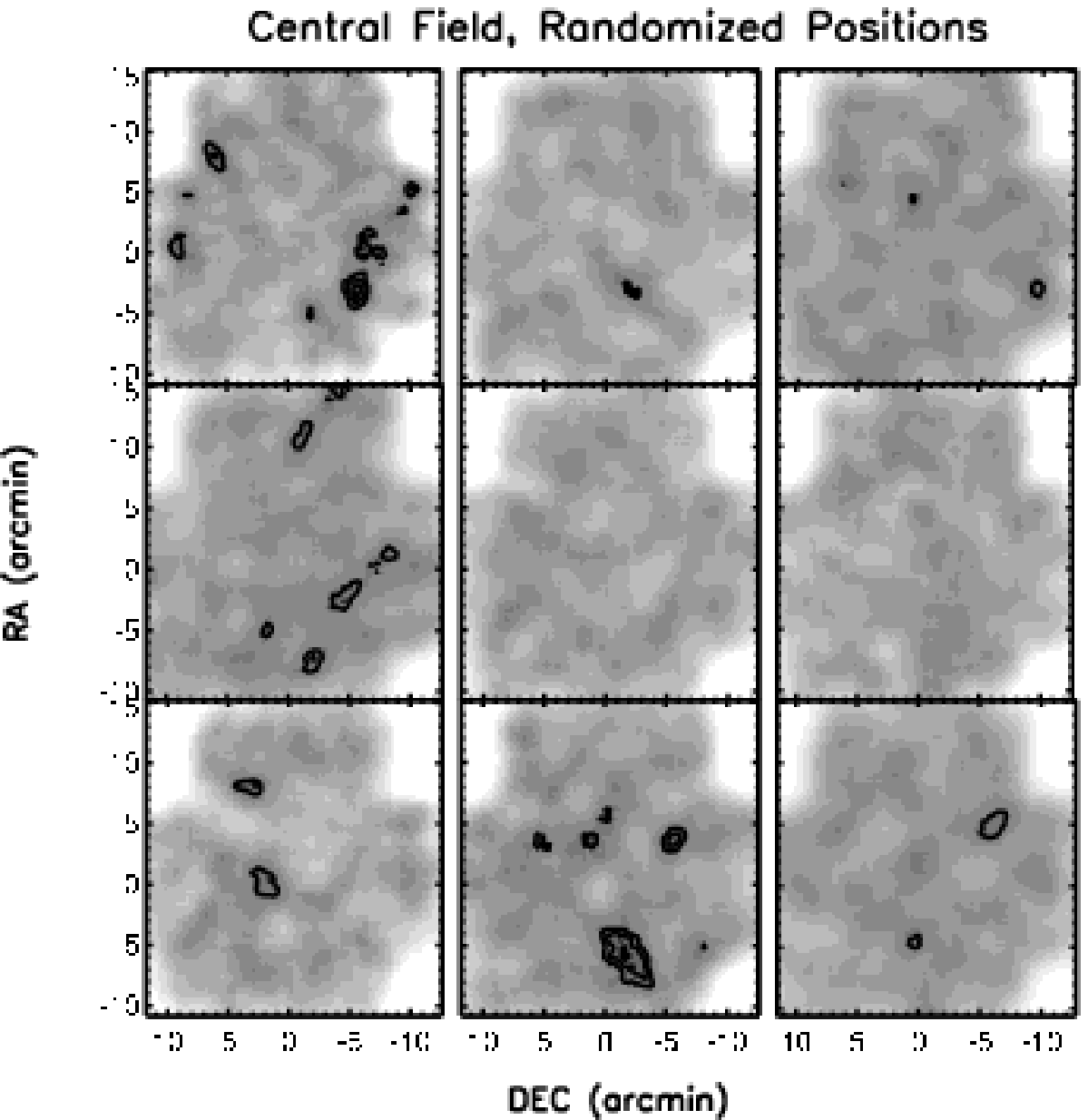}
\epsfysize=7.5cm \epsfbox{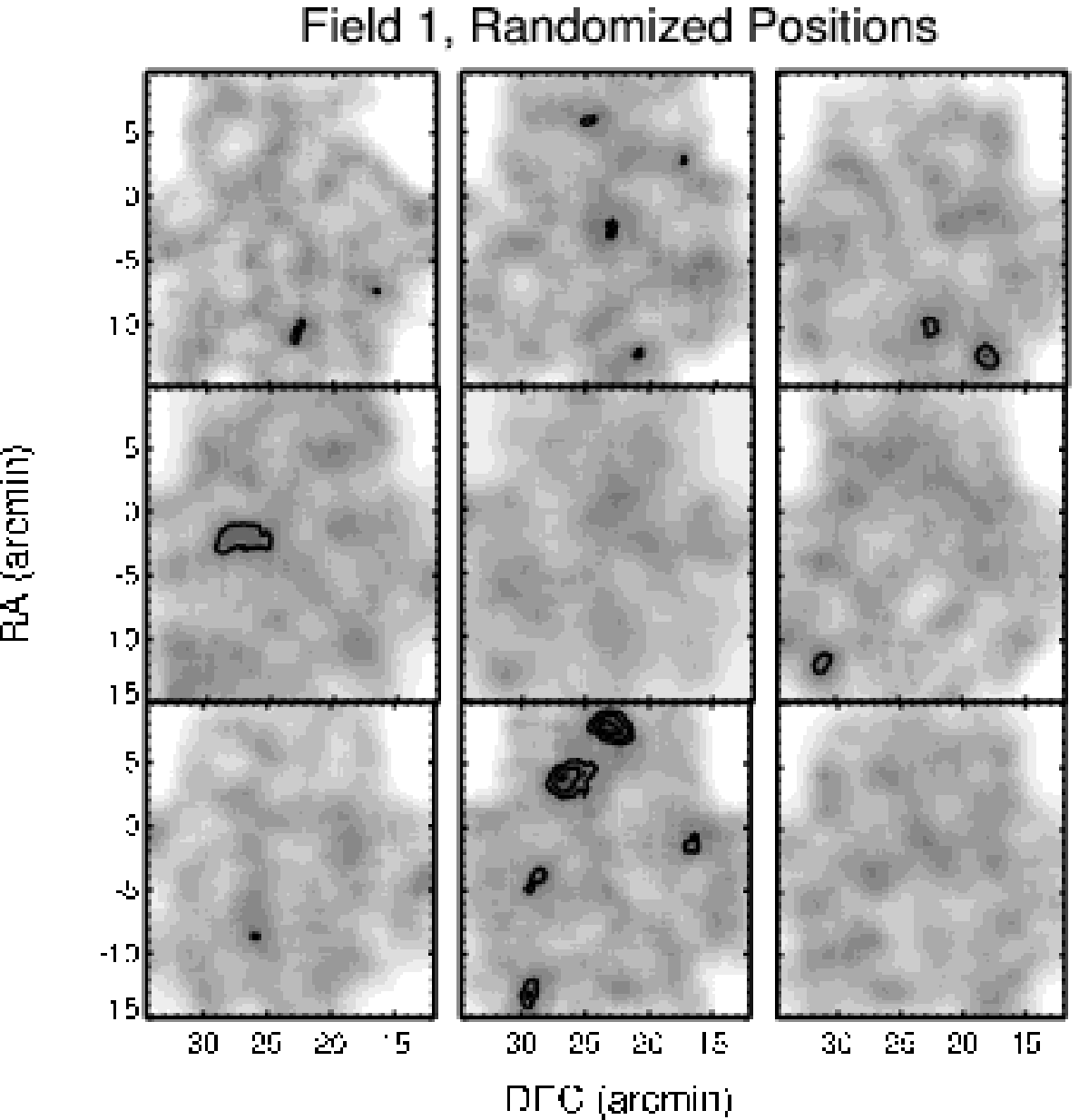}
}

\caption{ Smoothed contour plots, made as in Fig~\ref{fig:smoothherc},
of nine random realizations of Hercules stars where we have reassigned
star positions with ones drawn from a uniform distribution across the
LBT field of view.  The left panel shows results from our central
Hercules pointing photometry, while the right panel shows results
using our Field 1 photometry.  The contours show the 3, 4, and 5
$\sigma$ levels.  The plots show that 3$\sigma$ overdensities are
relatively common.
\label{fig:randcmds}}
\end{center}
\end{figure*}

Given our binning and smoothing to make Figure~\ref{fig:smoothherc},
how significant is any given overdensity?  To gauge this, we have used
the same photometry as the input catalog, but randomized the star
positions across each individual LBT field of view according to a
uniform distribution.  We then applied identical CMD cuts to identify
likely Hercules members, binned the data and smoothed it with a
Gaussian of $\sigma$=1.0 arcminute, identically to that done above.
In Figure~\ref{fig:randcmds}, we show nine such random realizations
for the central Hercules field and Field~1.  The random realizations
of our other fields have similar characteristics.  As can be seen,
$3-\sigma$ overdensities are relatively common, with occasional 4 or
even 5$\sigma$ peaks.  This will be kept in mind when examining our
Hercules overdensities.

\subsubsection{Inserting Artificial Remnants}\label{sec:fakeherc}
We will be dealing with small numbers of stars in particular regions
of these smoothed maps where there is an apparent overdensity of stars
consistent with the CMD of Hercules.  Visual inspection of the CMDs of
these overdense regions to confirm their similarity to the Hercules
CMD is difficult.  To help, we have developed tools that allow us to
use the SFH for Hercules determined in \S~\ref{sec:starform}, along
with our artificial star tests for each of our LBT pointings, to
generate artificial 'Hercules' CMDs, using the {\it testpop} program
within the StarFISH package.  These Hercules 'nuggets' can then be
injected into our Hercules photometry catalog with an arbitrary
spatial distribution, to see if our smoothing process can recover
them, and to determine the quality of the resulting Hess diagrams.  We
inject artificial Hercules nuggets with exponential profiles into our
Hercules catalog, for simplicity.  By varying the number of stars and
the half light radius, we can compare these artificial CMDs directly
with those associated with stellar overdensities in our smoothed maps
and measure our sensitivity to faint remnants associated with
Hercules.

\begin{figure*}
\begin{center}
\mbox{
\epsfysize=4.5cm \epsfbox{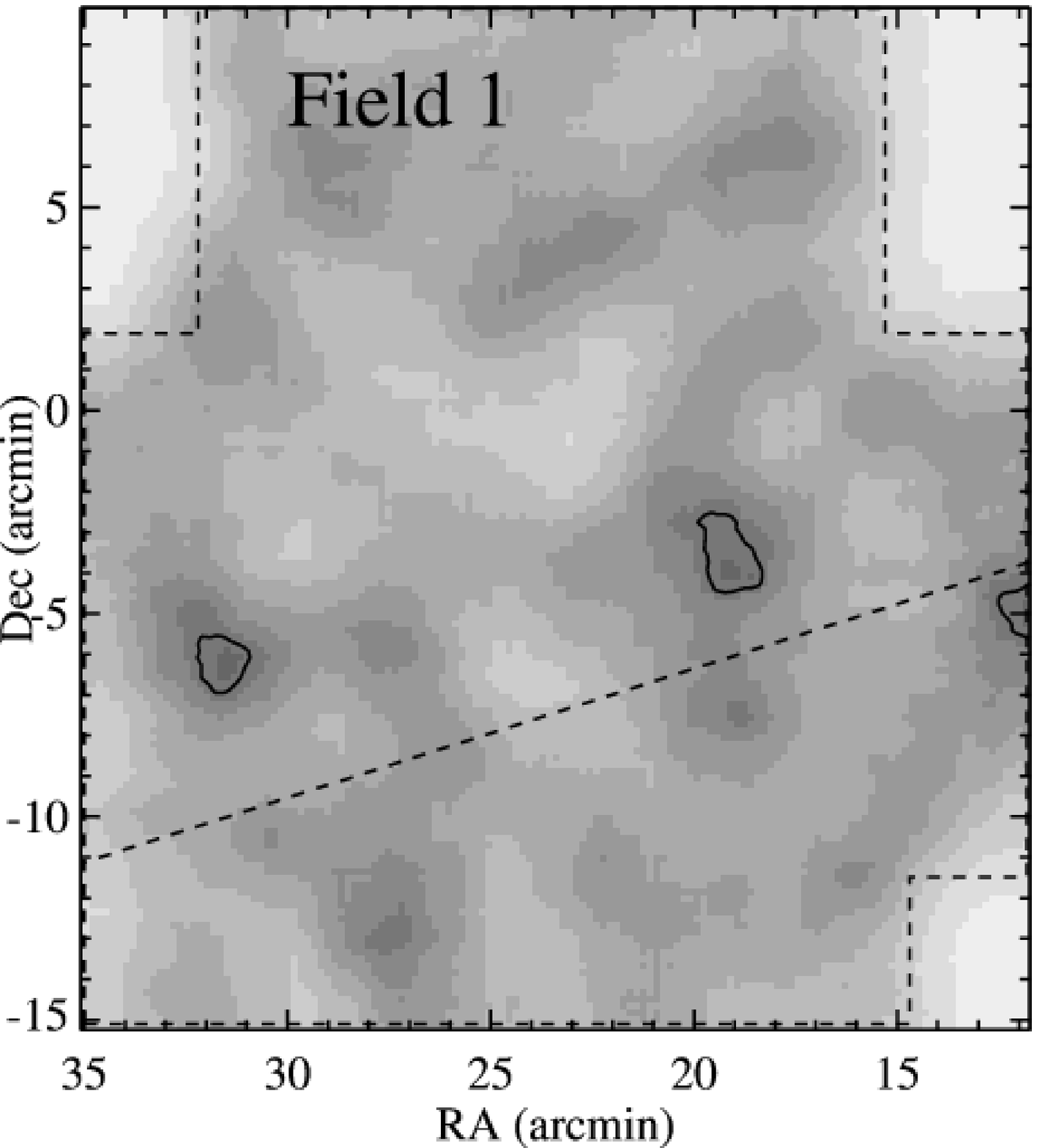}
\epsfysize=4.5cm \epsfbox{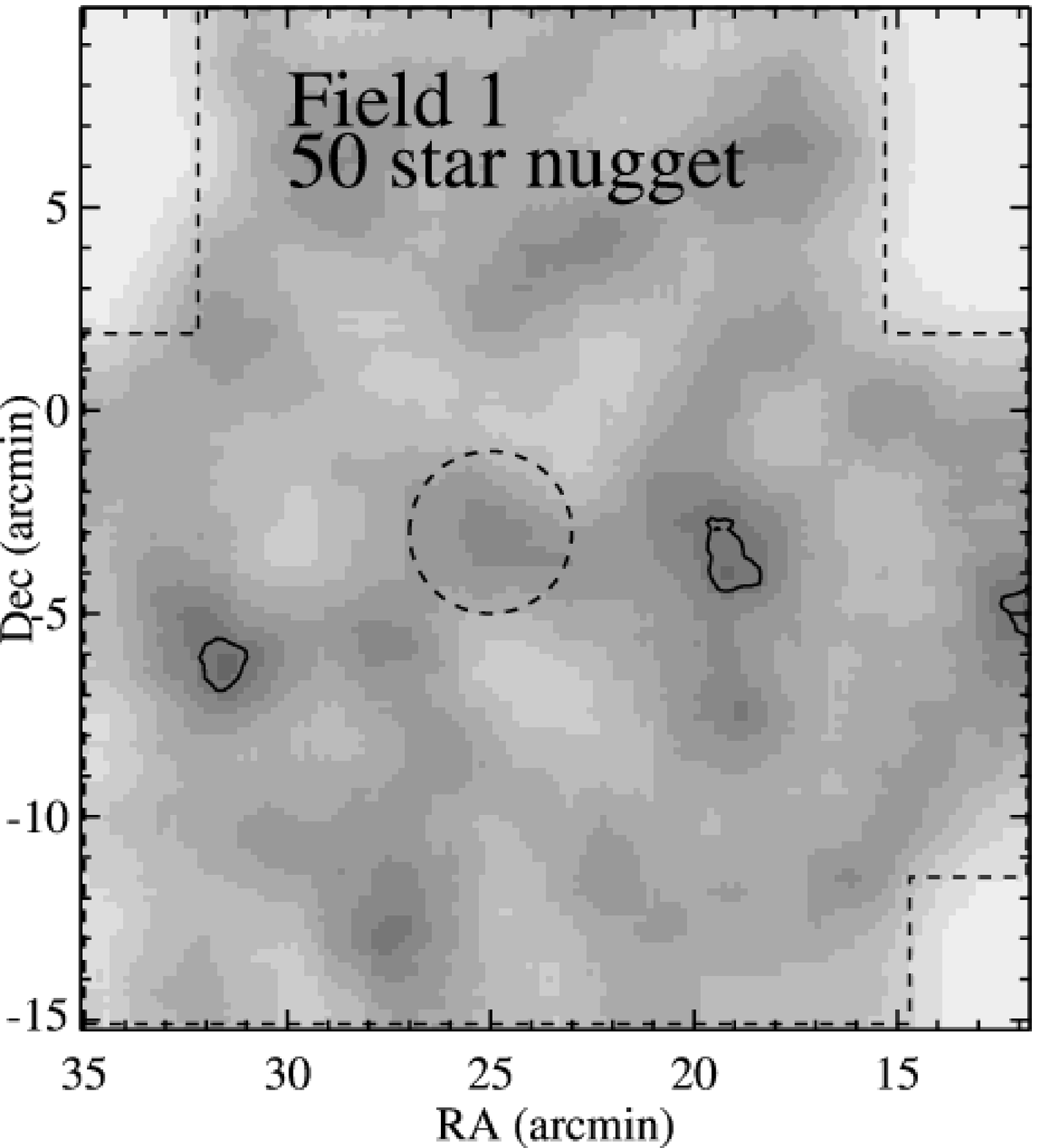}
\epsfysize=4.5cm \epsfbox{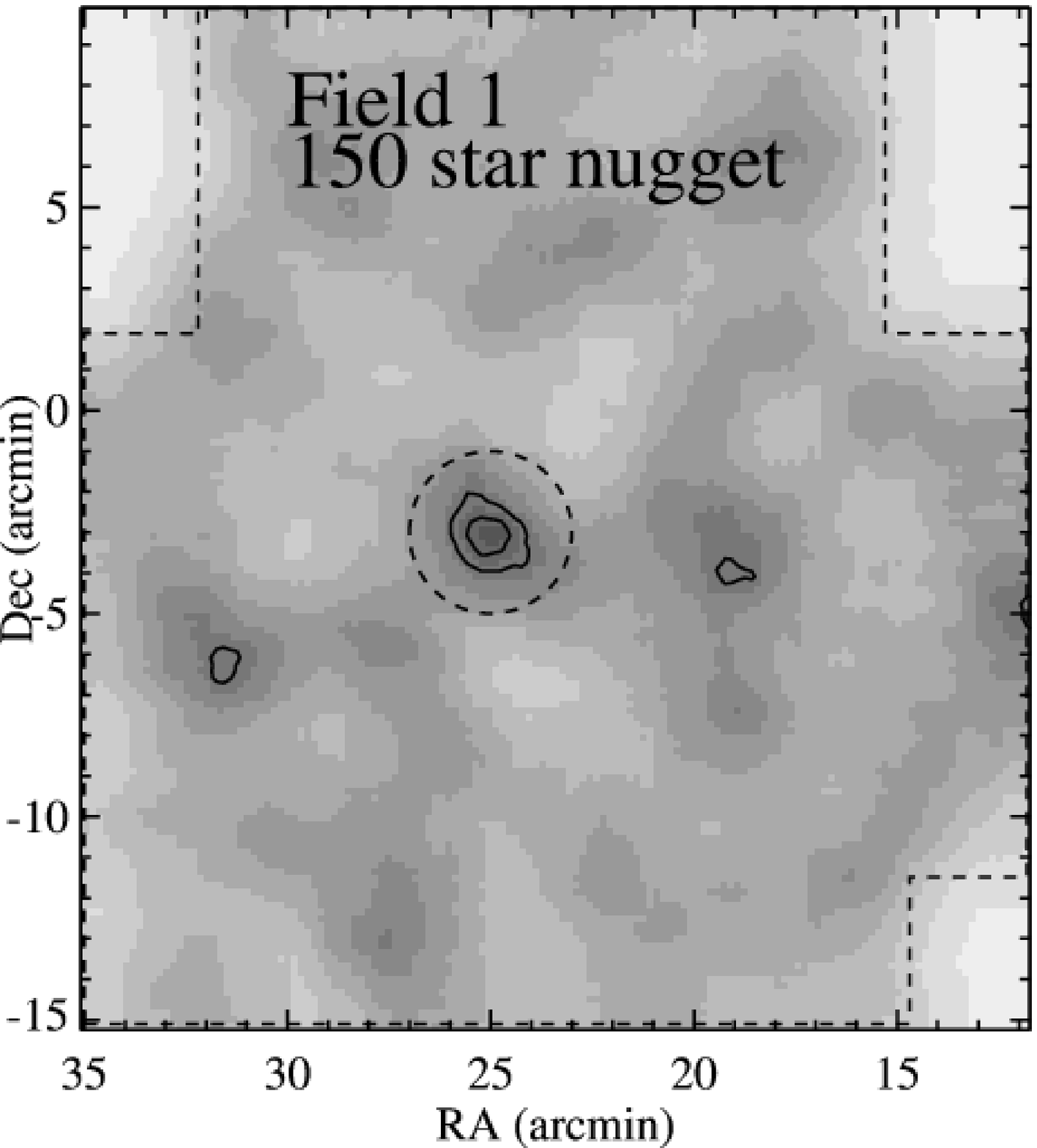}
}
\mbox{
\epsfysize=5.0cm \epsfbox{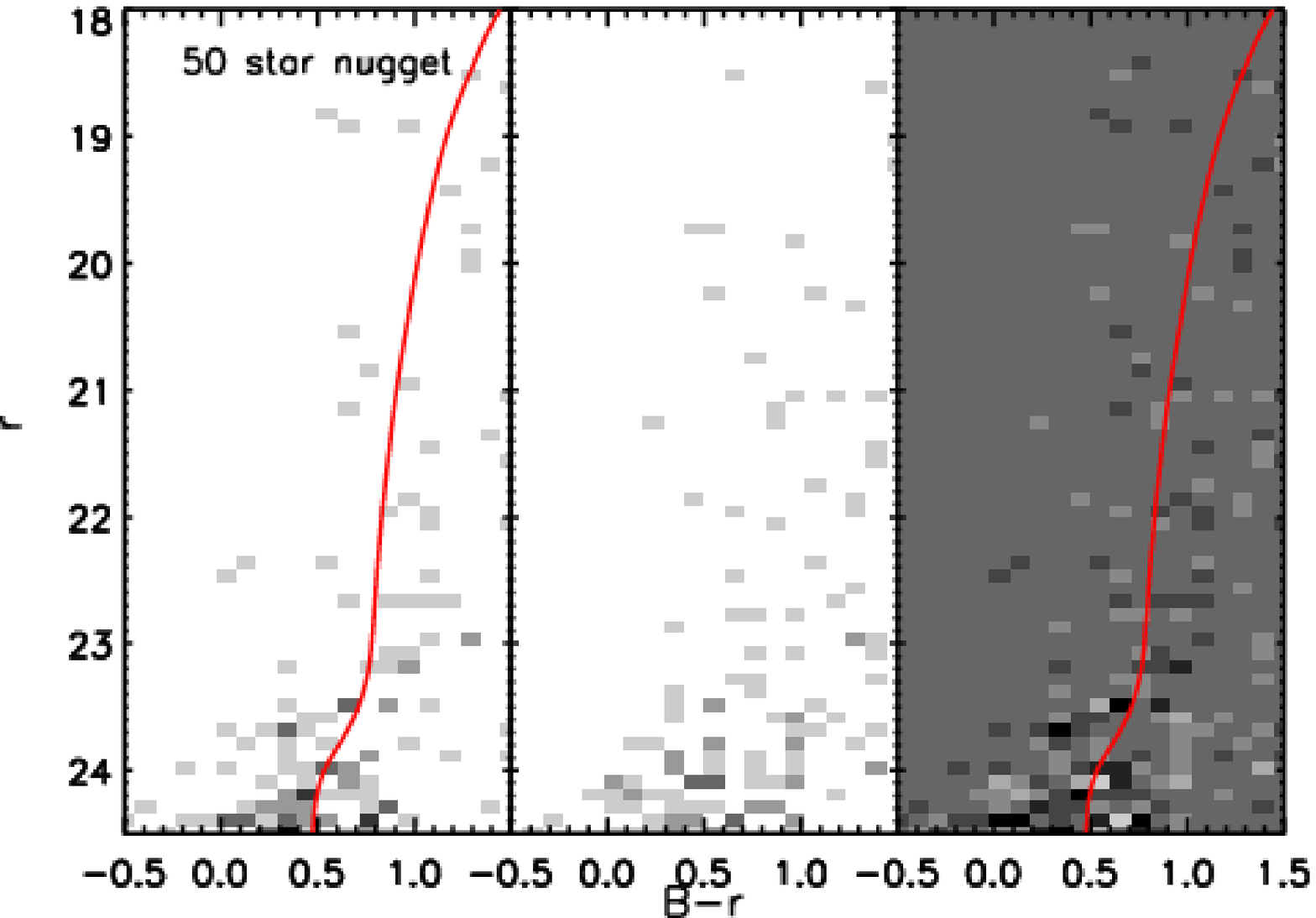}
\epsfysize=5.0cm \epsfbox{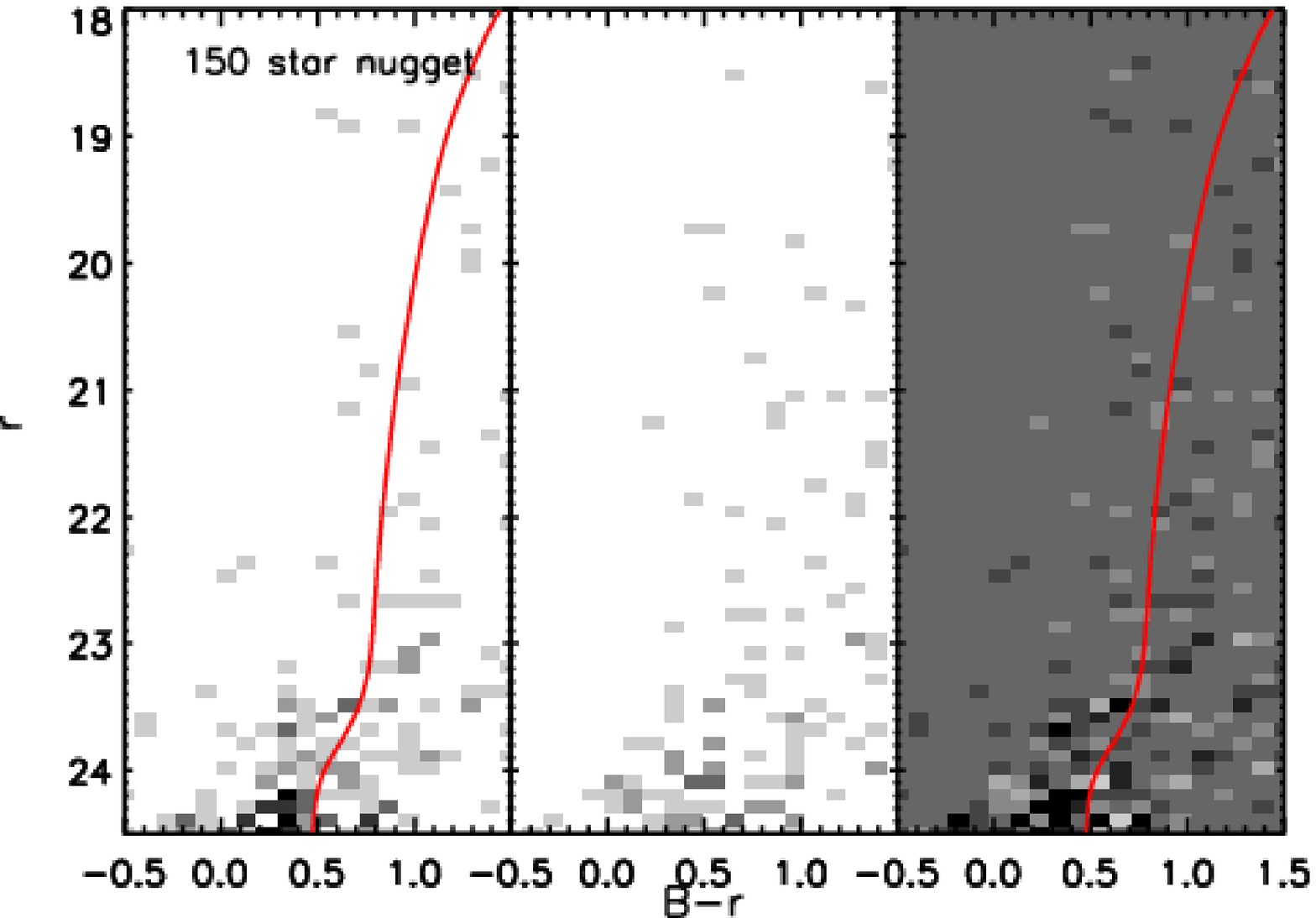}
}
\caption{An illustration of our technique for implanting fake Hercules
'nuggets' into our fields. Each of these 'nuggets' has the same SFH as
Hercules, taking into account the results of our artificial star tests
in this field.  On the top row, from left to right, we show first
Field 1 smoothed with a $\sigma=1$ arcminute Gaussian.  In the center
and right panel we show this same field after injecting 'nuggets' with
50 and 150 stars, respectively, distributed as an exponential with a
half light radius of 3 arcminutes (see Table~\ref{tab:fakeresult}).
These nuggets are similar in size as the real stellar overdensities in
the field.  Bottom -- Hess diagrams of the two nuggets inserted into
the image shown in the top row, along with the M92 isochrone shifted
to $(m-M)=20.625$.  These Hess diagrams are made identically to those
in \S~\ref{sec:hercover}, with an equal area annulus outside the
encircled region serving as the background CMD.  Note that, despite
the high significance of the 150 star nugget it is difficult to say
for certain that the CMD is similar to Hercules, although there is an
indication of the beginning of a main sequence around $r\sim$24.2.  It
is difficult to see such a feature in the 50 star nugget's
background-subtracted Hess diagram.
\label{fig:inject}}
\end{center}
\end{figure*}

As an illustration, Figure~\ref{fig:inject} shows two artificial
Hercules nuggets (of 50 and 150 stars) implanted into Field 1, the
field directly East of Hercules, with the properties shown in
Table~\ref{tab:fakeresult}.  They each have a SFH
identical to that of Hercules and have a distribution randomly drawn
from an exponential profile with a half light radius of 3.0
arcminutes.  The 50 and 150 star nuggets result in $\sim$2.2 and 4.8
$\sigma$ overdensities, after smoothing with a one arcminute Gaussian.
We extract these nuggets using a circular aperture with a radius of 2
arcminutes, as illustrated in the figure, with the background Hess
diagram being taken from an equal area annulus outside the aperture.
As can be seen from the resulting Hess diagrams in the bottom panels
of Figure~\ref{fig:inject}, it is difficult to say that the 50-star
nugget has a Hess diagram consistent with the Hercules CMD, although
the 150-star nugget has a hint of a main sequence around $r\sim24.2$.

In Table~\ref{tab:fakeresult} we show the results of five such tests
on each of our pointings (besides the central pointing of Hercules),
where we have fixed the exponential scale radius to 3 arcminutes (116
pc at the distance of Hercules), and have simply varied the number of
stars drawn from our fake Hercules CMDs generated for each pointing.
We then calculated the central surface brightness and total magnitude
of these nuggets as we did in \S~\ref{sec:absmag}.  Generally
speaking, it is only for those nuggets that result in overdensities
$\gtrsim$3$\sigma$ where the beginning of a main sequence can be seen
in the resulting Hess diagrams, and so it is this limit that we adopt
when investigating the candidate stellar overdensities in our fields.
Additionally, it is clear that the more stars that are used in
constructing our Hess diagrams, the clearer any signal will be, and so
we sum these overdensities when we can.

Table~\ref{tab:fakeresult} is only illustrative of our sensitivity to
external Hercules structure, as the detection of any given nugget is
subject to several random factors.  For instance, we are injecting
nugget stars drawn from an exponential profile with a 3 arcminute
scale radius, but are only using extraction apertures with a 2
arcminute radius, which is justified upon visual inspection of
Figure~\ref{fig:inject}.  Also, scattering of individual stars away
from our CMD detection threshold due to our incorporation of results
from our artificial star tests also adds a random component, leading
to a natural variance in the detectability of a given nugget.

\subsubsection{Hercules overdensities}\label{sec:hercover}

We have seen, both by randomizing the spatial positions of our input
photometry (Figure~\ref{fig:randcmds}) and by implanting fake
'Hercules' nuggets into our catalogs (\S~\ref{sec:fakeherc}), that it
will be extremely difficult to investigate and be reasonably assured
that any given 3$\sigma$ overdensity is truly associated with
Hercules.  We therefore focus only on the apparent 'stream' seen in
the central pointing emanating from Hercules to the Northwest,
coincident with the major axis, and those Hercules 3$\sigma$
overdensities which are nearly projected onto the position angle of
Hercules.  These could plausibly be high-density knots in a tidal
stream that is currently undetectable, keeping in mind that our fields
adjacent to Hercules are shallower by $\sim$1 mag than our central
pointing.

\begin{figure*}
\begin{center}
\mbox{
\mbox{\epsfysize=6.5cm \epsfbox{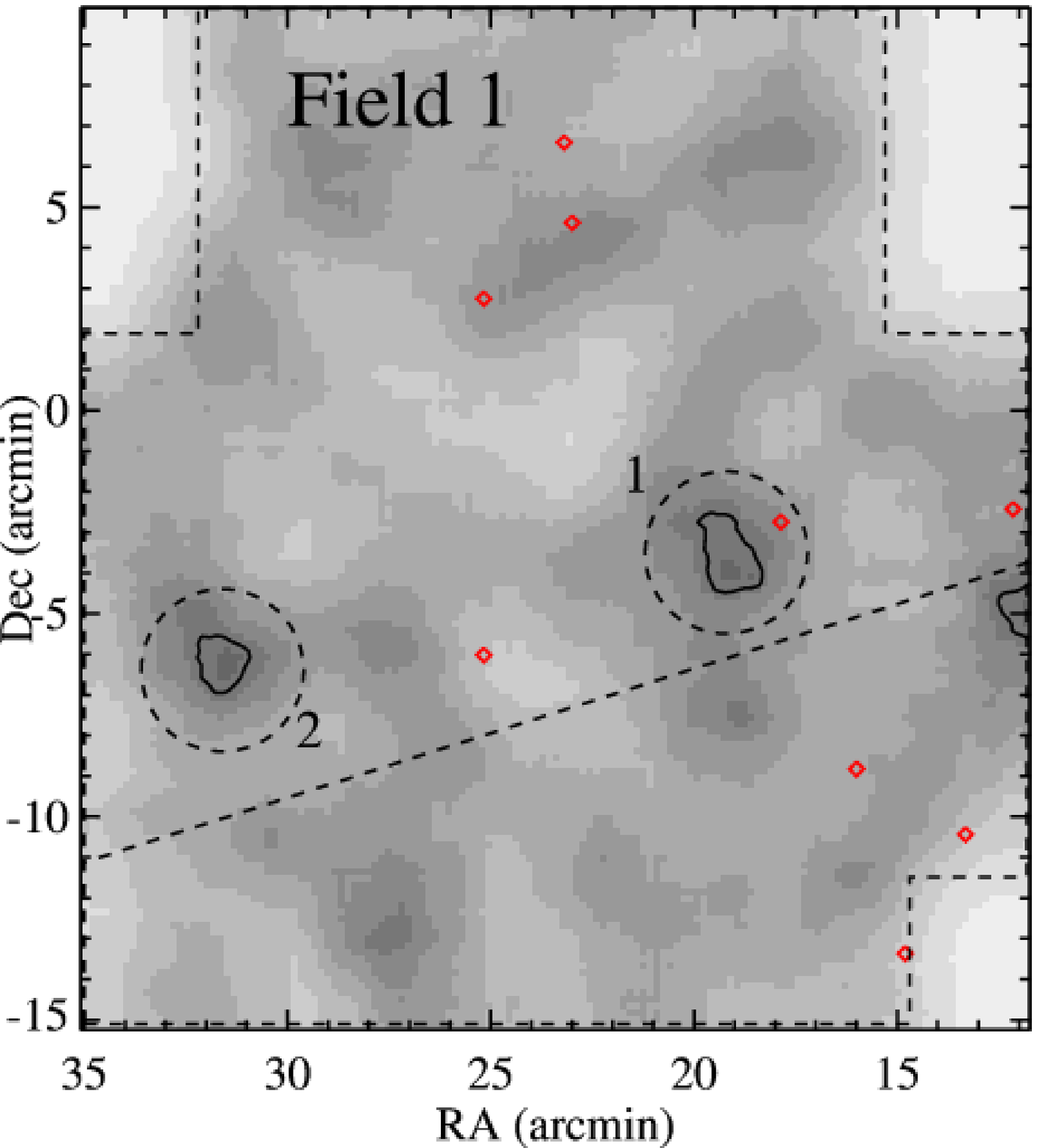}}
\mbox{\epsfysize=6.5cm \epsfbox{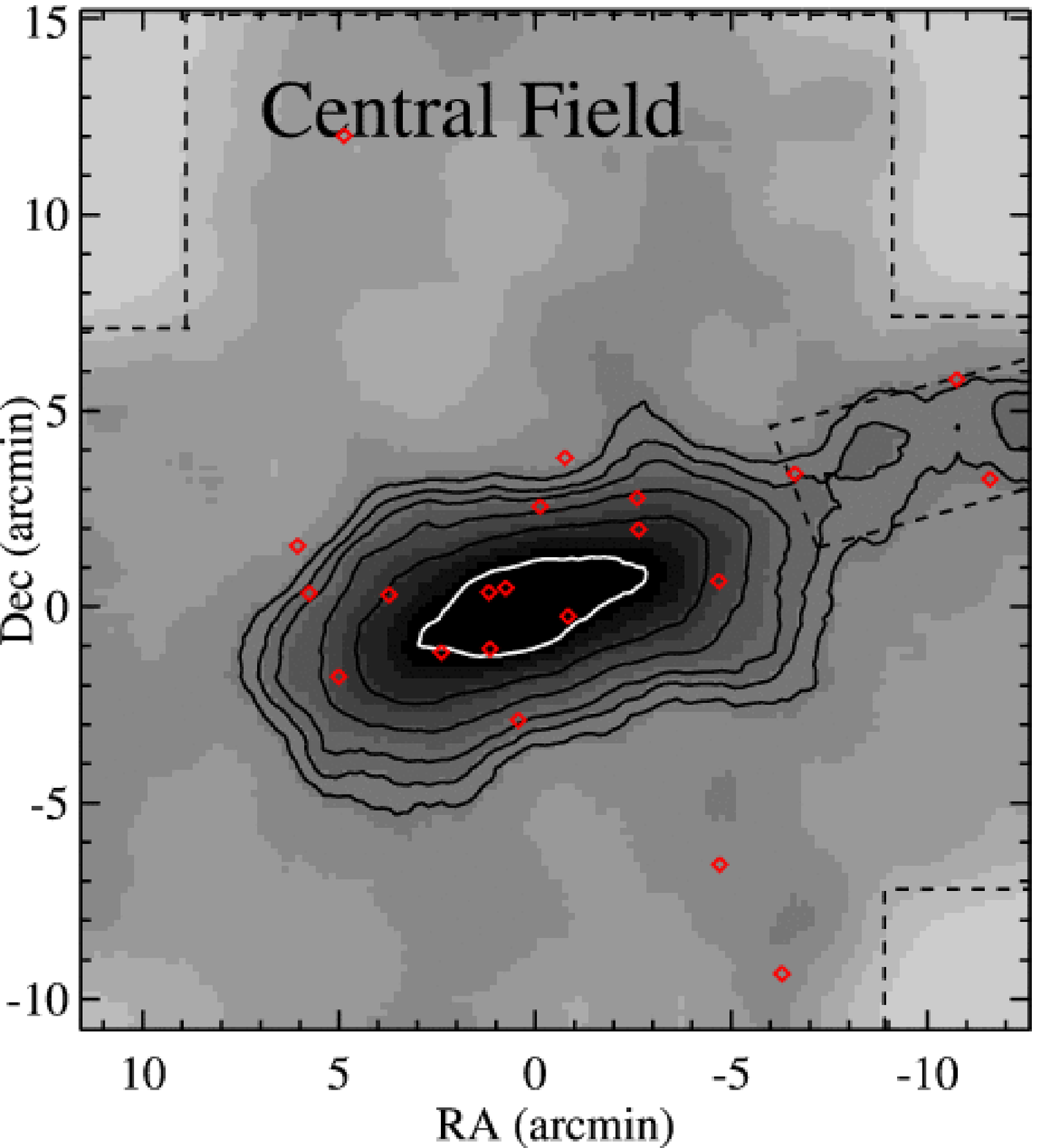}}
\mbox{\epsfysize=6.5cm \epsfbox{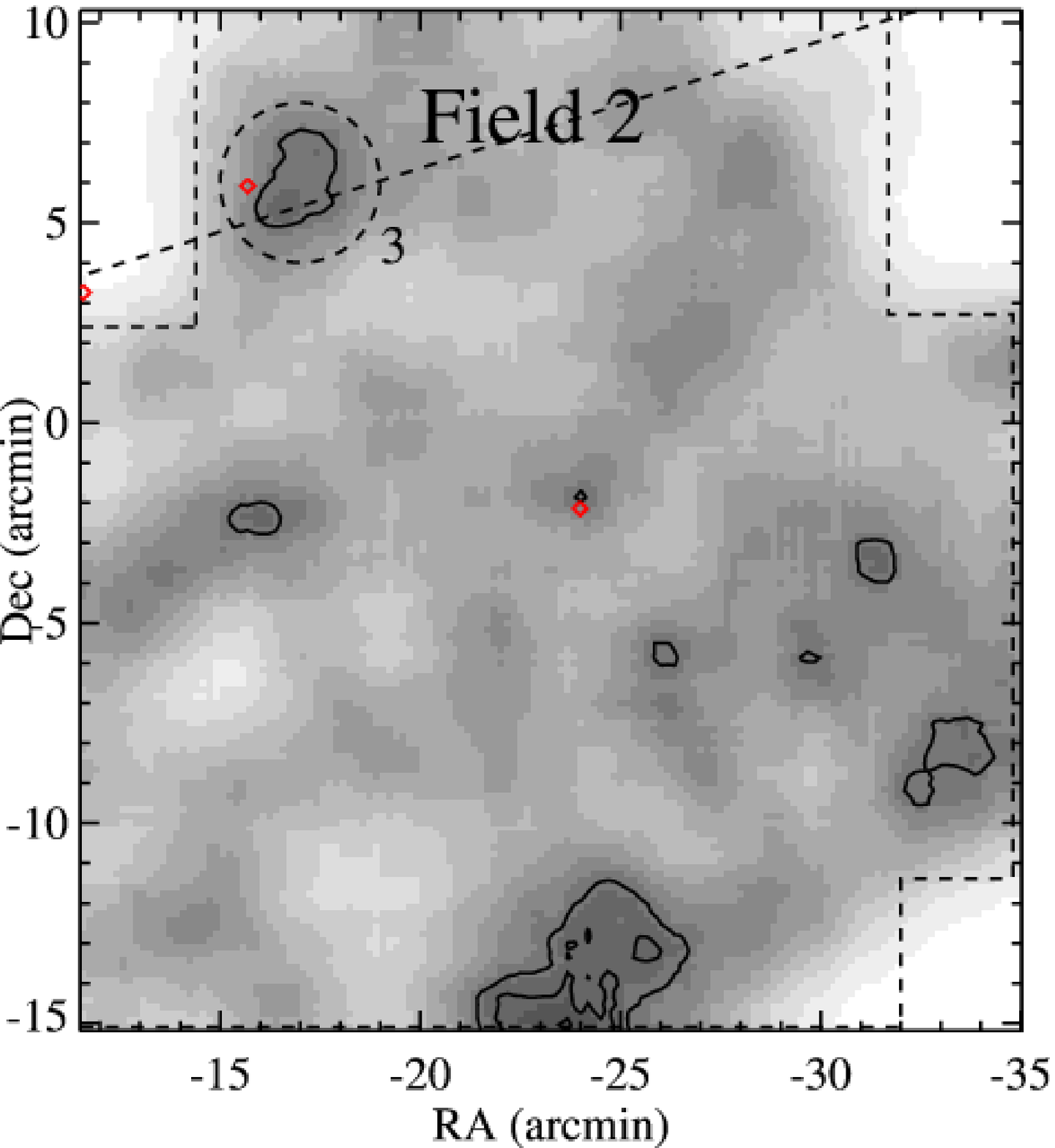}}
}
\mbox{
\mbox{\epsfysize=6.5cm \epsfbox{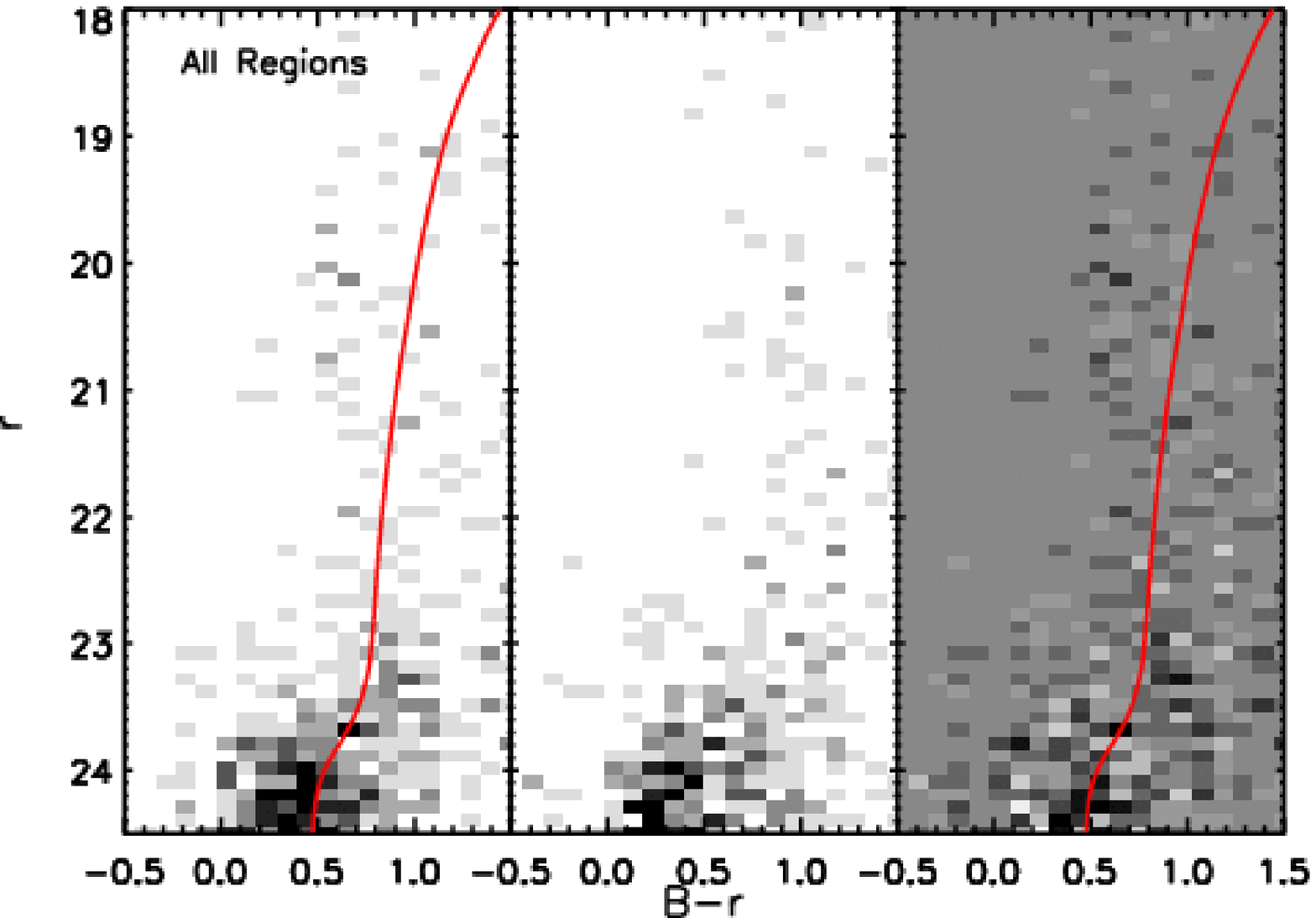}}
\mbox{\epsfysize=6.5cm \epsfbox{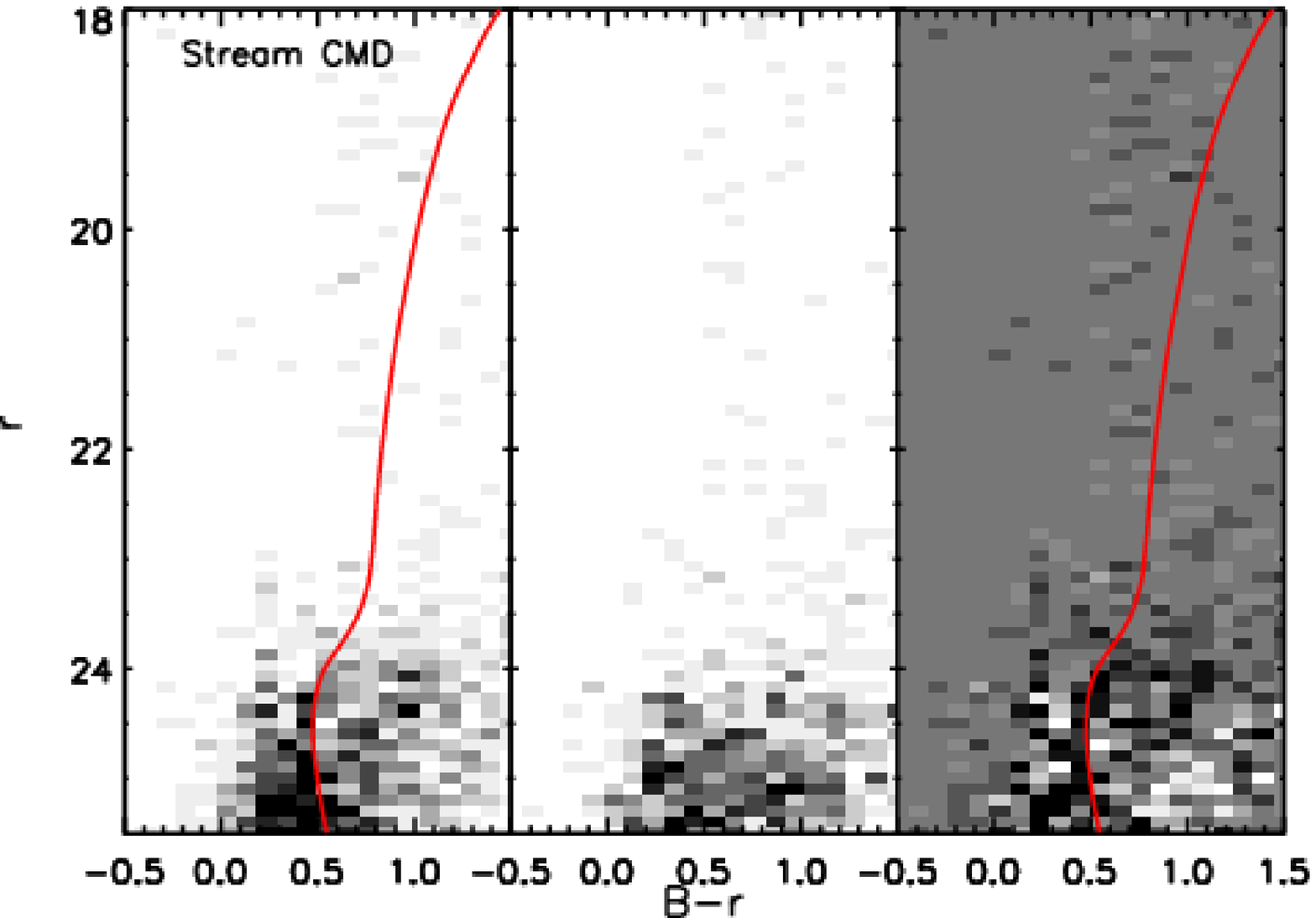}}
}
\caption{ Top Row -- Smoothed maps of Hercules and Fields 1 and 2,
highlighting possible external structure associated with Hercules.
The marked regions are used to make the two Hess diagrams shown in the
bottom row.  Candidate Hercules BHB stars are shown as red diamonds.
Bottom row -- Hess diagrams of the stream region in the central
pointing (Right) and the three nuggets projected nearly along the
major axis of Hercules (Left).  Both Hess diagrams have features that
are consistent with being Hercules stars, although the results should
be considered tentative given our randomized tests presented in
\S~\ref{sec:morphology}.
\label{fig:exten}}
\end{center}
\end{figure*}

We present the three relevant smoothed maps in Figure~\ref{fig:exten},
and have noted the regions of interest with dashed shapes.  In
addition to the smoothed map, we have also marked the position of
candidate blue horizontal branch stars -- those that are within
2$\sigma$ of the 14 Gyr, [Fe/H]=-2.3 \citet{Girardi04} isochrone -- as
red diamonds in Figure~\ref{fig:exten}.  In general, BHB stars suffer
from less foreground Galactic contamination than those on the red
giant branch or main sequence, and so can be a tracer of potential
external structure \citep{leov}.  Note that the stream has 3 potential
BHB stars, and that two out of the three proposed Hercules knots also
have a candidate BHB star.  There are two other BHB star candidates
along the position angle of Hercules, in Field 1, which are not
directly associated with either nugget.  Though by no means
definitive, the presence of these stars is encouraging.

First we make a background-subtracted Hess diagram of the extension to
the Northwest of Hercules, using stars in the dashed box in the middle
panel of the top row of Figure~\ref{fig:exten}, with an equal area
background taken from a region 10' north of Hercules (bottom right;
Figure~\ref{fig:exten}).  Indeed, there appears to be a main sequence
and perhaps a few RGB stars in this apparent stream, although the main
sequence is thinner than our expectation given our photometric errors.
Next, we have summed the stars in the three circular apertures in
Fields 1 and 2 (and stars in equal area background apertures), and
created a separate Hess diagram shown in the bottom left of
Figure~\ref{fig:exten}.  This CMD is reminiscent of those found when
implanting our artificial Hercules nuggets, with the beginning of a
main sequence coinciding with that of Hercules apparent.

While still tentative, it is possible we are detecting the highest
density features associated with a stream emanating from Hercules.  If
true, the 'nuggets' that we detect in our adjacent to Hercules fields
can plausibly be knots analogous to those seen in the tidal stream of
Palomar 5 \citep{odenkirchen03}.  We will discuss possible
implications of this scenario in \S~\ref{sec:discuss}.

\subsection{Distance modulus across the face of Hercules}\label{sec:DMxHerc}

Given its high ellipticity, radial velocity of $\sim$140 km/s away
from the MW (with respect to the Galactic standard of rest), and a
hint that it may contain kinematic substructure \citep{simongeha}, it
is worth exploring whether or not Hercules is significantly elongated
along the line of sight.  One complication is the fact that Hercules
appears to have an intrinsic spread in metallicity -- as seen in
spectroscopic work \citep{Kirby08} and confirmed by our SFH analysis
-- which presents a 'thicker' giant branch and main sequence than a
simple single stellar population does.  We keep this in mind as we go
through the following analysis.

\begin{figure*}
\begin{center}
\mbox{
\epsfysize=4.5cm \epsfbox{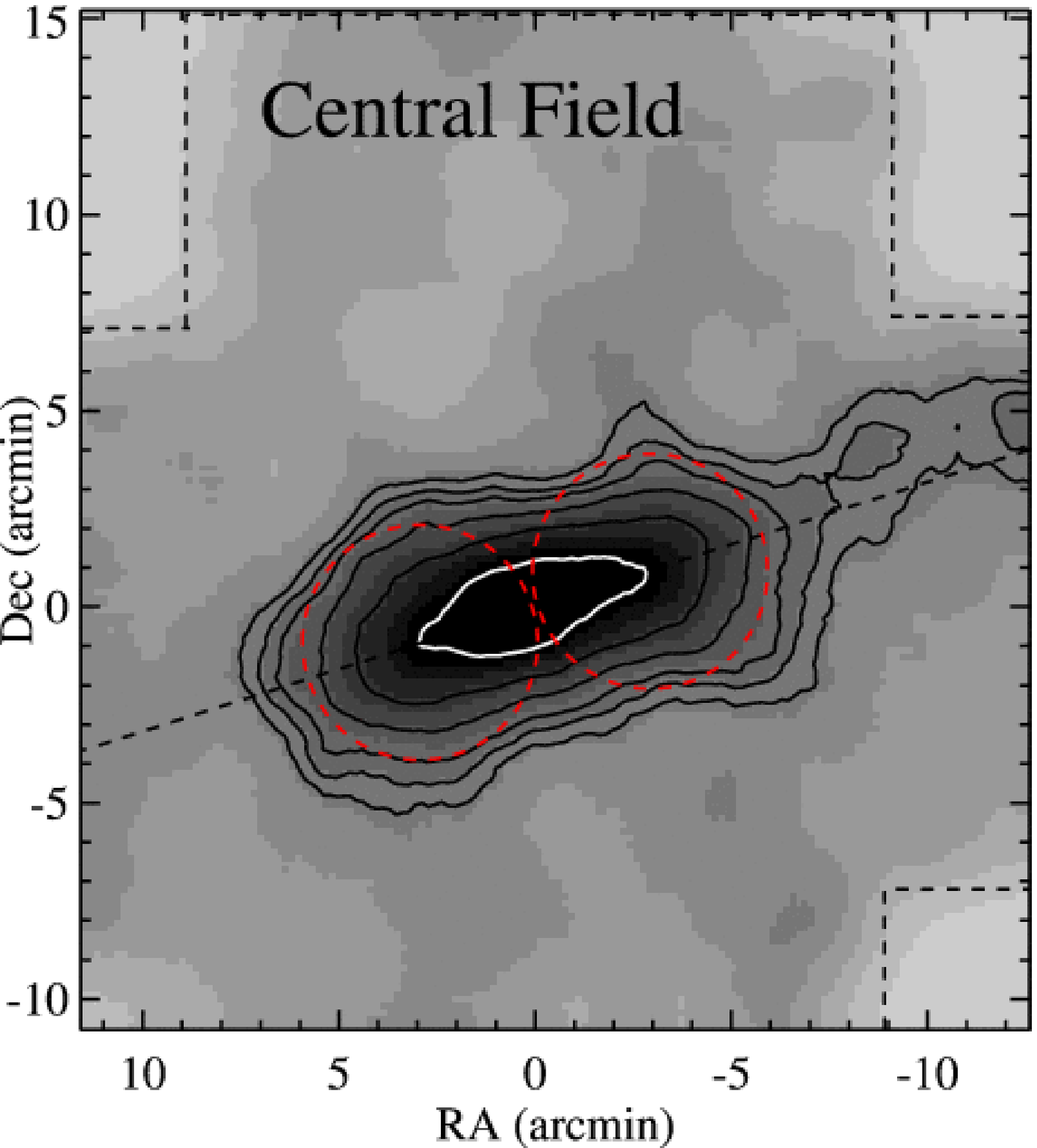}
\epsfysize=4.5cm \epsfbox{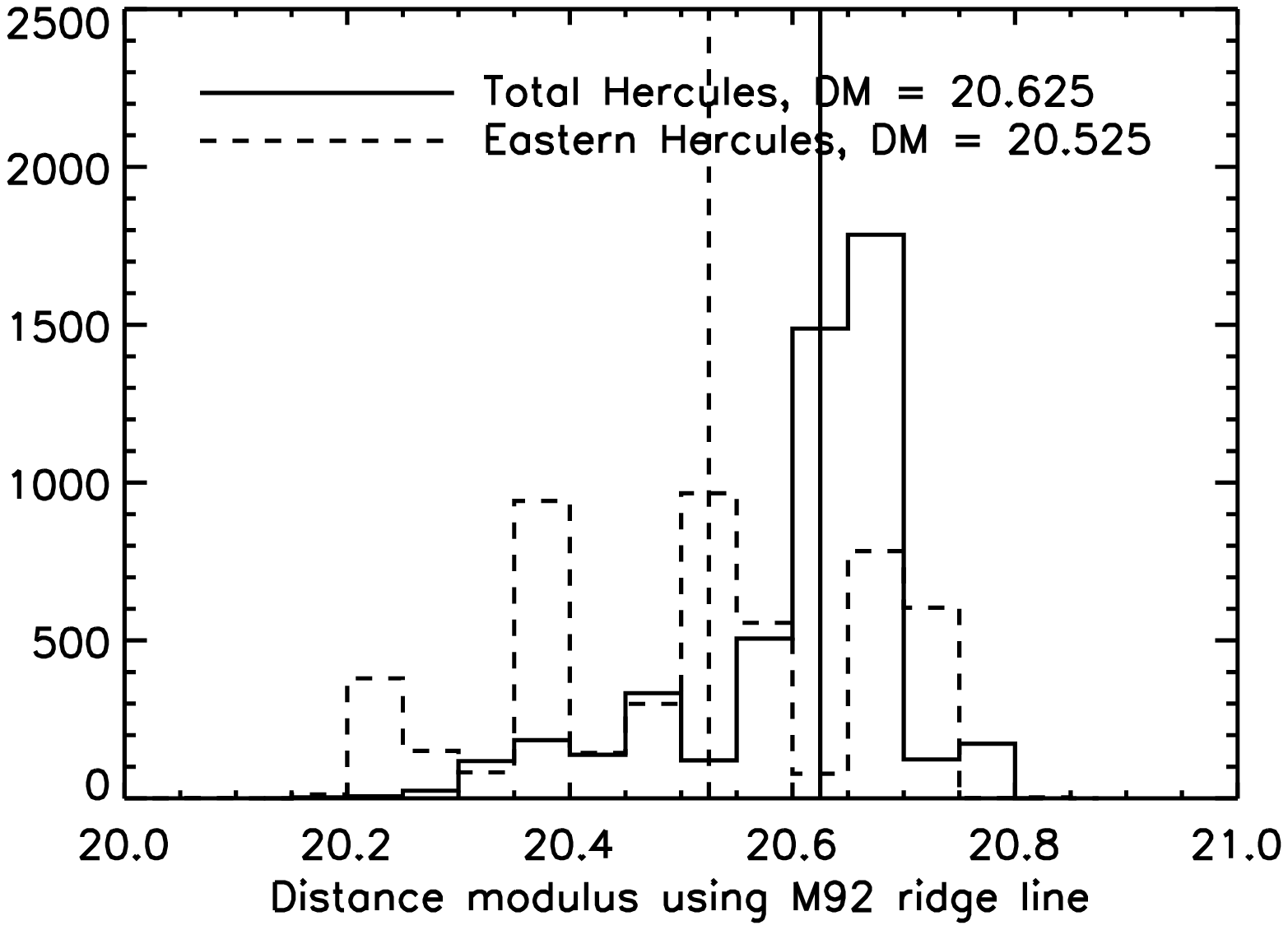}
\epsfysize=4.5cm \epsfbox{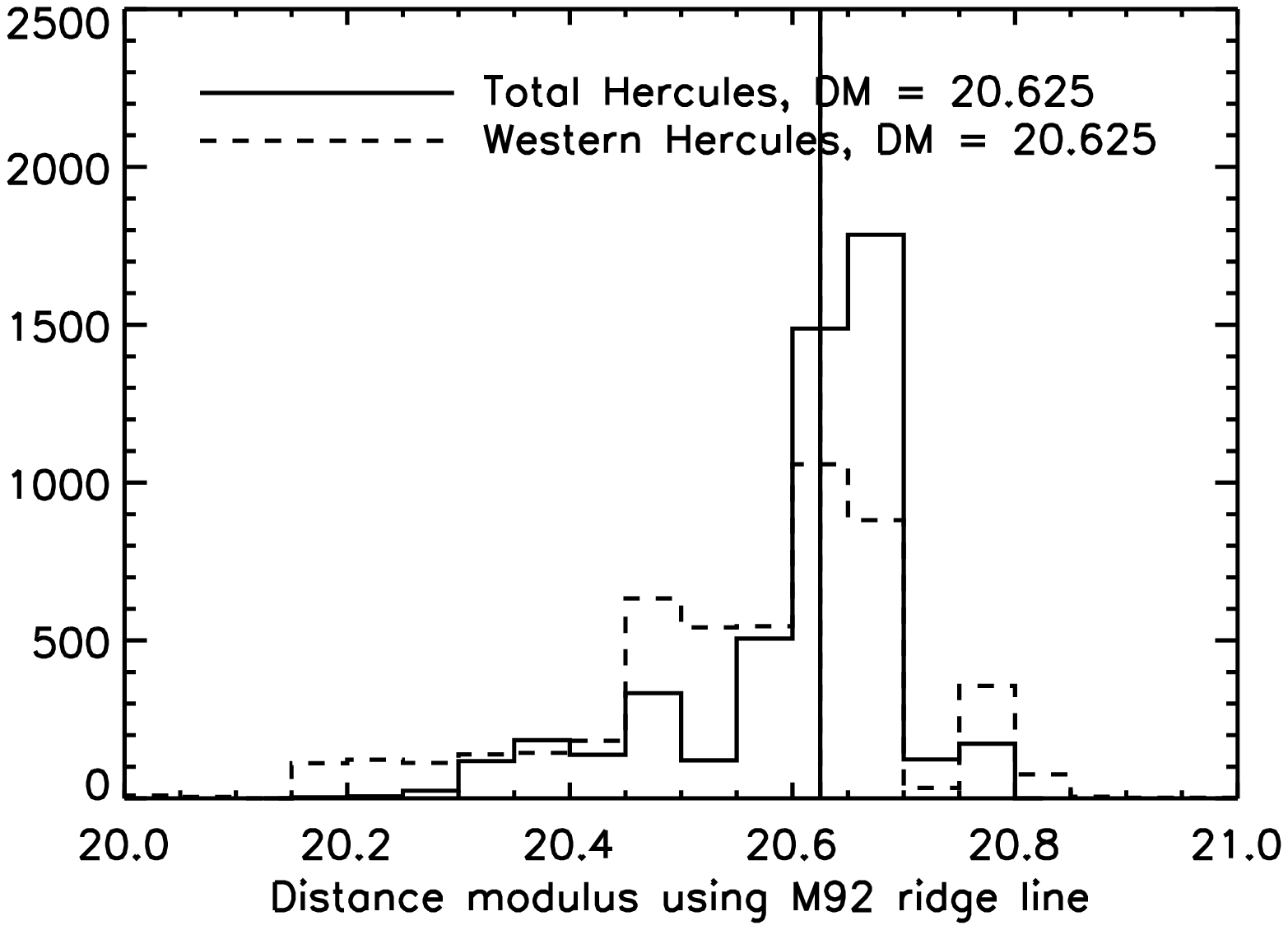}
}
\mbox{
\epsfysize=5.0cm \epsfbox{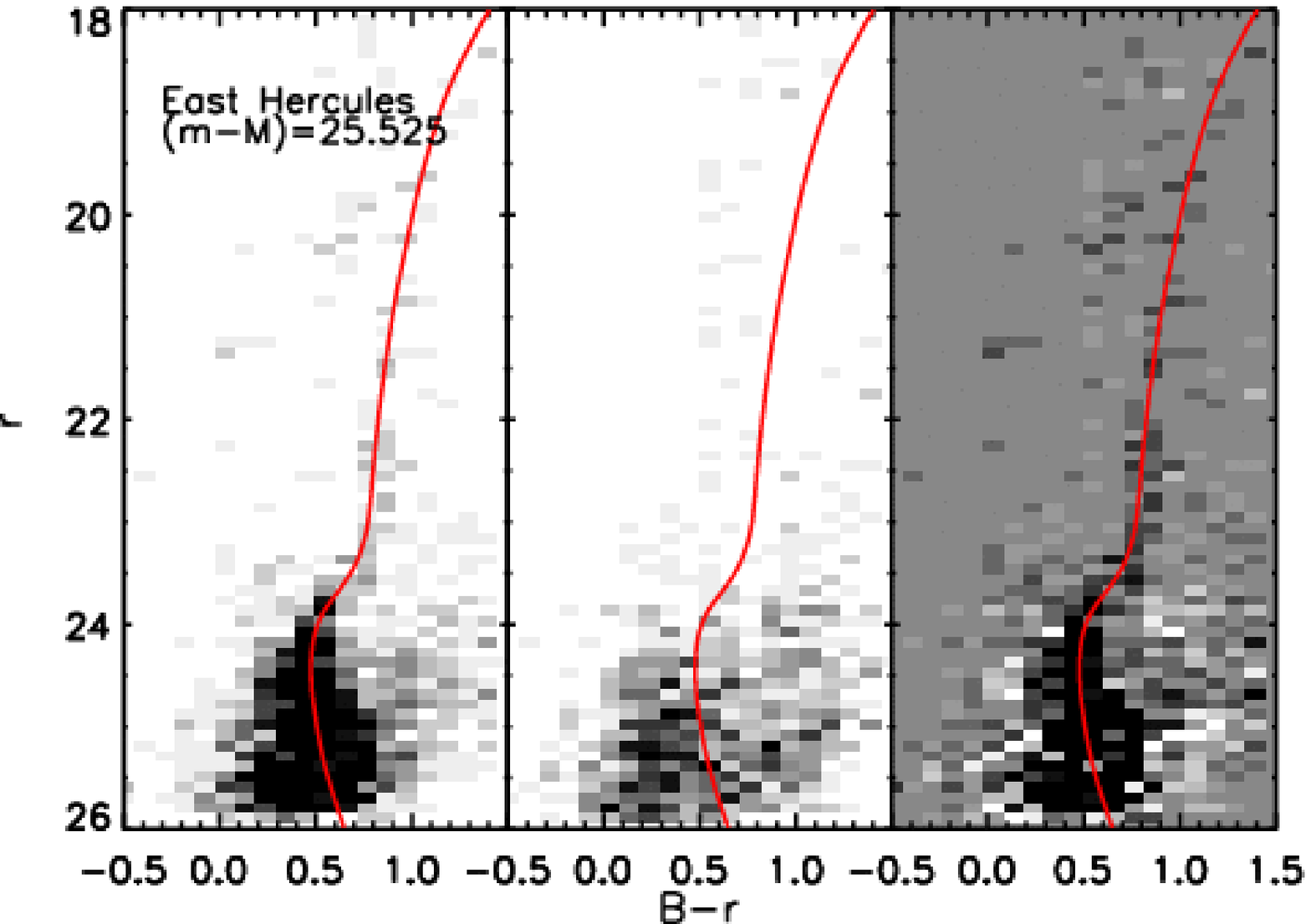}
\epsfysize=5.0cm \epsfbox{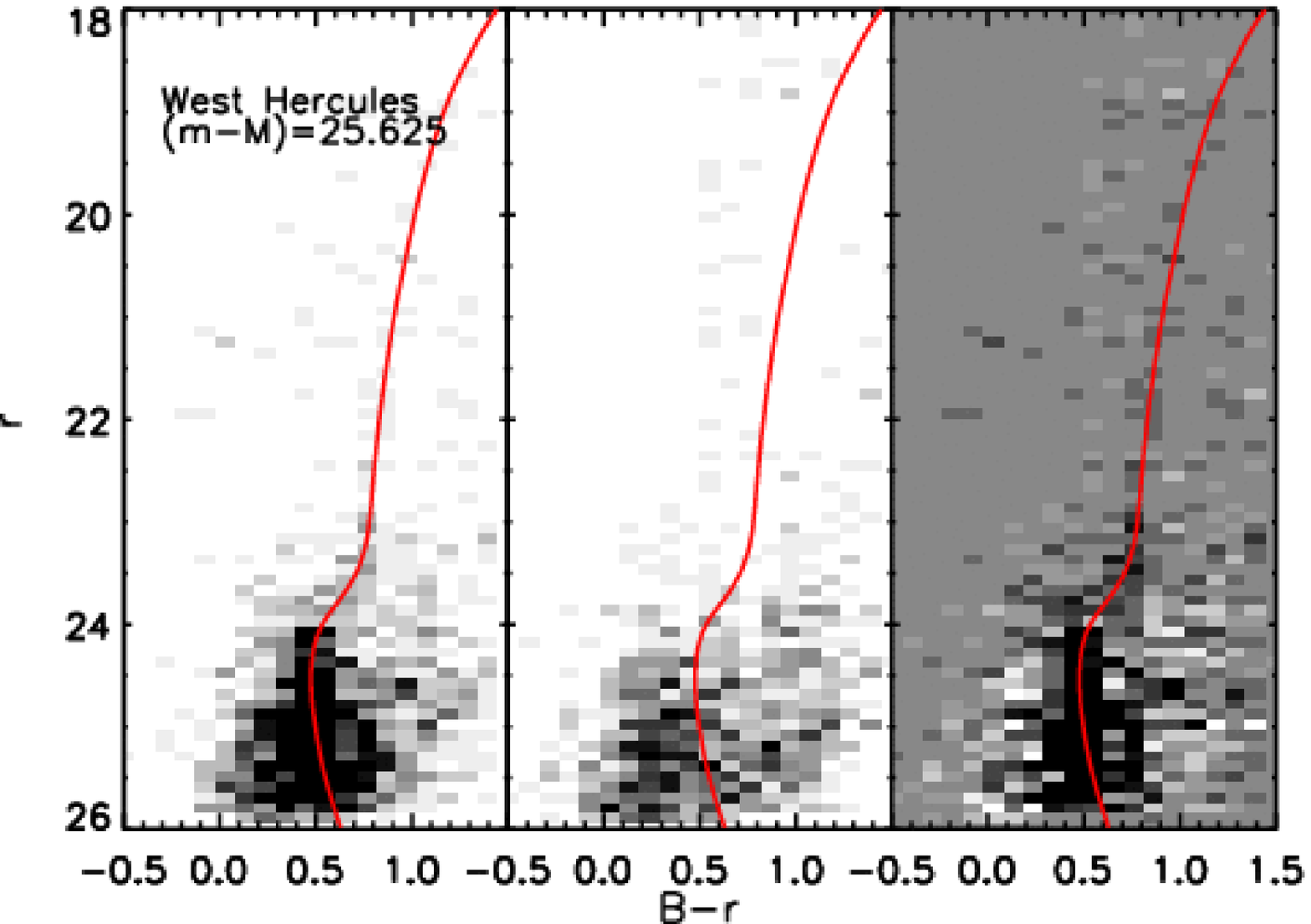}
}
\caption{ Upper Left -- An illustration of the two regions in which we
separately measured the distance modulus, as in \S~\ref{sec:dist},
which roughly covers the central half light radius (5.9 arcmin). Upper
center and right -- Bootstrap histograms of the distance modulus for
the eastern and western portions of Hercules, in comparison to that of
Hercules as a whole.  Note that the Eastern portions seems to be
closer (127 kpc) and has a very different histogram in comparison to
Hercules as a whole and its western portion.  Bottom left and right --
Hess diagrams of the East and West portions of Hercules, along with
the best-fitting M92 isochrone transformed to the appropriate distance
modulus.
\label{fig:xHerc}}
\end{center}
\end{figure*}

As a first test, we remeasure the distance modulus as in
\S~\ref{sec:dist}, but now in two circular regions with radius 3
arcminutes on either side of Hercules' central position; see
Figure~\ref{fig:xHerc} for an illustration.  We use a background
region with radius of 3 arcminutes situated 12 arcminutes north of the
centroid of Hercules to account for background/foreground
contamination.  We focus on the M92 fiducial, which was found to be
the best-fitting empirical isochrone in \S~\ref{sec:dist}.  Note that
splitting Hercules up into 3 circular regions with radius of 2
arcminutes give similar results, but the distance modulus bootstrap
histograms are even less clearly defined, due to small number
statistics.  We present the result of our bootstrap analysis in
Figure~\ref{fig:xHerc}, along with background subtracted Hess diagrams
of both portions of Hercules.  The eastern Hercules aperture indicates
that it is at $(m-M)$=20.525 mag (127 kpc), closer than the western
portion and of Hercules taken as a whole ($(m-M)$=20.625 mag; 133
kpc).  The eastern bootstrap histogram is broad, with multiple peaks,
unlike the histogram for the western portion of Hercules which is very
similar to that when Hercules is taken as a whole.  While this is
interesting, the result must be taken with caution -- despite their
differences, the eastern bootstrap histogram does have significant
overlap with both the western bootstrap and Hercules as a whole.

Intrigued by these results, we decided to fit a model with Hercules'
distance changing linearly as a function of major axis distance, and
with no binning of the data.  This model is appropriate if Hercules is
actually a stellar overdensity in a thin stream whose length we are
observing nearly along the line of sight.  We assume that the center
of Hercules is at $(m-M)=20.625$ (133.4 kpc), as found in
\S~\ref{sec:dist}, and allow the observer-Hercules distance to change
as a function of the major axis distance --

\begin{equation}
Distance(x_i) = mx_{i} +133.4 (kpc)
\end{equation}

\begin{figure*}
\begin{center}
\mbox{
\epsfysize=6.0cm \epsfbox{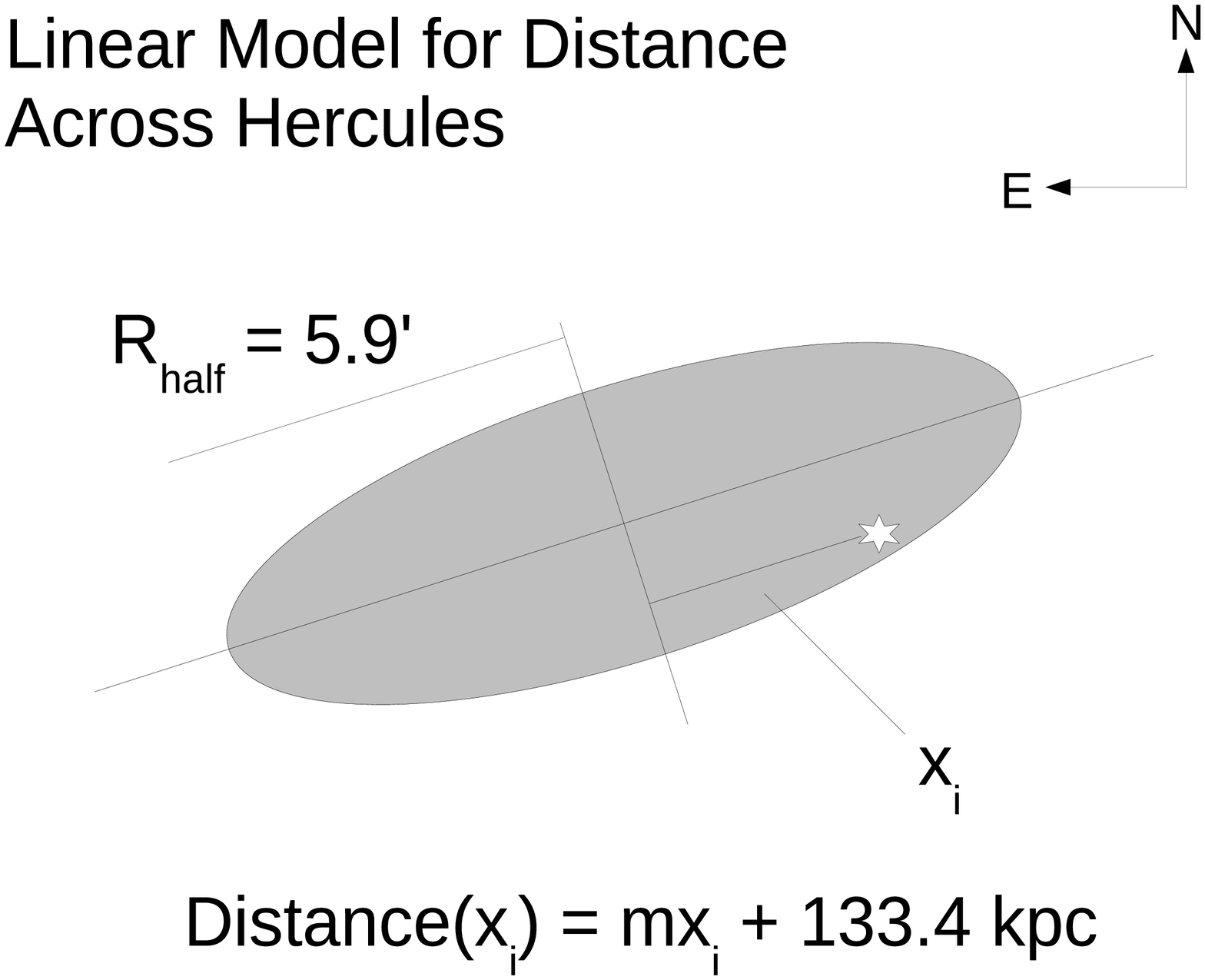}
}
\mbox{
\epsfysize=6.0cm \epsfbox{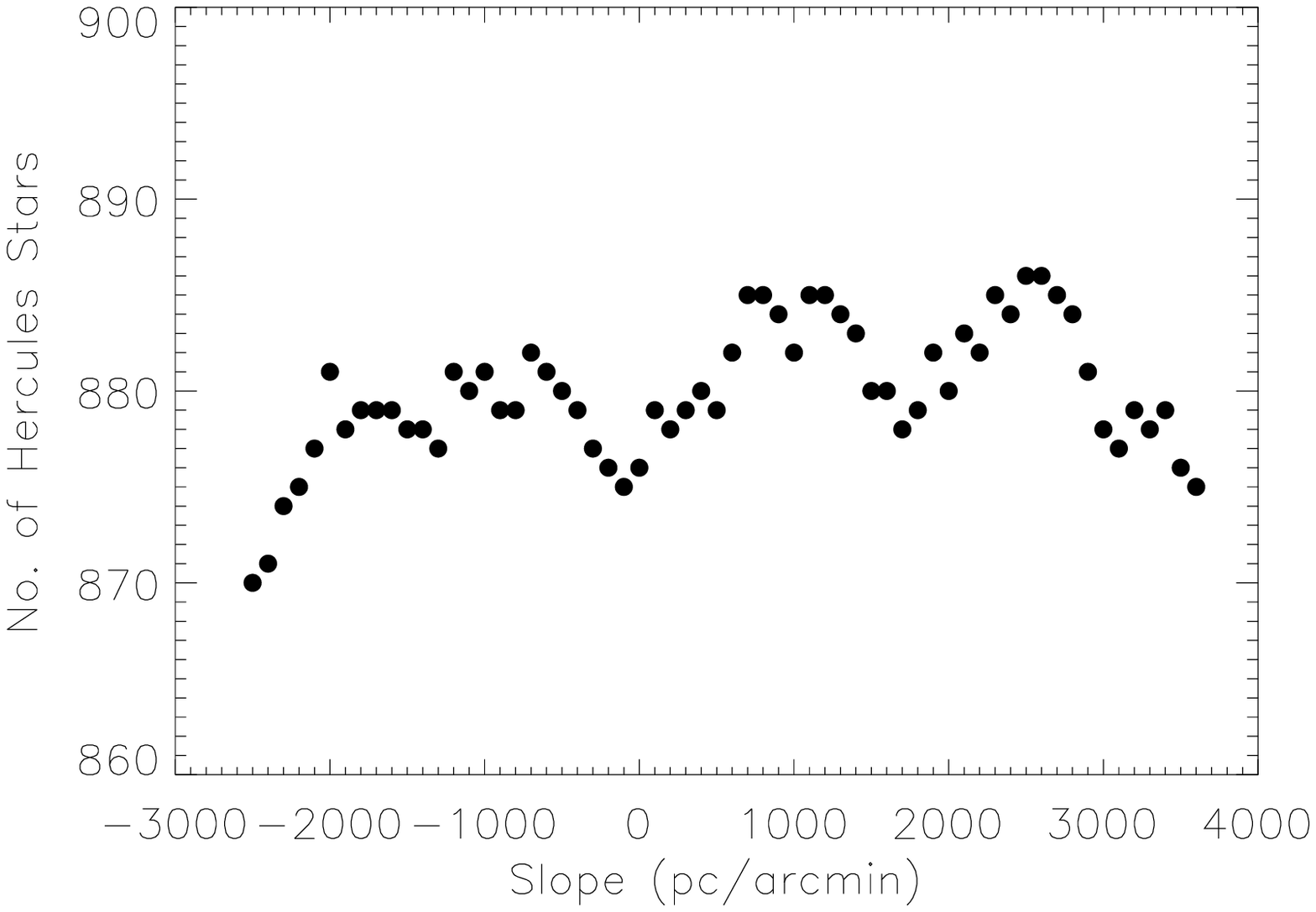}
\epsfysize=6.0cm \epsfbox{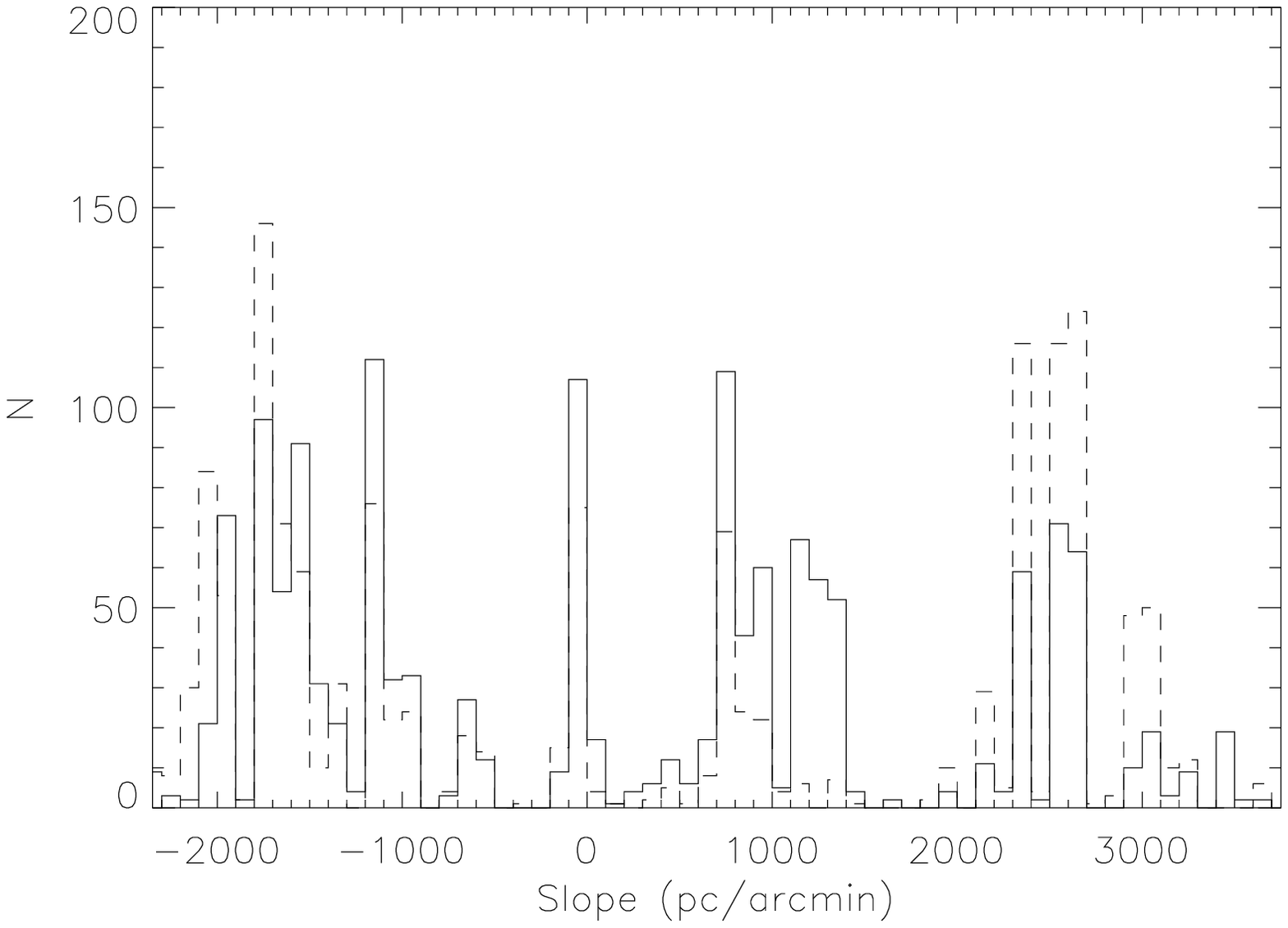}
}
\caption{Top -- A cartoon illustrating our model for the
observer-Hercules distance changing as a function of major axis
distance.  Bottom Left -- The number of Hercules stars, after
background subtraction, as a function of slope, $m$.  Bottom Right --
Histogram of the best-fitting slope with 1500 bootstrap resamples in
which we fit our linearly changing distance model.  The solid
histogram corrsponds to the best-fitting slope that maximized the
number of Hercules stars for each of our bootstrap resamples, while
the dashed line indicates which slope corresponded to the minimum
number of background stars.  There is no dominant, preferred slope and
the best-fitting slope very often corresponds to a minimum in the
background, rather than a maximum in the Hercules-centered ellipse.
\label{fig:xHerc2}}
\end{center}
\end{figure*}

\noindent For a given slope, $m$, and major axis distance for the
$i$-th star, $x_{i}$, the presumed distance to a Hercules member is
known and if its magnitude and color are consistent (to within
$1-\sigma$) with the isochrone of M92 transformed to that distance,
then it is tallied.  We restrict our analysis to the central
$r_h$=5.9', and use several identically shaped ellipses between 9-12'
north of the center of Hercules as our control.  Modeled in the same
way, we subtract the number of stars consistent with our linear model
in the background region from those found in Hercules.  We choose to
vary $m$ between -2500 and 3600 pc/arcmin.  The cartoon in
Figure~\ref{fig:xHerc2} illustrates our model, while the bottom panels
summarize the results over 1500 bootstrap resamples.

As can be seen from the bootstrap-derived histogram in the bottom
right of Figure~\ref{fig:xHerc2}, there is no clearly preferred slope.
Also, the dashed histogram -- which represents the slope where the
background star counts corresponded to a minimum -- shows that more
often than not we are really just measuring the slope that corresponds
to a minimum in the background rather than a true Hercules maximum.
Of course, this does not mean that Hercules has no depth along the
line of sight.  It may just mean that our linear model for distance
across Hercules is too simple and does not correspond to reality.

Despite our careful search, we find no conclusive evidence that
Hercules is elongated along the line of sight.  At best, we can
constrain any line of sight depth to be roughly the same size as the
difference between the East and West portions of Hercules, but even
then the uncertainties in these two measurements overlap at the
$1-\sigma$ level.  Within the half light radius, Hercules has at most
a difference in distance modulus of $\sim$0.1 mag between its Eastern
and Western portions, which corresponds to $\sim$6 kpc.  Note that
this limit is not particularly stringent, and is an order of magnitude
larger than the projected size of Hercules on the sky.

\section{Discussion \& Conclusions}\label{sec:discuss}

In this work, we have presented a comprehensive imaging study of the
Hercules MW satellite and surrounding regions.  With this, the first
very wide-field study of one of the recently discovered SDSS
satellites, we have determined the stellar population and structural
properties of Hercules, and have thoroughly searched for signs of
extended structure.

In utilizing a ML technique analogous to that presented by
\citep{Martin08}, we have fit the structure of Hercules to several
standard parameterized models -- an exponential, Plummer and King
profile.  We confirm that Hercules is extremely elliptical, with
$\epsilon=0.67$.  Our structural parameters are consistent with those
presented in the literature \citep[e.g.][]{Coleman07,Martin08},
however, we also demonstrate that data deeper than the discovery SDSS
data are essential for properly characterizing the structural
properties of the new satellites.  For instance, using SDSS data
identical to that used by \citet{Martin08}, along with bootstrap
resampling to determine our uncertainties, we find that Hercules has a
half-light radius of 7.65'$\pm$5.16'. This constraint tightens to
$r_{h}$=5.91'$\pm$0.50' with our LBT data set.  It is critical that
all of the satellites be followed up with deep, wide-field imaging in
order to properly characterize their structural properties.

With the CMD-fitting software package StarFISH, we find the stellar
population of Hercules to be old ($>$12 Gyr) and metal poor
($[Fe/H]\sim-2.0$), albeit with a spread in metallicity.  It is
interesting to compare the SFH of Hercules both with the other UF
satellites and with all satellites that are $>$100 kpc from the MW.
Of the new UF dwarf spheroidals (excluding Leo T, which appears to be
a different class of object), only Ursa Major II and Can Ven I have
clear, multiple epochs of star formation \citep{sdsssfh}.  That said,
formal CMD-fitting has not been performed on data deeper than the SDSS
\citep[again excluding Leo T; ][]{leot}, and so it remains to be seen
if many of the new MW satellites harbor small, young populations.
Nonetheless, if the current picture of the new satellites holds,
Hercules is in the mainstream of these objects, with an old and metal
poor population.

The SFH of the classical dSph's as a function of MW distance has long
been thought to provide clues as to the relative importance of
environmental processes (e.g.~ionizing radiation, ram pressure
stripping, and supernova feedback) in the galaxy formation process
\citep[e.g.][]{vandenbergh94}.  For instance, of the eight classical
dwarf spheroidals (excluding the disrupting Sagittarius), the four
nearest ($\lesssim$90 kpc) all have primarily ancient stellar
populations, while the four furthest all have extended SFHs \citep[for
a review of the SFHs of the classical satellites, see][]{Dolphin05}.
Two interpretations are possible when looking at the classical dSph's
alone, assuming that the present MW distance of a given satellite is
representative of its average MW distance \citep[which is not likely
to be the case for all the satellites, e.g. Leo I;][]{M08}.
First, this could signal that environment plays a large role, where
tidal and ram pressure stripping of gas or local ionizing radiation
serves to truncate SF in the nearest dwarfs.  Alternatively,
pericentric passages trigger SF, and the nearest systems -- which have
had more such passages -- exhaust their gas quickly
\citep{HZ04,ZH04}.  In this scenario, the new UF satellites would
exhibit a similar dichotomy in their SFHs as a function of MW distance
as the classical dSphs.  The dual nature of the classical dSph's star
formation as a function of MW distance could also simply be a function
of the size of the initial baryonic reservoir, since the four nearest
are also the least luminous.  In this instance, the UF satellites
would all have primarily ancient stellar populations.  While SFHs of
the new dwarfs must be derived with deeper data, the presence of
multiple epochs of SF in both UMaII (D=30 kpc) and Can Ven I (D=218
kpc), and our result that Hercules (D=133 kpc) has solely an ancient
stellar population seems to muddy any picture where either environment
or initial baryonic reservoir {\it solely} determine a satellite's
SFH.

As we have mentioned previously, Hercules is a prime candidate for
wide-field followup due to its extended morphology and possible hint
of kinematic substructure.  To investigate this possible kinematic
substructure, J. Simon and M. Geha kindly provided their kinematic
data on Hercules, which we have overplotted onto our smoothed map of
the central field (Figure~\ref{fig:velstruct}).  Red diamonds indicate
the location of possible 'substructure' stars -- those with velocities
between 41 and 43 $km s^{-1}$ -- while the blue diamonds are the
position of the other stars identified with Hercules.  There is no
obvious spatial correlation between Hercules structural features and
the spatial position of the candidate 'substructure' stars.  Further
kinematic work will be necessary to confirm any non-Gaussian features
in Hercules' velocity histogram.

Can we find a plausible orbit for Hercules that explains its current
structure and possible orbital debris?  Hercules' great distance (133
kpc), and high velocity away from the MW (145 km s$^{-1}$) mean that it
likely has a fairly elongated orbit.  \citet{klessen03} explored the
consequences of a purely tidal model, with no dark matter and radial
orbits, for the dwarf spheroidal galaxy Draco.  The hope was that
depth along the line of sight in that system would explain Draco's
properties.  While this model was unsuccessful in that instance, they
did show that slight variations in viewing angle along the 'barrel' of
a tidal stream could produce the appearance of a highly elongated
morphology -- similar to that seen in Hercules (see Figure 3 of
Klessen et al. 2003).  In \S~\ref{sec:DMxHerc}, we searched for and
found no evidence for depth along the line of sight, which suggests
that a scenario in which we are looking at a tidal stream nearly down
its length will have some difficulty explaining Hercules' unusual
elongation -- although our limit to the line of sight depth of $\sim$6
kpc is not particularly stringent yet.  However, if our tentative
evidence that Hercules is embedded in a larger stellar stream seen in
the plane of the sky is correct, it is still worth considering that
Hercules had a significant encounter with the MW at perigalacticon.

As an exercise, we calculated a series of orbits for a point-like
Hercules model in a static, multicomponent Galactic potential
identical to \citet{Johnston95} with reasonable assumed values of the
tangential velocity along both the minor and major axis of Hercules.
The resulting orbits can then be compared with the observed orbital
properties of the other MW satellites.  We explore eight different
cases, with four having a proper motion vector in the Galactic Rest
Frame (GRF) along Hercules' major axis and to the East (cases 1-4),
while four have GRF tangential motion along Hercules' minor axis and
to the North.  Orbital elements for our eight principal cases are
presented in Table~\ref{table:orbit}. Cases 1 through 4 all exhibit
roughly polar orbits, a common feature among the MW satellites
\citep[e.g.][]{Palma02}.  The pericentric radius of cases 1 and 2 make
it unlikely that Hercules would have survived as a bound object over a
Hubble time, while case 3 would likely have also caused damage
\citep[see e.g. ][]{Mayer01,Mateomess}.  Case 4 would probably allow
for Hercules to survive a Hubble time.  If case 3 is close to
Hercules' true orbit, then it would lie near the orbital pole grouping
of some of the other dwarf spheroidals claimed by \citet{Kroupa05}.
In cases 5 through 8, where the GRF tangential motion is along
Hercules' minor axis and to the North, the inclination approaches the
equatorial case.  Only case 8 would allow long-term survival of
Hercules.  These models are all speculative at this point, but if
Hercules has similar orbital properties as the other MW satellites, we
would expect future proper motion measurements to be nearest our case
3, due to the polar orbit, the orbital pole in the vicinity of the
other satellites, and an intermediate perigalacticon distance.  With
the refurbishment of HST and the installation of WF3, it will be
possible to measure the proper motion of Hercules and derive its
orbit, which will be critical for understanding the true nature of
this object.

There is more wide-field imaging work to be done on Hercules.  It
would be worth obtaining more complete sky coverage in its vicinity,
and to go to greater depth.  For instance, our western pointings were
not optimally placed to search for external structure in that
direction and deeper data in our eastern fields (comparable to that
obtained in our central pointing) would shed light on whether or not
the eastern nuggets are truly remnants of Hercules.  Data to fill in
regions to the north and south of Hercules would also be of interest.

While the nature of Hercules remains elusive, we now have the tools in
place to study all of the new MW satellites in detail.  There is a
critical need for more in depth study of all of the new MW satellites
to understand their structure, star formation history and dynamical
state -- and to ultimately put them into context with respect to the
Cold Dark Matter paradigm for structure formation.

\acknowledgments

Many thanks to the LBC SDT team and the Arizona LBT Queue run
observers.  John Hill and Olga Kuhn were essential in our obtaining
our Hercules data.  We would like to thank Ben Weiner and Michael
Cooper for taking extra care when performing some of these
observations.  We are grateful to Josh Simon and Marla Geha for
providing their kinematic data on Hercules for this paper.  Also, we
are grateful to John Moustakas who provided an initial version of his
LBC reduction code.  EO was partially supported by NSF grants
AST-0505711 and 0807498.  DZ acknowledges support from NASA LTSA award
NNG05GE82G and NSF grant AST-0307492.

\bibliographystyle{apj}
\bibliography{apj-jour,mybib}

\clearpage

\clearpage

\begin{inlinefigure}
\begin{center}
\resizebox{\textwidth}{!}{\includegraphics{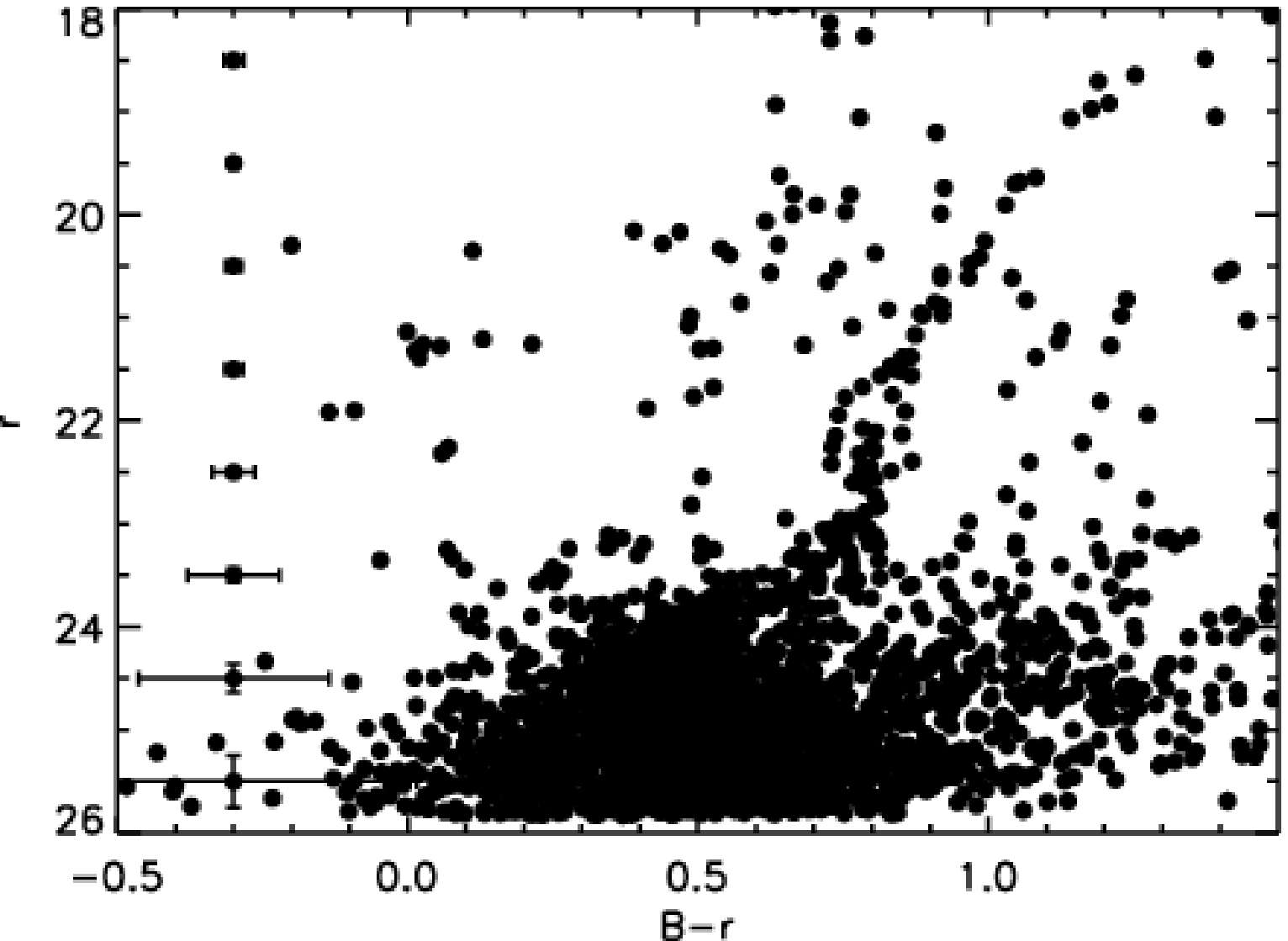}}
\end{center}
\figcaption{$B-r$ versus $r$ color magnitude diagram of stars within a
5.9' elliptical radius (the exponential profile half light radius) of
the center of Hercules.  Error bars showing the color and magitude
uncertainties as a function of $r$ are overplotted.  
\label{fig:cent_cmds}}
\end{inlinefigure}

\clearpage

\clearpage

\begin{inlinefigure}
\begin{center}
\resizebox{\textwidth}{!}{\includegraphics{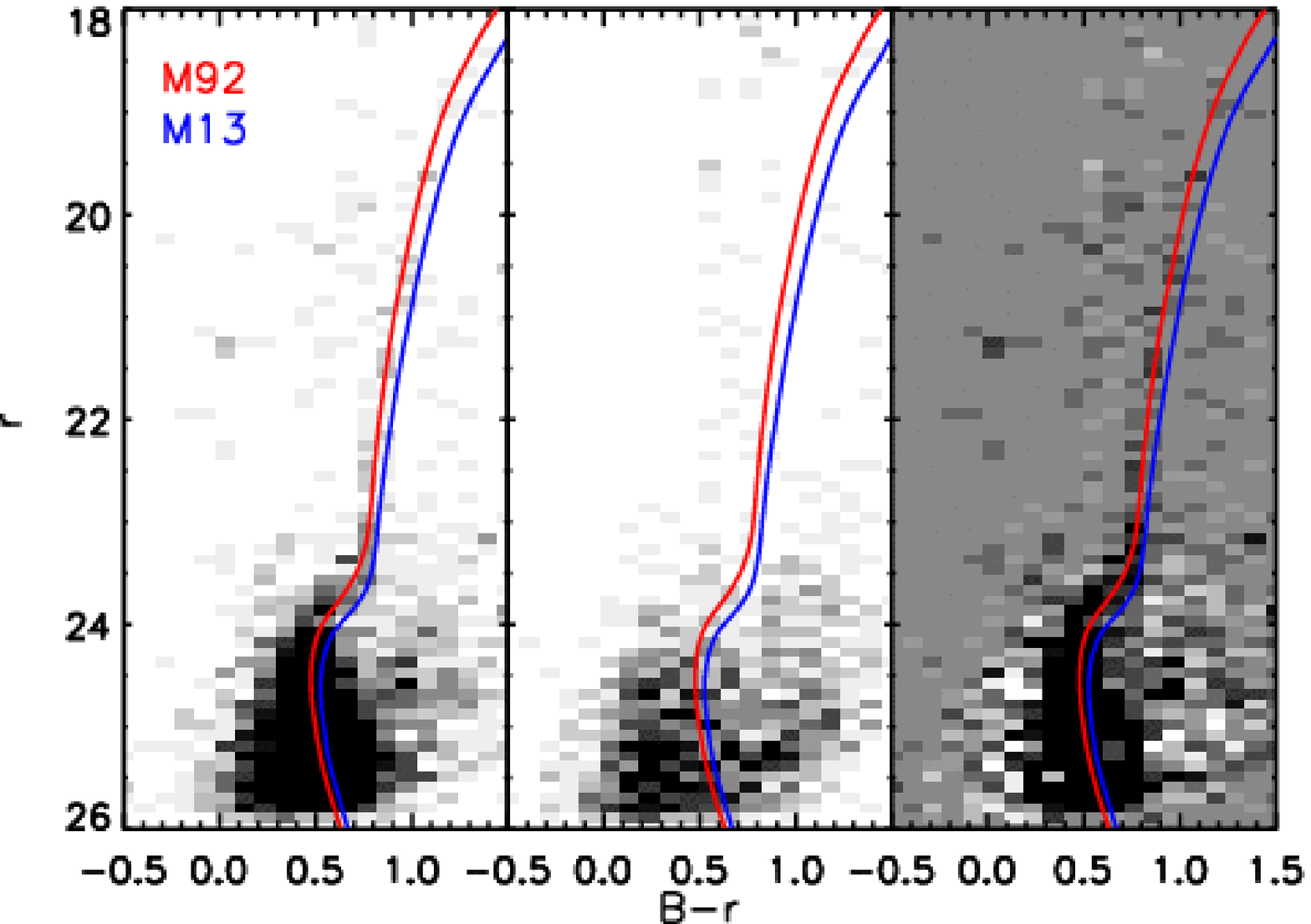}}
\end{center}
\figcaption{Hess diagram with both M92 ($[Fe/H]=-2.4$;red) and M13
($[Fe/H]=-1.57$;blue) at a distance modulus of 20.625.  While the M92
isochrone produces the best match to Hercules' CMD, there may be lower
metallicity components to its stellar population.  Given the zeropoint
uncertainties of $\delta r \sim$0.03 and $\delta B
\sim$0.05, our CMDs are also consistent with a more moderate metal
abundance.
\label{fig:DMwm13}}
\end{inlinefigure}

\clearpage

\begin{inlinefigure}
\begin{center}
\resizebox{\textwidth}{!}{\includegraphics{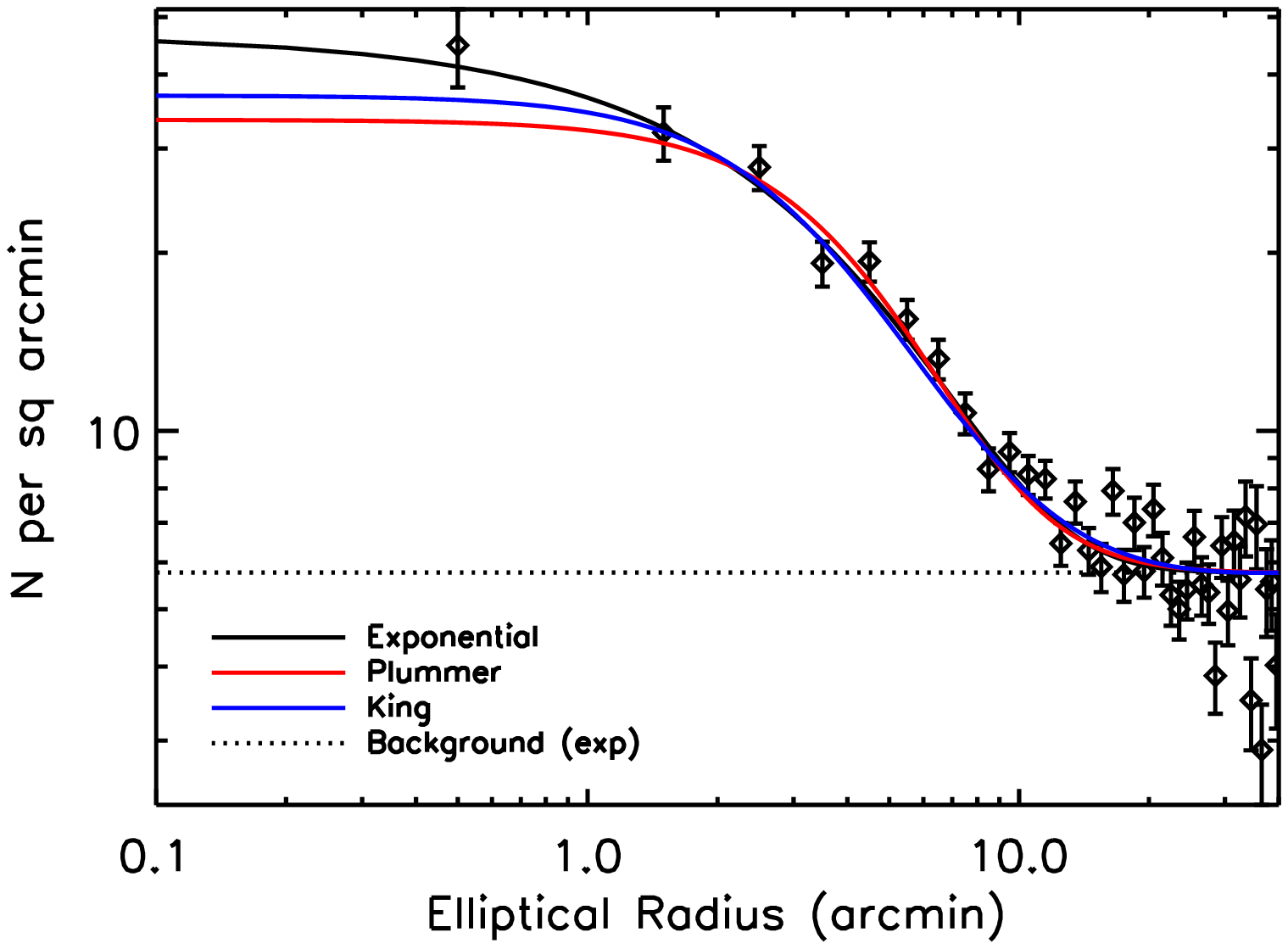}}
\end{center}
\figcaption{Stellar profile of Hercules.  The data points are the
binned star counts for all stars in our central pointing which are
consistent with the ($m-M$)=20.625 M92 fiducial.  The plotted lines
show the best fit one-dimensional exponential, Plummer and King
profiles.  Note that in deriving these best fits, we are not fitting
to the binned data, but directly to the stellar distribution.
\label{fig:stellarprofile}}
\end{inlinefigure}

\clearpage

\begin{inlinefigure}
\begin{center}
\resizebox{\textwidth}{!}{\includegraphics{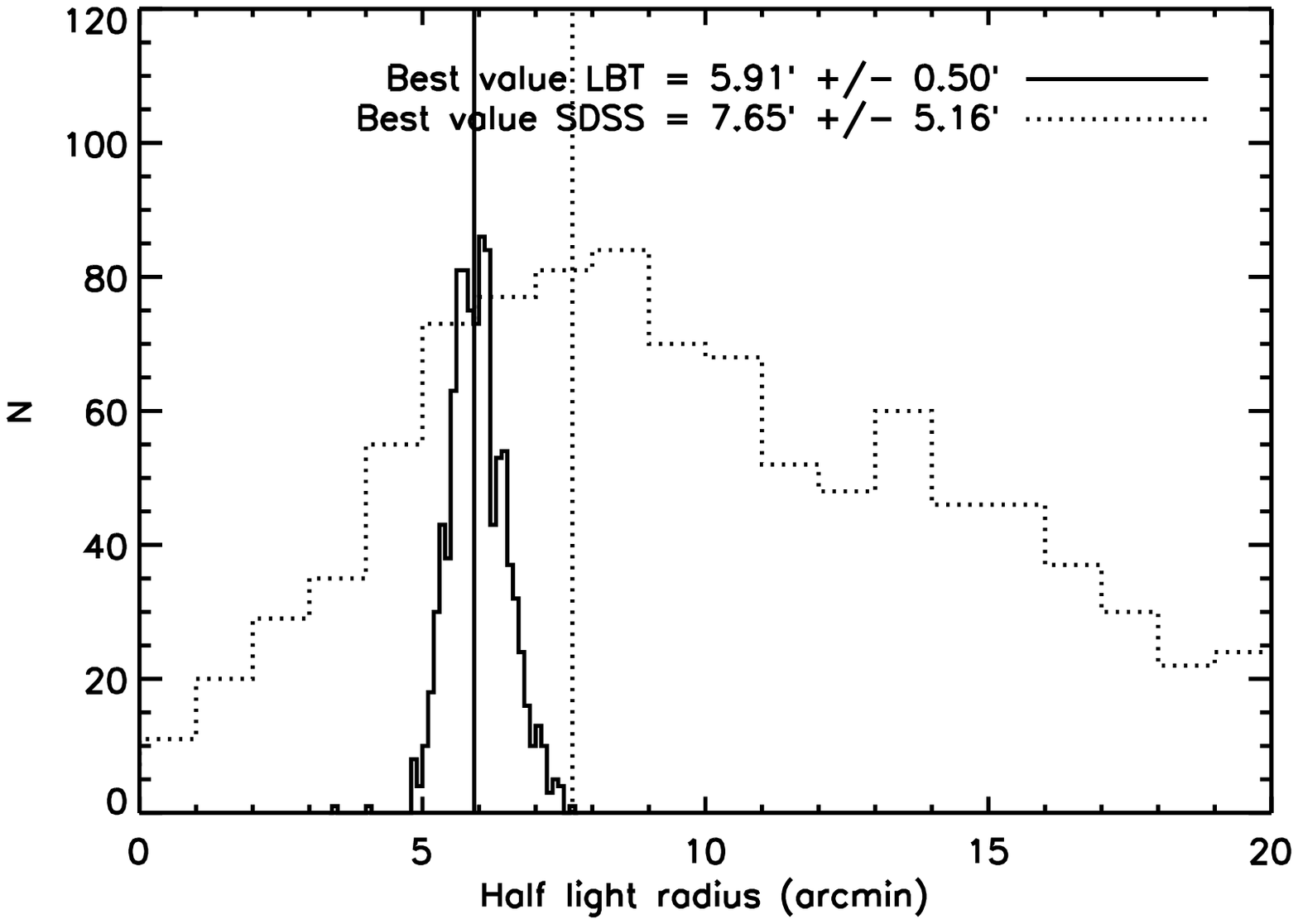}}
\end{center}
\figcaption{Histogram showing the results of our bootstrap resampling
when measuring $r_h$, using both the LBT data and that taken from
SDSS.  While the SDSS and LBT data are generally in good agreement
within the uncertainties, this clearly illustrates the need for deep
data for the new MW satellites. \label{fig:sdsscomp}}
\end{inlinefigure}

\clearpage

\begin{inlinefigure}
\begin{center}
\resizebox{\textwidth}{!}{\includegraphics{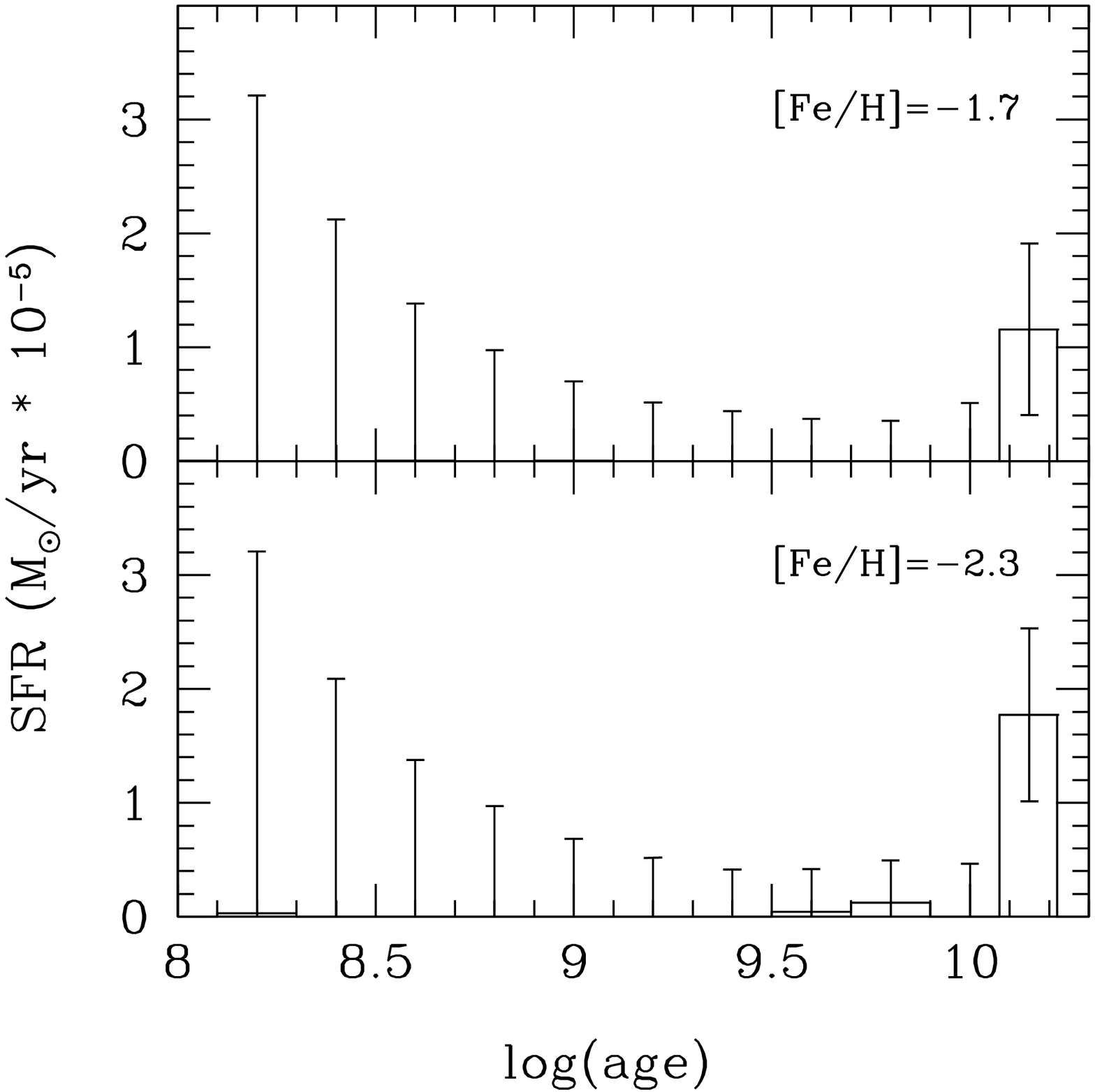}}
\end{center}
\figcaption{SFH solution from the StarFISH fit. Only the [Fe/H]=$-$2.3
and $-$1.7 bins contributed to the solution and so are the only ones
plotted here.  Hercules consists of an old metal poor population, with
some indication of metallicity spread.  Error bars with no
accompanying histogram are upper limits.\label{fig:sfh}}
\end{inlinefigure}

\clearpage

\clearpage

\clearpage

\clearpage

\clearpage

\clearpage

\clearpage

\clearpage

\begin{inlinefigure}
\begin{center}
\resizebox{\textwidth}{!}{\includegraphics{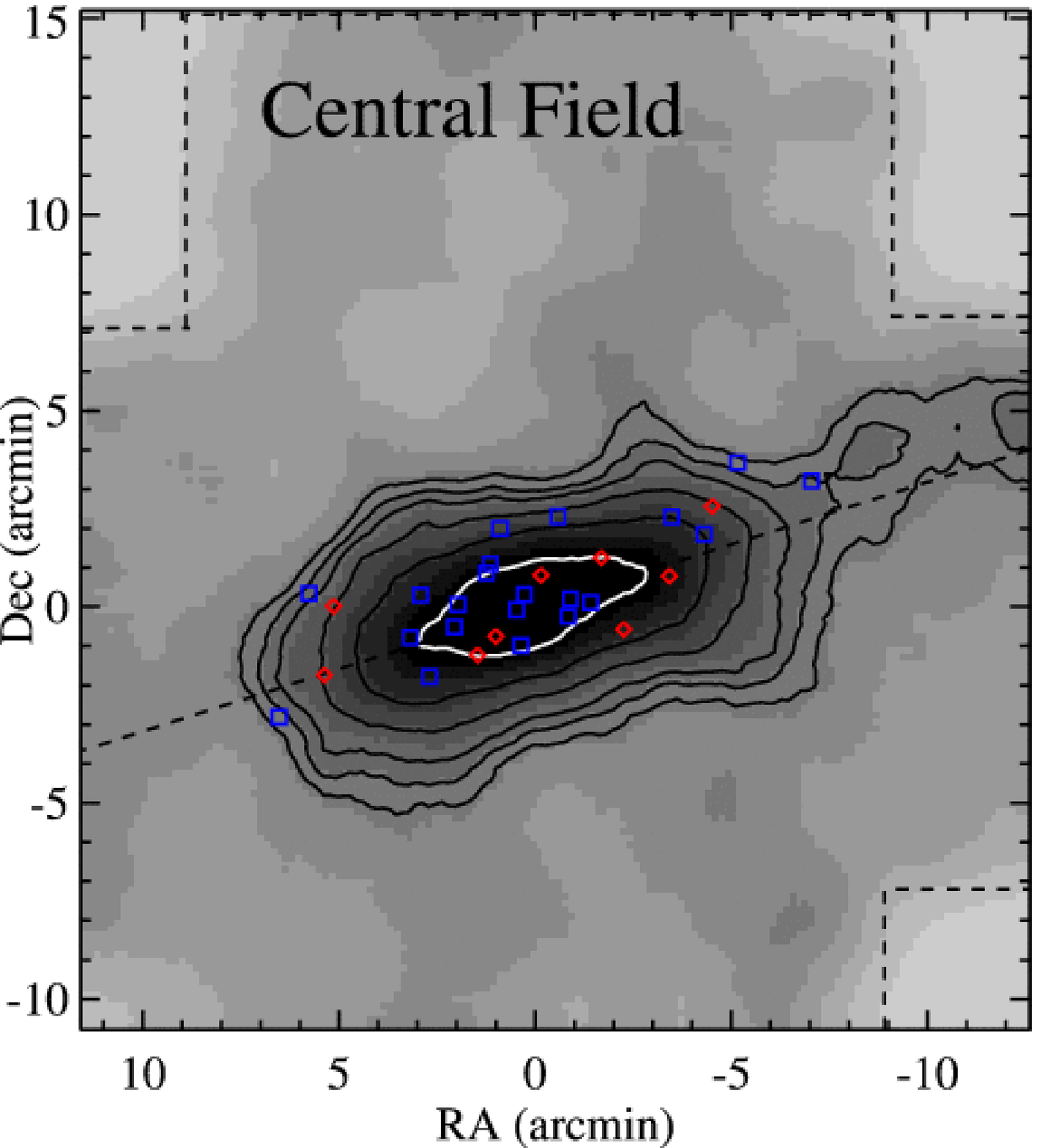}}
\end{center}
\figcaption{The positions of stars with kinematic measurements
overplotted onto the smoothed map of Hercules.  The red diamonds
coincide with the possible 'substructure' stars identified by
\citet{simongeha} between 41 and 43 $km s^{-1}$, while the blue boxes
represent the other Hercules members with velocities.  There is no
indication of spatial segregation of the kinematically interesting
points, and nor do they coincide with any particular features in
Hercules.
\label{fig:velstruct}}
\end{inlinefigure}

\clearpage

\begin{deluxetable}{llcccccccc}
\rotate
\tablewidth{0pt}
\tablecolumns{10}
\tablecaption{Summary of LBT observations and completeness by field \label{table:obsinfo}}

\tablehead{
\colhead{Pointing} & \colhead{ UT Date} & \colhead{$\alpha$} & \colhead{$\delta$} & \colhead{Filter} &\colhead{Exposure} & \colhead{Seeing} \tablenotemark{a}& \colhead{50\%} & \colhead{90\%} & \colhead{95\%}\\
\colhead{} & \colhead{} & \colhead{(J2000.0)} & \colhead{(J2000.0)} & \colhead{}&\colhead{Time (sec)} & \colhead{(arcsec)} & \colhead{Comp. (mag)} & \colhead{Comp. (mag)} & \colhead{Comp. (mag)}}
\startdata

%my pointing 9
Central & 2007 Mar 17 & 16:31:01.99 & +12:47:30.12 & $B$ &5$\times$300 & 0.8 & 26.1 & 23.9 & 21.6 \\
& 2007 Mar 17 &&&$V$ &4$\times$300 & 1.0 & 26.0 &24.2 & 21.8\\
&2008 June  1 &&&$r$ &5$\times$300 & 0.9 & 25.4 & 23.5 & 21.1\\

%my pointing 1
1& 2008 May 29 & 16:32:37.44 &+12:43:11.28 & $B$ & 6$\times$300 & 1.3 & 25.2 & 23.0 & 21.6\\
&2008 May 29 & &&$r$ & 5$\times$300 & 0.9 & 24.6 & 22.5 & 21.0 \\

%my pointing 3
2& 2008 May 30 &16:29:26.64 & +12:43:11.10 & $B$ &6$\times$300 & 1.4 & 25.2 & 23.9 & 22.0 \\
&2008 May 30 &&&$r$ & 5$\times$300 & 0.9 & 24.6 & 22.9 & 21.8 \\

%my pointing 5
3& 2008 June 2-3 &16:34:12.719 & +12:47:30.84 & $B$ &6$\times$300 & 1.2 & 25.0 & 23.4 & 21.4 \\
&2008 May 30 &&&$r$ & 6$\times$300 & 1.1 & 24.5 & 23.1 & 20.9 \\

%my pointing 7
4& 2008 June 03, May 31 &16:27:51.36 & +12:47:30.98 & $B$ &5$\times$300 & 1.2 & 25.1 & 23.0 & 21.7 \\
&2008 May 31 &&&$r$ & 5$\times$300 & 0.9 & 24.6 & 22.6 &21.2 \\
\enddata
\tablenotetext{a}{Seeing value is that of the center of the field of the combined frame. }
\end{deluxetable}

\clearpage

\begin{deluxetable}{lcccccccccccc}
\rotate
\tablewidth{0pt}
\tablecolumns{13}
\tablecaption{Hercules Photometry -- Central Field\label{table:centphot}}
\tablehead{
\colhead{Star No.} & \colhead{$\alpha$} & \colhead{$\delta$} & \colhead{$B$} &\colhead{$\delta B$} & \colhead{$A_{B}$} & \colhead{$V$} &\colhead{$\delta V$} & \colhead{$A_{V}$} & \colhead{$r$} &\colhead{$\delta r$} & \colhead{$A_{r}$} & \colhead{SDSS or LBT}\\
\colhead{} & \colhead{(deg J2000.0)} & \colhead{(deg J2000.0)} & \colhead{(mag)}&\colhead{(mag)} & \colhead{(mag)} & \colhead{(mag)} & \colhead{(mag)} & \colhead{(mag)}& \colhead{(mag)} & \colhead{(mag)} & \colhead{(mag)} & \colhead{}}
\startdata
0&247.74718&12.79046&20.16&0.02&0.26&19.17&0.02&0.20&18.81&0.01&0.17&SDSS\\
1&247.74373&12.78842&18.88&0.01&0.26&18.28&0.01&0.20&18.12&0.01&0.17&SDSS\\
2&247.74659&12.80174&18.87&0.01&0.26&18.30&0.01&0.20&18.14&0.01&0.16&SDSS\\
3&247.74539&12.78312&18.25&0.01&0.27&17.60&0.01&0.20&17.41&0.01&0.17&SDSS\\
4&247.75413&12.77563&17.59&0.01&0.27&16.78&0.01&0.21&16.52&0.01&0.17&SDSS\\
5&247.77166&12.80192&17.09&0.01&0.26&16.40&0.01&0.20&16.19&0.01&0.16&SDSS\\
6&247.77027&12.78519&17.65&0.01&0.27&16.98&0.01&0.20&16.77&0.01&0.17&SDSS\\
7&247.77106&12.78756&18.36&0.01&0.27&17.78&0.01&0.20&17.62&0.01&0.17&SDSS\\
8&247.73778&12.77706&16.71&0.01&0.27&15.91&0.01&0.20&15.65&0.01&0.17&SDSS\\
9&247.73598&12.79375&20.41&0.02&0.26&18.97&0.02&0.20&18.39&0.01&0.17&SDSS\\
\enddata
\tablenotetext{a}{See electronic edition for complete data table.}
\end{deluxetable}

\clearpage

\begin{deluxetable}{lcccccccccc}
\rotate
\tablewidth{0pt}
\tablecolumns{11}
\tablecaption{Hercules Photometry -- Adjacent Fields\label{table:fieldphot}}
\tablehead{
\colhead{Star No.} & \colhead{$\alpha$} & \colhead{$\delta$} & \colhead{$B$} &\colhead{$\delta B$} & \colhead{$A_{B}$} & \colhead{$r$} &\colhead{$\delta r$} & \colhead{$A_{r}$} & \colhead{Field No.} &\colhead{SDSS or LBT}\\
\colhead{} & \colhead{(deg J2000.0)} & \colhead{(deg J2000.0)} & \colhead{(mag)}&\colhead{(mag)} & \colhead{(mag)} & \colhead{(mag)} & \colhead{(mag)} & \colhead{(mag)} & \colhead{}&\colhead{}}
\startdata
0&248.14455&12.72711&18.77&0.01&0.24&17.94&0.01&0.15&1&SDSS\\
1&248.17113&12.73422&18.53&0.01&0.24&18.02&0.01&0.15&1&SDSS\\
2&248.15911&12.69833&17.52&0.01&0.25&16.45&0.01&0.15&1&SDSS\\
3&248.14109&12.71092&20.49&0.02&0.24&18.89&0.02&0.15&1&SDSS\\
4&248.13436&12.73258&16.44&0.01&0.24&15.51&0.01&0.15&1&SDSS\\
5&248.12870&12.71724&17.23&0.01&0.24&16.56&0.01&0.15&1&SDSS\\
6&248.15911&12.69651&20.15&0.01&0.25&18.76&0.01&0.15&1&SDSS\\
7&248.16960&12.69793&18.84&0.01&0.25&18.13&0.01&0.16&1&SDSS\\
8&248.15837&12.74721&18.62&0.01&0.24&17.49&0.01&0.15&1&SDSS\\
9&248.18539&12.70580&19.45&0.01&0.25&18.56&0.01&0.16&1&SDSS\\
\enddata
\tablenotetext{a}{See electronic edition for complete data table.}
\end{deluxetable}

\clearpage

\begin{deluxetable}{lccc}
\tabletypesize{\small}
\tablecolumns{10}
\tablecaption{Hercules structure -- parameterized fits \label{table:paramfits}}

\tablehead{
\colhead{Parameter} & \colhead{Measured} & \colhead{Uncertainty} & \colhead{Bootstrap median}\\
}
\startdata

$(m-M)_{Empirical}$ & 20.625& 0.1 & 20.625\\
$Distance_{Empricial}$ & 133.4 & 6.1 & 133.4\\

$(m-M)_{Dotter}$ & 20.60& 0.05 & 20.60 \\
$Distance_{Dotter}$ (kpc) & 131.8 & 3.0 & 131.8 \\

$(m-M)_{Girardi}$ & 20.65& 0.1 & 20.575 \\
$Distance_{Girardi}$ & 134.9 & 6.2 & 130.3 \\

$M_{V}$ & $-5.3$ & 0.4 & $-$5.3 \\
$\mu_{0,V}$ & 27.7 & 0.4 & 27.7 \\ 
\hline
Exponential Profile \\
\hline\hline
RA (h~m~s)&16:31:03.00&$\pm12''$&16:31:02.00\\
DEC (d~m~s)&+12:47:13.77&$\pm5''$&+12:47:13.83\\
%in degrees -- 247.76252 12.787157
$r_{h}$ (arcmin) &5.91&0.50&5.97\\
(pc) & 229.3 & 19.4 & 231.7 \\
$\epsilon$&0.67&0.03&0.67\\
$\theta$ (degrees)&$-$72.36&1.65&$-$72.35\\
\hline
Plummer Profile \\
\hline\hline
RA (h~m~s)&16:31:03.12&$\pm14''$&16:31:03.50\\
DEC (d~m~s)&+12:47:14.01&$\pm6''$&+12:47:15.21\\
%in degrees 247.76299 12.787224
$r_{h}$ (arcmin) &6.27&0.53&6.17\\
(pc) &243.3&20.6&239.421\\
$\epsilon$&0.67&0.03&0.67\\
$\theta$ (degrees)&$-$72.59&1.72&$-$72.47\\
\hline
King Profile \\
\hline\hline
RA (h~m~s)&16:31:03.22&$\pm14''$&16:31:02.50\\
DEC (d~m~s)&+12:47:14.11&$\pm6''$&+12:47:14.11\\
%in degrees 247.76345 12.787252
$r_c$ (arcmin)&3.59&0.44&3.80\\
(pc)&139.3&17.1& 147.5\\
$r_t$ (arcmin)&37.45&8.97&35.25\\
(pc)&1453.2&348.1&1367.8\\
$\epsilon$&0.68&0.03&0.68\\
$\theta$&$-$72.32&1.70&$-$72.34\\
\hline\hline
\\

\enddata \tablenotetext{a}{All transverse distances are reported using
the $(m-M)_{Empirical}$ = 20.625 distance modulus.}
\tablenotetext{b}{Absolute magnitude and central surface brightness are calculated using the exponential profile fit.}
\end{deluxetable}

\clearpage

\begin{deluxetable}{lcccccc}
\tabletypesize{\small}
\tablecolumns{10}
\tablecaption{Input Hercules 'Nuggets' and Detections \label{tab:fakeresult}}

\tablehead{
\colhead{Pointing} & \colhead{No. of stars} & \colhead{$M_{r}$} & \colhead{$\mu_{0,r}$} & \colhead{$M_{B}$} & \colhead{$\mu_{0,B}$} & \colhead{Peak $\sigma$} \\
}
\startdata
%1& 25 & -2.4 & 30.4 & -1.7 & 31.1 & 0.1 \\
1&50 & -2.9 & 29.8 & -2.2 & 30.6 & 2.2 \\
&100 & -4.1 & 28.7 & -3.5 & 29.2 & 2.0 \\
& 150 & -4.0 & 28.8 & -3.6 & 29.2 & 4.8 \\
& 200 & -4.5 & 28.3 & -4.3 & 28.5 & 6.1 \\
%2& 25 & -2.0 & 30.8 & -1.9 & 29.9 & 0.4 \\
2&50 & -3.6 & 29.2 & -2.2 & 30.6 & 1.8 \\
&100 & -3.7 & 29.0 & -3.3 & 29.5 & 3.3 \\
& 150 & -4.5 & 28.3 & -4.0 & 28.8 & 5.3 \\
& 200 & -4.8 & 28.0 & -4.1 & 28.7 & 8.3 \\
%3& 25 & -2.1 & 30.7 & -0.8 & 32.0 & 0.2 \\
3&50 & -3.3 & 29.5 & -2.0 & 30.8 & 2.1 \\
&100 & -3.4 & 29.4 & -3.7 & 29.1 & 3.7 \\
& 150 & -4.4 & 28.4 & -3.5 & 29.3 & 7.1 \\
& 200 & -4.7 & 28.1 & -4.0 & 28.8 & 6.9 \\
%4& 25 & -2.2 & 30.6 & -1.1 & 31.7 & 0.3 \\
4&50 & -2.6 & 30.2 & -2.8 & 29.9 & 1.6 \\
&100 & -3.7 & 29.1 & -3.1 & 29.7 & 3.8 \\
& 150 & -4.5 & 28.3 & -3.6 & 29.2 & 4.0 \\
& 200 & -4.9 & 27.9 & -4.1 & 28.7 & 9.0 \\
\enddata \tablenotetext{a}{All nuggets have an exponential profile with half light radius of 3 arcminutes.}

\end{deluxetable}

\clearpage

\begin{deluxetable}{lccccccccc}
%\tabletypesize{\small}
\tablecolumns{11}
\tablecaption{Representative orbital parameters for Hercules
\label{table:orbit}}

\tablehead{
\colhead{Case} & \colhead{$V_{GSR,rad}$} & \colhead{$V_{GSR,tan}$} &\colhead{$R_{peri}$} & \colhead{$R_{apo}$} & \colhead{$T$}  & \colhead{$e$}& \colhead{Inclination} & \colhead{$\mu_{\alpha}$} & \colhead{$\mu_{\delta}$} \\
&\colhead{(km s$^{-1}$)} & \colhead{(km s$^{-1}$)} & \colhead{(kpc)} &\colhead{(kpc)} & \colhead{(Gyr)}&\colhead{} & \colhead{degree} & \colhead{(marcsec/cent)} & \colhead{(marcsec/cent)}
}
\startdata
1&142&20&5&167&2.2&0.95&83&-19.64&-24.2\\
2&141&50&13&168&2.3&0.86&89&-15.11&-25.7\\
3&140&105&36&176&2.6&0.66&92&-6.83&-28.3\\
%3'&145&96&32&179&2.6&0.70&83&-36.97&-18.71\\
4&137&204&86&245&4.0&0.48&93&8.23&-33.1\\
5&143&11&3&167&2.2&0.97&135&-21.18&-21.24\\
6&144&40&11&169&2.3&0.88&141&-19.74&-16.71\\
7&146&95&32&179&2.6&0.70&141&-17.11&-8.43\\
8&150&195&79&249&4.0&0.52&141&-12.32&6.63
\enddata

\end{deluxetable}

\end{document}